\RequirePackage{fix-cm}
\documentclass[twocolumn,epjc3]{svjour3}  
\smartqed  % flush right qed marks, e.g. at end of proof
\RequirePackage{graphicx}
\journalname{Eur. Phys. J. C}
\usepackage{cite} %A: para contraer referencias
\usepackage{amsmath}%A
\usepackage{hyperref}
\usepackage{cite}%A
\usepackage{amsmath,amssymb}%A
\usepackage{color}%A
\usepackage{subfigure}
\usepackage{float}

\begin{document}
\title{Stellar models with like--Tolman IV complexity factor}


\author{J. Andrade
\thanksref{        addr3} 
        \and
        E. Contreras\thanksref{e3,addr3} 

}
\thankstext{e3}{e-mail: 
\href{mailto:econtreras@usfq.edu.ec}{\nolinkurl{econtreras@usfq.edu.ec}}}

\institute{Departamento de F\'isica, Colegio de Ciencias e Ingenier\'ia, Universidad San Francisco de Quito,  Quito, Ecuador.\label{addr3}
}
\date{Received: date / Accepted: date}

\maketitle

\begin{abstract}
In this work, we construct stellar models ba-\break
sed on the complexity factor as a supplementary condition
which allows to close the system of differential equations arising from the Gravitational Decoupling. The assumed complexity is a generalization of the one obtained from the well known Tolman IV solution. We use Tolman IV, Wyman IIa, Durgapal IV and Heintzmann IIa as seeds solutions. Reported compactness parameters of SMC X-1 and Cen X-3 are used to study the physical acceptability of the models. Some aspects related to the density ratio are also discussed. 

\end{abstract}
%%%%%%%%%%%%%%%%%%%%%%%%%%%%%%%%%%%%%%%%%%

\section{Introduction}\label{intro}
For a long time, stellar models 
were considered to be supported by 
Pascalian fluids (equal principal stresses); approximation which resulted to be appropriate to describe a variety of circumstances. However, now is well--known that for certain ranges of the density there are some
physical phenomena which might take place leading to local anisotropy in the configuration. (see Refs. \cite{CHH,14, hmo02, 04, LH-C3, hod08, hsw08, p1, p2, anis1, anis2, anis3, anis4bis,n1,n2,n3}, for discussions on this point). Among all these possibilities, we could mention: i) intense magnetic field observed in compact objects such as white dwarfs, neutron stars, or magnetized strange quark stars  (see, for example, Refs. \cite{15, 16, 17, 18, 19,23, 24, 25, 26}) and ii) viscosity (see \cite{Anderson, sad, alford, blaschke, drago, jones, vandalen, Dong} and references therein).  Besides, it  has  been recently proven that the presence  of dissipation, energy  density inhomogeneities and  shear yield the isotropic pressure condition becomes unstable  \cite{LHP}. Based on these points, the renewed  interest in the study of fluids not satisfying the isotropic condition is clear and justifies our present work on the construction of anisotropic models \cite{Lopes:2019psm, Panotopoulos:2020zqa, Bhar:2020ukr, Tello-Ortiz:2020svg, Tello-Ortiz:2020nuc}.\\

The strategies to construct anisotropic solutions are many but recently, the well known Gravitational Decoupling (GD) \cite{ovalle2017} by the Minimal Geometric Deformation approach (MGD)  (for implementation in $3+1$ and $2+1$ dimensional spacetimes see
\cite{ovalle2014,ovalle2015,Ovalle:2016pwp,ovalle2018,ovalle2018bis,estrada2018,ovalle2018a,lasheras2018,estrada,rincon2018,ovalle2019a,ovalleplb,tello2019,lh2019,estrada2019,gabbanelli2019,sudipta2019,linares2019,leon2019,casadioyo,tello2019c,arias2020,abellan20,tello20,rincon20a,jorgeLibro,Abellan:2020dze,Ovalle:2020kpd,Contreras:2021yxe,Heras:2021xxz,contreras-kds,tello2021w,darocha1,darocha2,darocha3,sharif1,sharif2,sharif3,estrada2021}. For recent developments see \cite{Zubair:2021zqs, Azmat:2021qig, Muneer:2021lfz, Zubair:2020lna, Maurya:2021jod, Meert:2021khi, Sultana:2021cvq, Maurya:2021fuy }) has been broadly implemented to extend isotropic solutions to anisotropic domains. In static and spherically symmetric spacetimes, the are only three independent Einstein field equations but five unknown, namely two metric functions, the density energy and the radial and tangential pressures. However,
the GD demands the assumption of a seed solution which
allows to decrease the number of degrees of freedom 
and, as a consequence, only one extra condition is required to close the system. In this sense, a key point in the implementation of MGD is to provide such an auxiliary condition which could be the mimic constraint for the pressure and the density, regularity condition of the anisotropy function, barotropic equation of state, among others. In this work we use the recently introduced definition of complexity  for self--gravitating fluids \cite{complex1} and, in particular, we propose a like-Tolman IV complexity factor.\\

This work is organized as follows. The next section is devoted to reviewing the main aspects of GD by MGD. In section \ref{complexityF} we introduce the concept of complexity and obtain an expression for the complexity factor in GD. Then, in section \ref{stellar} we calculate and generalize the complexity factor from the Tolman
IV solution and implement this result with the aim to construct extension of Tolman IV, Wyman IIa, Durgapal IV and Heintzmann IIa. Section \ref{discussion} is devoted to interpreting and discussing our results and some comments and final remarks are given in the last section.

\section{Gravitational decoupling}\label{GD}
In this section we introduce the GD by MGD (for more details, see \cite{ovalle2017}). Let us start with the Einstein field equations (EFE)
\begin{eqnarray}\label{EFE}
G_{\mu\nu}=R_{\mu\nu}-\frac{1}{2}g_{\mu\nu}R=8\pi \tilde{T}_{\mu\nu},
\end{eqnarray}
with
\begin{equation}\label{energy-momentum}
    \tilde{T}_{\mu\nu} = T^{(s)}_{\mu\nu} + \alpha\theta_{\mu\nu}\;,
\end{equation}
where $T^{(s)}_{\mu\nu}$ represents the matter content of a known solution \footnote{In this work we shall use $c=G=1$.}, namely the {\it seed} sector,  and $\theta_{\mu\nu}$ describes an extra source coupled through the parameter $\alpha$. Note that, since the Einstein tensor fulfills the Bianchi identities, the total energy--momentum tensor satisfies
\begin{equation}\label{divergencia-cero-total}
    \nabla_{\mu} \tilde{T}^{\mu\nu} = 0\;.
\end{equation}
It is important to point out that, whenever 
$\nabla_\mu T^{\mu\nu(s)} = 0$, the condition
\begin{equation}\label{divergencia-cero-theta}
    \nabla_\mu \theta^{\mu\nu} = 0\;,
\end{equation}
is automatic and as a consequence, there is no exchange of energy-momentum  between the seed solution and the extra source $\theta^{\mu\nu}$ so that the interaction is entirely gravitational.\\

In a static and spherically symmetric spacetime 
sourced by 
\begin{eqnarray}
T^{\mu(s)}_{\nu}&=&\textnormal{diag}(\rho^{(s)},-p_{r}^{(s)},-p^{(s)}_{t},-p^{(s)}_{t})\label{tmunu}\\
\theta^{\mu}_{\nu}&=&\textnormal{diag}(\theta_{0}^{0},\theta_{1}^{1},\theta_{2}^{2},\theta_{2}^{2}),\label{thetamunu}
\end{eqnarray}
and a metric given by 
\begin{eqnarray}\label{metric}
ds^{2}=e^{\nu}dt^{2}-e^{\lambda}dr^{2}
-r^{2}(\theta^{2}+\sin^{2}\theta d\phi^{2}),
\end{eqnarray}
Eqs. (\ref{EFE}) and (\ref{energy-momentum}) lead to
\begin{eqnarray}
      \tilde{\rho} &=& \frac{1}{8\pi}\left[\frac{1}{r^2} +
        e^{-\lambda}\!\left(\frac{\lambda'}{r} - \frac{1}{r^2}\right)\right] \!,\label{mgd05}\\
           \tilde{P}_{r} &=& \frac{1}{8\pi}\left[-\frac{1}{r^2} +
        e^{-\lambda}\!\left(\frac{\nu'}{r} + \frac{1}{r^2}\right)\!\right],\label{mgd06}\\
           \tilde{P}_{t} &=& \frac{1}{32\pi}e^{-\lambda}\!\left(2\nu'' + {\nu'}^2 - \lambda'\nu' + 2\frac{\nu'-\lambda'}{r} \right),\!\label{mgd07}
\end{eqnarray}
where we have defined \footnote{Note that the matter sector has dimensions of a length squared in the units we are using}
\begin{eqnarray}
    \tilde{\rho} &=& \rho^{(s)} + \alpha\theta_{0}^{0} \;,\label{mgd07a}\\
    \tilde{P}_{r} &=& p_r^{(s)}-\alpha\theta_{1}^{1}   \;,\label{mgd07b}\\
    \tilde{P}_{t} &=& p_{t}^{(s)} -\alpha\theta_{2}^{2} \;.\label{mgd07c}
\end{eqnarray}
Is clear that given the non-linearity of Einstein’s equations, the decomposition (\ref{energy-momentum}) does not lead to two set of decoupled equations; one for each source involved. Nevertheless, contrary to the broadly belief, such a decoupling is possible, to some extent, in the context of MGD as we shall demonstrate in what follows.\\

Let us introduce a geometric deformation in the metric functions given by 
\begin{eqnarray}
    \nu\;\; &\longrightarrow &\;\; \xi + \alpha g,\label{comp-temporal} \\
    e^{-\lambda}\;\; &\longrightarrow &\;\; e^{-\mu} + \alpha f\;,\label{comp-radial}
\end{eqnarray}
where $\{f,g\}$ are the so-called decoupling functions and $\alpha$ is the same free parameter that ``controls'' the influence of $\theta_{\mu\nu}$ on $T_{\mu\nu}^{(s)}$ in Eq. (\ref{energy-momentum}). In this work we shall concentrate in the particular case $g = 0$ and $f \ne 0$. Now, replacing (\ref{comp-temporal}) and (\ref{comp-radial}) in the system (\ref{mgd05}-\ref{mgd07}), we are able to split the complete set of differential equations into two subsets: one describing a seed sector sourced by the conserved energy-momentum tensor, $T_{\mu\nu}^{(s)}$
    \begin{eqnarray}
     \rho^{(s)} &=& \frac{1}{8\pi}\left[\frac{1}{r^2} +
        e^{-\mu}\!\left(\frac{\mu'}{r} - \frac{1}{r^2}\right)\right]\!,\label{mgd13}
        \\
     p_r^{(s)} &=& \frac{1}{8\pi}\left[-\frac{1}{r^2} +
        e^{-\mu}\!\left(\frac{\nu'}{r} + \frac{1}{r^2}\right)\right]\!,\label{mgd14}
        \\
       p_{t}^{(s)} &=& \frac{1}{32\pi}e^{-\mu} \!\! \left( 2\nu'' + {\nu'}^2 
    - \mu' \nu' + 2\frac{\nu'-\mu'}{r} \right) \!,\nonumber\\\label{mgd15}
\end{eqnarray}
and the other set corresponding to quasi-Einstein field equations sourced by $\theta_{\mu\nu}$
 \begin{eqnarray}
    \theta^0_0 &=& \frac{1}{8\pi}\left[- \frac{f}{r^2} -\frac{f'}{r}\right]\,,\label{mgd16}\\
    \theta^1_1 &=& \frac{1}{8\pi}\left[-f
        \left(\frac{\nu'}{r} + \frac{1}{r^2}\right)\right]\!,\label{mgd17}\\
    \theta^2_2 &=& \frac{1}{8\pi}\left[-\frac{f}{4} \left( 2\nu'' + 
    {\nu'}^2 + 2\frac{\nu'}{r}\right)
    %\nonumber\\
    %& & \hspace{.5cm} 
    -\frac{f'}{4} \left( \nu' + \frac{2}{r} \right)\right]\!.\nonumber\\\label{mgd18}
    \end{eqnarray}
As we have seen, the components of $\theta_{\mu\nu}$ satisfy the conservation equation $\nabla_\mu \theta^{\mu}_{\nu}=0$, namely
\begin{eqnarray}\label{consthe}
    \theta'^1_{1} 
     - \frac{\nu'}{2} (\theta^0_0 - \theta^1_1)      - \frac{2}{r}(\theta^2_2 - \theta^1_1)=0.
\end{eqnarray}

In this work, we consider that the interior configuration is surrounded by the Schwarzschild vacuum so that, on the boundary surface $\Sigma$, we require
\begin{eqnarray}
    e^{\nu}\Big|_{\Sigma^{-}} &=& \left(1 - \frac{2M}{r}\right)\Bigg|_{\Sigma^{+}}\,,\label{mgd11a}\\
    e^{\lambda}\Big|_{\Sigma^{-}} &=& \left(1 - \frac{2M}{r}\right)\Big|_{\Sigma^{-}}\,,\label{mgd11b}\\
     \tilde{P}_r(r) \Big|_{\Sigma^{-}} &=&      \tilde{P}_r(r) \Big|_{\Sigma^{+}}=0\,,\label{mgd11c}
\end{eqnarray}
which corresponds to the continuity of the first and second fundamental form across that surface of the star. \\

To conclude this section, we emphasize the importance of GD as a useful tool to find solutions of EFE. As it is well known, in static and spherically symmetric spacetimes sourced by anisotropic fluids, EFE reduce to three equations given by  (\ref{mgd05}), (\ref{mgd06}) and (\ref{mgd07}) and five unknowns, namely $\{\nu,\lambda,\tilde{\rho},\tilde{P}_{r},\tilde{P}_{t}\}$. In this regard, two auxiliary conditions must be specified, namely metric conditions, equations of state, etc. Nevertheless, as in the context of GD a seed solution must be given, the number of degrees of freedom reduces from five to four and, as a consequence, only one extra condition is required. In general, this condition have been implemented in the decoupling sector given by Eqs. (\ref{mgd16}), (\ref{mgd17}) and (\ref{mgd18}) as some equation of state which leads to a differential equation for the decoupling function $f$. In this work, we take an alternative route to obtain the decoupling function; namely, the complexity factor that we shall introduce in the next section as a supplementary condition of the total solution.

\section{Complexity of compact sources}\label{complexityF}
Recently, a new definition of complexity for self--\break gravitating fluid distributions has been introduced in
Ref. \cite{complex1} which is based on the idea that the least complex gravitational system is the one supported by a homogeneous energy density distribution and isotropic pressure. In this direction, there is a scalar associated with the orthogonal splitting of the Riemann tensor \cite{LH-C2} in static and spherically symmetric space--times which encodes the intuitive idea of complexity, namely
\begin{eqnarray} \label{YTF2}
Y_{TF} = 8\pi \Pi - \frac{4\pi}{r^3}\int^{r}_{0} \tilde{r}^3 \rho' d\tilde{r},
\end{eqnarray}
with $\Pi\equiv P_{r}-P_{t}$. Also, it can be demonstrated that in terms of Eq. (\ref{YTF2}) the Tolman mass reads
\begin{eqnarray} \label{m_T}
m_{T} = (m_{T})_{\Sigma}\left(\frac{r}{r_{\Sigma}}\right)^3 + r^3\int^{r_{\Sigma}}_{r} \frac{e^{( \nu + \lambda )/2}}{{\tilde{r}}} Y_{TF} d\tilde{r},
\end{eqnarray}
so that $Y_{TF}$ enclose the modifications on the active gravitational mass produced by the energy density inhomogeneity and the anisotropy of the pressure. \\

It is worth noticing that the vanishing complexity condition ($Y_{TF}=0$) can be satisfied
not only in the simplest case of isotropic and homogeneous system but in all the cases where
\begin{eqnarray}
\Pi=\frac{1}{2r^{3}}\int\limits_{0}^{r}\tilde{r}^{3}\rho' d\tilde{r},
\end{eqnarray}
which provides a non--local equation of state that can be used as a complementary condition to close the system of EFE (for a recent implementation, see \cite{casadioyo}, for example). However, given that this condition could fail
in some cases in the construction of specific stellar models,
non--vanishing values of $Y_{TF}$ must be supplied. An example of how this can be achieved can be found in \cite{casadioyo}.\\

In this work we shall use the complexity factor as a supplementary condition for the total sector so  we replace  (\ref{comp-temporal}), (\ref{comp-radial}) in (\ref{YTF2}) and use (\ref{mgd05}), (\ref{mgd06}) and (\ref{mgd07}) to obtain
\begin{eqnarray}
\frac{\alpha \xi'}{4}f'&+&\frac{\alpha}{2}\left(\xi''-\frac{\xi'}{r}+\frac{\xi'^2}{2}\right)f\nonumber\\
&& +\frac{e^{-\mu}}{2}\left(\xi''-\frac{\xi'}{r}+\frac{\xi'^2}{2}-\frac{\mu'\xi'}{2}\right) + Y_{TF}=0.\label{diffeq}
\end{eqnarray}
Note that as far as the pair $\{\xi,\mu\}$ is specified (the seed solution), Eq. (\ref{diffeq}) becomes
a differential equation for the decoupling function $f$ when a value of $Y_{TF}$ is specified. 

\section{Stellar models with like Tolman IV complexity}\label{stellar}
In this work, we construct interior solutions based  on Tolman IV, Wyman IIa, Durgapal IV and Heintzmann IIa as isotropic seeds in the framework of GD by using the complexity factor as supplementary condition. At first sight, the vanishing complexity seems to be straightforward but it can be demonstrated that such a condition fails for the seeds under consideration in this work. As an alternative, we  generalize the complexity factor of the well--known Tolman IV solution.\\

As it is well--known, Tolman IV reads \cite{Tolman1939}
\begin{eqnarray}
e^{\nu}&=&B_{0}^2\left(1+\frac{r^2}{A_{0}^2}\right)\label{mtolmanradial}\\
e^{-\lambda}&=&\frac{(C_{0}^2-r^2)(A_{0}^2+r^2)}{C_{0}^2(A_{0}^2+2r^2)}\label{mtolmantemporal}\\
\rho&=&\frac{3A_{0}^4+A_{0}^2(3C_{0}^2+7r^2)+2r^2(C_{0}^2+3r^2)}{8\pi C_{0}^2(A_{0}^2+2r^2)^2}\label{rhoTIV}\\
p&=&\frac{C_{0}^2-A_{0}^2-3r^2}{8\pi C_{0}^2(A_{0}^2+2r^2)}\label{prTIV},
\end{eqnarray}
where $A_{0}$ and $C_{0}$ are constants with dimension of a length and $B_{0}$ is a dimensionless constant.
Now, replacing (\ref{mtolmanradial}), (\ref{mtolmantemporal}), (\ref{rhoTIV}) and (\ref{prTIV}) in (\ref{YTF2}) we arrive at

\begin{eqnarray}\label{ytft4}
Y_{TF}=\frac{r^2(A_{0}^2+2C_{0}^2)}{C_{0}^2(A_{0}^2+2r^2)^2},
\end{eqnarray}
which has dimensions of the inverse of a length squared. Note that Eq. (\ref{ytft4}) can be easily generalized to  \begin{eqnarray}\label{YTFTIV}
Y_{TF}= \frac{a_{1}r^2}{(a_{2}+a_{3}r^2)^2}
\end{eqnarray}
where $a_{1}$ and $a_{3}$ are arbitrary dimensionless constants and $a_{2}$ must be a constant with dimension of a length squared. It should be emphasized that reason for introducing the set $\{a_{1},a_{2},a_{3}\}$ is nothing but to generalize the complexity factor (\ref{ytft4}). 
In what follows we shall consider Eq. (\ref{YTFTIV}) as the condition to close the system and generate anisotropic models from isotropic seeds. 

\subsection{Model 1: like--Tolman IV solution}
Replacing (\ref{mtolmanradial}) and (\ref{mtolmantemporal}) in
(\ref{diffeq}) and using (\ref{YTFTIV}) we arrive at
\begin{eqnarray}\label{fmodel1}
f&=&\left(A_{0}^2+r^2\right)\bigg[ c_1 + \frac{1}{\alpha}\bigg(\frac{a_1}{a_3\zeta(r)}+
\nonumber\\
&&\hspace{3.5cm}-\frac{A_{0}^2+2C_{0}^2}{2C_{0}^2(A_{0}^2 + 2r^2)}\bigg)\bigg],
\end{eqnarray}
where $c_{1}$ is an integration constant with dimensions of the inverse of a length squared  and $\zeta(r)$ is an auxiliary function with dimensions of a length squared
(see Appendix, section \ref{appendix-m2}). 

It can be shown that to ensure regularity in the matter sector the constant $c_{1}$ must satisfy
\begin{eqnarray}\label{c1model1}
c_1= \frac{-2 a_1 A_{0}^2 C_{0}^2 + a_2 a_3 A_{0}^2 + 2 a_2 a_3 C_{0}^2}{2 \alpha a_2 a_3 A_{0}^2 C_{0}^2}.
\end{eqnarray}
Replacing (\ref{fmodel1}) in (\ref{comp-radial}) and using (\ref{c1model1})
we find
\begin{equation}\label{explambda1}
e^{-\lambda}=\frac{\left(A_{0}^2+r^2\right) \left(2 a_1 C_{0}^2+ a_3 \zeta(r)\left(2\alpha c_1 C_{0}^2-1\right)\right)}{2 a_3 C_{0}^2 \zeta(r)}.
\end{equation}

Now, from (\ref{mgd05}), (\ref{mgd06}), (\ref{mgd07}) we arrive at
\begin{eqnarray}
\tilde{\rho}&=&\frac{1}{8\pi a_2 A_{0}^2\zeta(r)^2}\bigg[a_1 A_{0}^4\left(3a_2+a_3 r^2\right)\nonumber\\
&&\hspace{1.3cm} + a_1 A_{0}^2 r^2\left(5a_2+3a_3 r^2\right) -3a_2\zeta(r)^2\bigg]\label{densitymodel1}\\
\tilde{P}_{r}&=&\frac{3 a_2 \zeta(r)- a_1 A_{0}^2(A_{0}^2 + 3 r^2)}{8\pi a_2 A_{0}^2 \zeta(r)}\label{radialpmodel1}\\
\tilde{P}_{t}&=&\frac{3 a_2 \zeta(r)^2-a_1 a_2 A_{0}^4 - a_1 A_{0}^2 r^2 \left(5 a_2+3 a_3 r^2\right)}{8\pi a_2 A_{0}^2 \zeta(r)^2},\label{tangentialpmodel1}
\end{eqnarray}

Finally, the continuity of the first and the second fundamental form leads to
\begin{eqnarray}
a_1 &=& \frac{3 a_2 \zeta(R)}{A_{0}^2(A_{0}^2 + 3 R^2)}.\\
\frac{A_{0}^2}{R^2} &=& \frac{R-3M}{M}\label{A0}\\
B_{0}^2 &=& 1-\frac{3M}{R}\label{B0}.
\end{eqnarray}
Note that the free parameters are $\{R,M,a_{2},a_{3}\}$ (see Appendix \ref{appendix-m2} where $a_{3}$ appears explicitly). It is worth mentioning that from (\ref{A0}) and (\ref{B0}) 
is clear that compactness satisfies $M/R<1/3$, which corresponds to a 
more stringent condition when compared to the the Buchdahl's limit ($M/R<4/9$) for isotropic solutions. More precisely, the solutions allowed with this model should be less compact given that the interval $1/3\le M/R < 4/9$ is forbidden.\\

As we shall see later, the strategy to explore the feasibility of our solution will be to specify the compactness parameter associated with SMC X-1 and Cen X-3 in order to set suitable values for $a_{2}$ and $a_{3}$ (see section \ref{discussion} for details)

\subsection{Model 2: like--Wyman IIa solution}
In this case we consider the Wyman IIa metric \cite{delgaty1998physical, wyman1949radially} with $n=2$ as a seed solution which reads
\begin{eqnarray}
e^{\xi(r)}&=&(A-Br^2)^2\label{temporalwyman}\\
e^{-\mu(r)}&=& 1 + C r^2(A-3Br^2)^{-2/3}\label{radialwyman},
\end{eqnarray}
where $A$ is a dimensionless constant and
$B$ and $C$ are constants with dimensions of the inverse of a length squared. It is worth mentioning that all the results below will be written in terms of auxiliary functions $\{\zeta,\chi,\varrho_{1},
\mathcal{P}_{1},\mathcal{P}_{2}\}$ which are defined in Appendix, section \ref{appendix-m2}.\\

Following the same procedure that in the previous section we obtain

\begin{eqnarray}
f&=& \frac{r^2}{2\alpha a_3}\bigg[\frac{a_1(a_3 A + a_2 B)}{a_2 B \zeta(r)} -\frac{2 a_3 C}{(A-3B r^2)^{2/3}}\bigg]\nonumber\\
&& -\frac{a_1}{2\alpha a_{3}^2}\chi(r),
\end{eqnarray}
from where
\begin{eqnarray}\label{radialmetricmodel2}
e^{-\lambda}&=&\frac{1}{2 a_{3}^2}\bigg[\frac{a_3 r^2\left(a_1 a_3 A + a_2 B \left(a_1 + 2 a_{3}^2\right)\right)}{a_2 B \zeta(r)}
\nonumber\\
&&\hspace{3.4cm} +\frac{2 a_2 a_3^2}{\zeta(r)} -a_1\chi(r)\bigg],
\end{eqnarray}

From \ref{mgd05}, \ref{mgd06} and \ref{mgd07} the matter sector reads
\begin{eqnarray}
\tilde{\rho}&=&\frac{a_1}{16\pi a_{3}^2}\left[\frac{\chi(r)}{r^2}-\frac{a_3 \varrho_{1}(r)}{a_2 B \zeta(r)^2}\right]\label{densitymodel2}\\
\tilde{P}_{r}&=&\frac{a_3 r^2\mathcal{P}_{1}(r) -a_1 a_2 B \left(5 B r^2-A\right) \zeta(r)\chi(r)}{16\pi a_2 a_{3}^2 B r^2 \left(B r^2-A\right)\zeta(r)}\label{radialpmodel2}\\
\tilde{P}_{t}&=&\frac{a_3\mathcal{P}_{2}(r) -4 a_1 a_2 B^2 \zeta(r)^2 \chi(r)}{16\pi a_2 a_{3}^2 B\left(B r^2-A\right)\zeta(r)^2}\label{tangentialpmodel2}.
\end{eqnarray}

Continuity of the first and the second fundamental form leads to
\begin{eqnarray}
a_1 &=&\frac{8 a_2 a_{3}^2 B^2 R^2\left(A-5 B R^2\right)^{-1}\zeta(R)}{\left[ a_3 R^2 (a_3 A + a_2 B)-a_2 B \zeta(R)\chi(R)\right]}\\
A^2&=& \frac{B^2 (5 M - 2 R)^2 R^4}{M^2}\label{A2}\\
B^2&=&\frac{M^2}{4 R^5 (R-2 M)}\label{B2}.
\end{eqnarray}
As in the previous case, the free parameters are\break $\{R,M,a_{2},a_{3}\}$.
Note that, as $\{\zeta,\varrho_{1},\mathcal{P}_{2}\}$ have dimensions of a length squared and $\{\chi,\mathcal{P}_{1}\}$ are dimensionless (see Appendix \ref{appendix-m2}) , all the expressions above are dimensionally correct. It
is worth noticing that form (\ref{B2}), $R>2M$ which is in accordance with the restriction that any stable configuration should be greater than its Schwarzschild radius. Furthermore, (\ref{A2}) leads to $M/R\ne 2/5$ or the metric becomes degenerated, namely $g_{tt}=0,\ \forall r$.

\subsection{Model 3: like--Durgapal IV solution}
In this case we consider the Durgapal IV metric \cite{delgaty1998physical, durgapal1982class} as a seed solution, which reads
\begin{eqnarray}
e^{\xi}&=&A(C r^2+1)^4\label{temporaldurg}\\
e^{-\mu}&=&\frac{7-10Cr^2-C^2r^4}{7(C r^2+1)^2}\nonumber\\
&&+\frac{Cr^2}{(C r^2+1)^2(1+5Cr^2)^{2/5}}\label{radialdurg}
\end{eqnarray}
where $A$ is a dimensionless constants and $C$ is a constant with dimensions of the inverse of a length squared. In what follows, we shall use the auxiliary functions 
$\{\zeta,\eta_{1},\beta_{1},\beta_{4},\beta_{5},\beta_{8},\beta_{11},\varrho_{2},\mathcal{P}_{3}$ $,\mathcal{P}_{4}\}$ defined in the Appendix, section \ref{appendix-m2}.\\

In this case,
\begin{eqnarray}
f&=& \frac{r^2}{56\alpha a_{3}^3 C \beta_{1}(r)^2}\bigg[ 7 a_1\bigg(-a_3 C^3 r^2 + \frac{2 (a_2 C-a_{3})^3}{a_2 \zeta(r)}\nonumber\\
&& + 2 C^2(2 a_2 C-3 a_3)\bigg)\nonumber\\
&& + 8 a_{3}^3 C^2 \bigg(C r^2 + 10 -\frac{7B}{\left(5 C r^2+1\right)^{2/5}}\bigg)\bigg]\nonumber\\
&& - \frac{3 a_1 (a_3-a_2 C)^2 \chi(r)}{4\alpha a_{3}^4 \beta_{1}(r)^2},
\end{eqnarray}
and the radial metric reads
\begin{eqnarray}\label{radialmetricmodel3}
e^{-\lambda}=\frac{a_3\eta_{1}(r) -6 a_1 a_2 C (a_3-a_2 C)^2 \zeta(r)\chi(r)}{8 a_2 a_{3}^4 C\beta_{1}(r)^2 \zeta(r)}.
\end{eqnarray}
From (\ref{mgd05}), (\ref{mgd06}), (\ref{mgd07}) we obtain
\begin{eqnarray}
\tilde{\rho}&=&\frac{1}{64\pi a_2 a_{3}^4 C r^2 \beta_{1}(r)^3 \zeta(r)^2}\bigg[a_3 r^2\varrho_{2}(r)\nonumber\\
&&\hspace{0.8cm}- 6 a_1 a_2 C \beta_{4}(r) (a_3-a_2 C)^2 \zeta(r)^2\chi(r) \bigg]\label{densitymodel3}\\
\tilde{P}_{r}&=&\frac{\mathcal{P}_{3}(r)}{64\pi a_2 a_{3}^4 C r^2\beta_{1}(r)^3 \zeta(r)}\label{radialpmodel3}\\
\tilde{P}_{t}&=&\frac{\mathcal{P}_{4}(r)}{32\pi a_2 a_{3}^4 C\beta_{1}(r)^3 \zeta(r)^2},\label{tangentialpmodel3}
\end{eqnarray}
and the matching conditions lead to
\begin{eqnarray}
a_1&=&\frac{8 a_2 a_{3}^4 C^2 R^2 \left(6-C R^2 \beta_{5}(R)\right)\zeta(R)\beta_{8}(R)^{-1}}{\left[a_3 R^2\beta_{11}(R)+6 a_2 C (a_3-a_2 C)^2 \zeta(R)\chi(R)\right]},\\
A^2&=&\frac{1-\frac{2 M}{R}}{\left(\frac{M}{4 R-9 M}+1\right)^4}\label{A3}\\
C^2 &=& \frac{M}{R^2 (4 R-9 M)}\label{C3}.
\end{eqnarray}
Note that, as $\{\zeta,\eta_{1},\varrho_{2},\mathcal{P}_{3}\}$ have dimensions of a length squared and $\{\beta_1,\beta_4,\beta_5,\beta_8,\beta_{11}\}$ are dimensionless (see Appendix \ref{appendix-m2}), all the expressions above have the correct dimensions. In this case, it is clear that from Eqs. (\ref{A3}) and (\ref{C3}), the solution must satisfy $M/R<4/9$, which corresponds to the Buchdahl's limit for isotropic solutions.

\subsection{Model 4: like--Heintzmann IIa}
In this section we consider the Heintzmann IIa solution \cite{delgaty1998physical, heintzmann1969new} which metric components are
\begin{eqnarray}
e^{\xi}&=&A^2(1+Br^2)^3 \label{temporalheint}\\
e^{-\mu}&=&1-\frac{3Br^2}{2}\frac{1+(1+4Br^2)^{-1/2}}{1+Br^2}\label{radialheint}
\end{eqnarray}
where $A$ is a dimensionless constant and  $B$ is a constant with dimensions of the inverse of a length squared. In this section we shall use the list of auxiliary functions $\{\zeta,\eta_{2},\gamma_{1},\gamma_{3},\gamma_{6},\gamma_{7},\gamma_{9},\varrho_{3},\mathcal{P}_{5}$ $,\mathcal{P}_{6}\}$ defined in the Appendix, section \ref{appendix-m2}
.\\

In this case we have
\begin{eqnarray}
f&=& \frac{r^2}{6\alpha B \gamma_{1}(r)}\bigg[\frac{9 B^2 C}{\sqrt{4 B r^2+1}}+3 B^{2}\nonumber\\
&& -\frac{2 a_1\left(2 a_{2}^2 B^2+a_2 a_3 B \gamma_{2}(r)+a_{3}^2\right)}{a_2 a_{3}^2 \zeta(r)}\bigg]\nonumber\\
&& - \frac{2 a_1 (a_3-a_2 B)\chi(r)}{3 a_{3}^3 \alpha\gamma_{1}(r)}
\end{eqnarray}
from where, the radial metric results
\begin{eqnarray}\label{radialmetricmodel4}
e^{-\lambda}=\frac{a_3 \eta_{2}(r) +2 a_1 a_2 B (a_2 B-a_3) \zeta(r)\chi(r)}{3 a_2 a_{3}^3 B\gamma_{1}(r)\zeta(r)}.
\end{eqnarray}

From (\ref{mgd05}), (\ref{mgd06}) and (\ref{mgd07}) we arrive at
\begin{eqnarray}
\tilde{\rho}&=&\frac{2 a_1 a_2 B\gamma_{3}(r)(a_2 B-a_3) \zeta(r)^2 \chi(r)+a_3 r^2\varrho_{3}(r)}{24\pi a_2 a_{3}^3 B r^2 \gamma_{1}(r)^2 \zeta(r)^2}\label{densitymodel4}\\
\tilde{P}_{r}&=&\frac{\mathcal{P}_{5}(r)}{24\pi a_2 a_{3}^3 B r^2 \gamma_{1}(r)^2 \zeta(r)}\label{radialpmodel4}\\
\tilde{P}_{t}&=&\frac{\mathcal{P}_{6}(r)}{24\pi a_2 a_{3}^3 B \gamma_{1}(r)^2 \zeta(r)^2}\label{tangentialpmodel4}
\end{eqnarray}

Finally, matching conditions lead to
\begin{eqnarray}
a_1&=&-\frac{3 a_2 a_{3}^3 B^2 R^2 \gamma_{7}(R)\zeta(R)\gamma_{6}(R)^{-1}}{a_3 R^2\gamma_{9}(R)-2 a_2 B (a_2 B-a_3) \zeta(R)\chi(R)}\\
A^2&=&-\frac{(7 M-3 R)^3}{27 R (R-2 M)^2} \\
B^2 &=&\frac{M^2}{R^4 (3 R-7 M)^2}.
\end{eqnarray}

Note that, as in all the previous cases the free parameters are $\{R,M,a_{2},a_{3}\}$. Note that, as 
$\{\zeta,\eta_2,\varrho_3,\mathcal{P}_5,\mathcal{P}_6\}$ have dimensions of a length squared and $\{\gamma_1,\gamma_3,\gamma_6,\gamma_7,\gamma_9\}$ are dimensionless (see Appendix \ref{appendix-m2}), all the expressions above are dimensionally correct. It should be emphasized that, $M/R<3/7<4/9$ which corresponds to less compact solutions than the allowed by the Buchdahl's limit.

\section{Discussion}\label{discussion}
In this section we analyze the results obtained previously in order to verify the physical acceptability of the models \cite{ivanov2017analytical}. 
To this end, we shall use the compactness parameters given in Table
\ref{table1} and set suitable values of $a_{2}$ and $a_{3}$ in order to discuss to what extend our solutions are suitable to describe the SMC X-1 and Cen X-3 systems.
\begin{table*}[h]
\small
\centering
\begin{tabular}{cccccc}
\hline
 Compact start & $M/M_{\odot}$ & $R(km)$ & $u=M/R$ & $\rho(0)/\rho(R)$ & $Z(R)$\\ \hline
SMC X-1 \cite{rawls2011refined} & 1.04 & 8.301 & 0.19803 &  1.4659 \cite{maurya2018exact} & 0.286776 \\    
Cen X-3 \cite{torres2019anisotropic} & 1.49 & 10.8 & 0.2035 & 1.915 \cite{prasad2019relativistic} & 0.298592\\
\hline         
\end{tabular}
\caption{\small{Physical parameters for the compact starts SMC X-1 and Cen X-3.}}
\label{table1}
\end{table*}
\subsection{Metrics }
\begin{figure}[htb!]
  \begin{center}
    \subfigure[]{
        \includegraphics[scale=0.223]{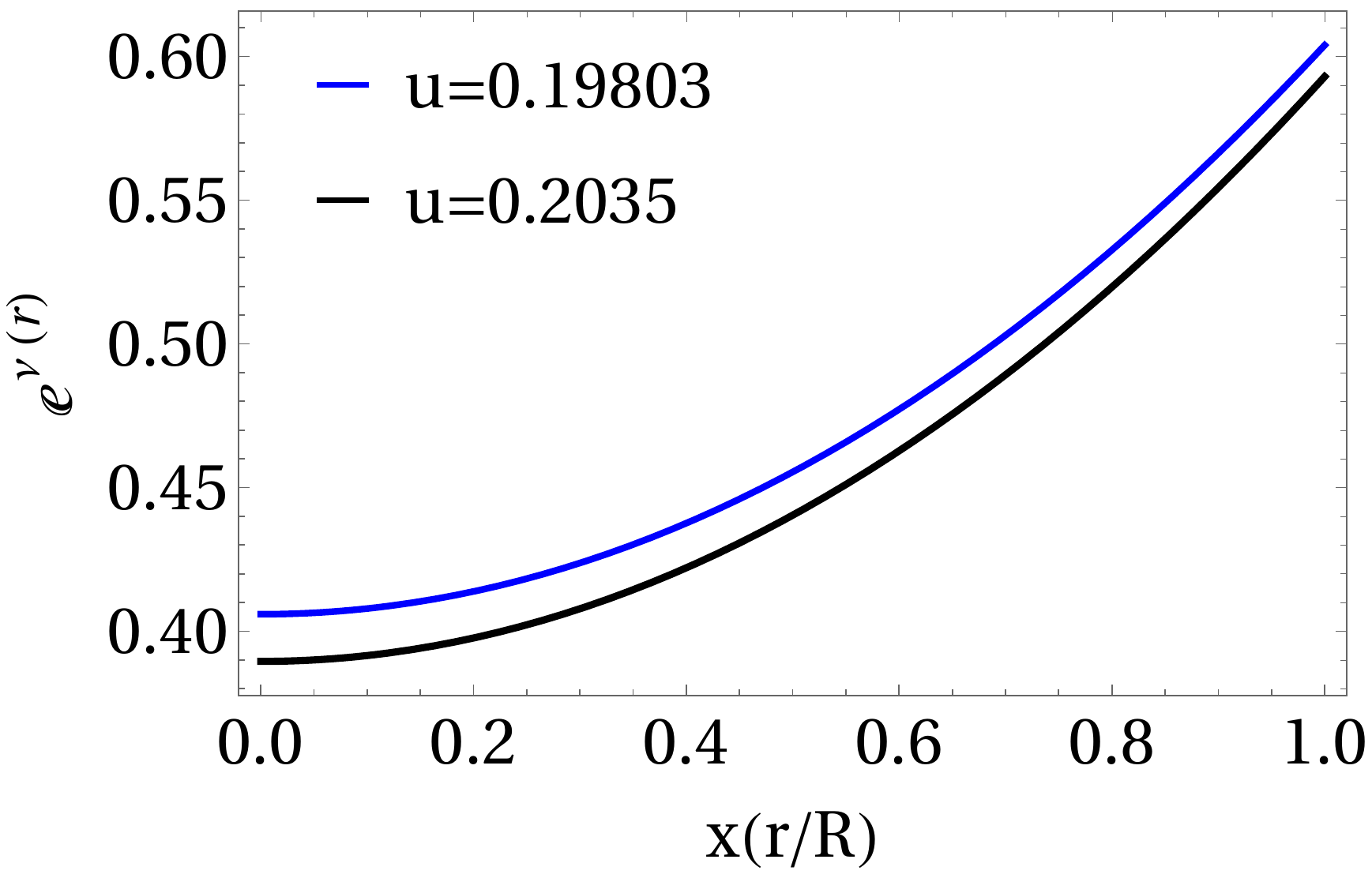}
        \label{mtemp1}}
    \subfigure[]{
        \includegraphics[scale=0.223]{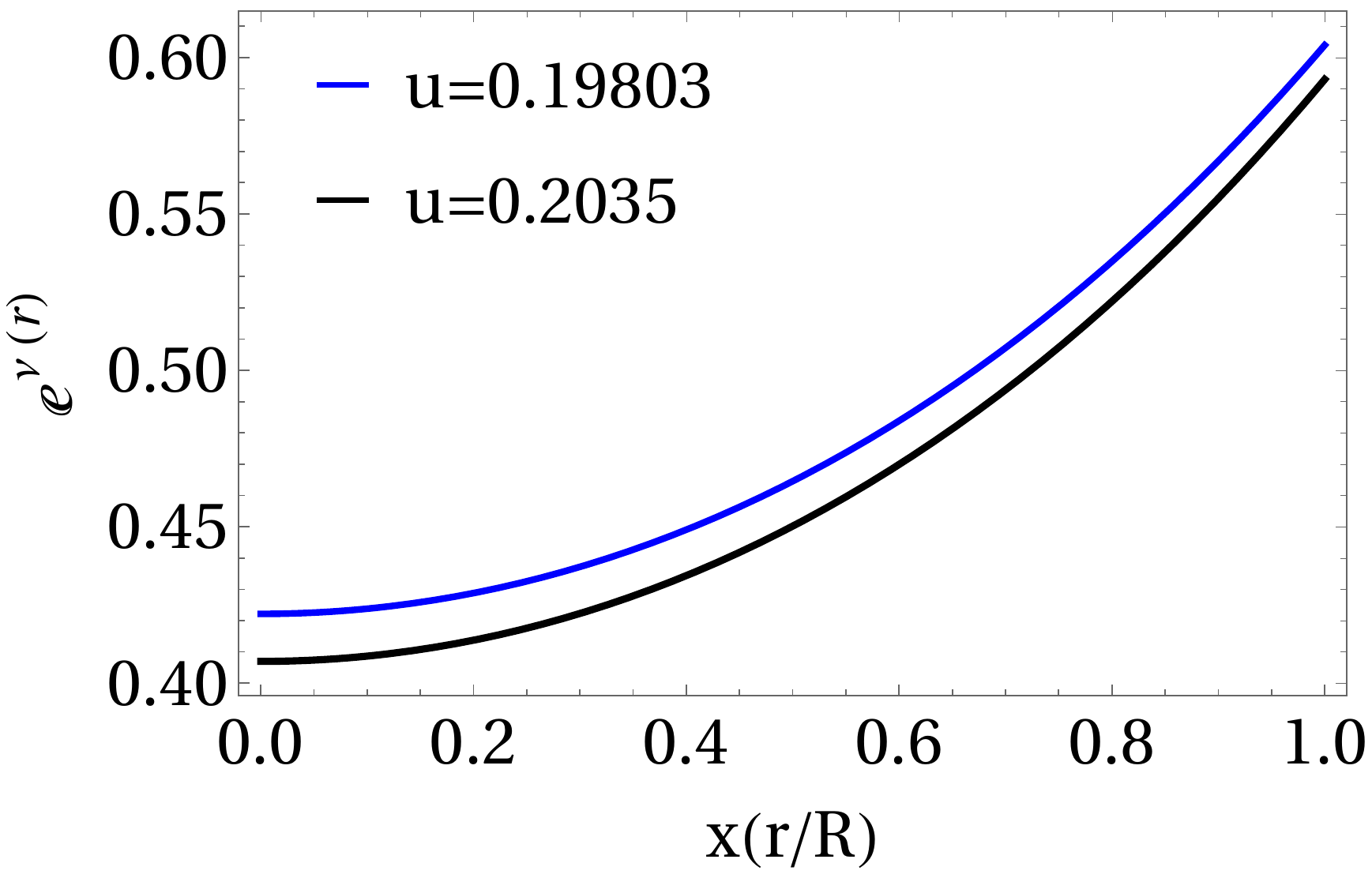}
        \label{mtemp2}}
\subfigure[]{
        \includegraphics[scale=0.223]{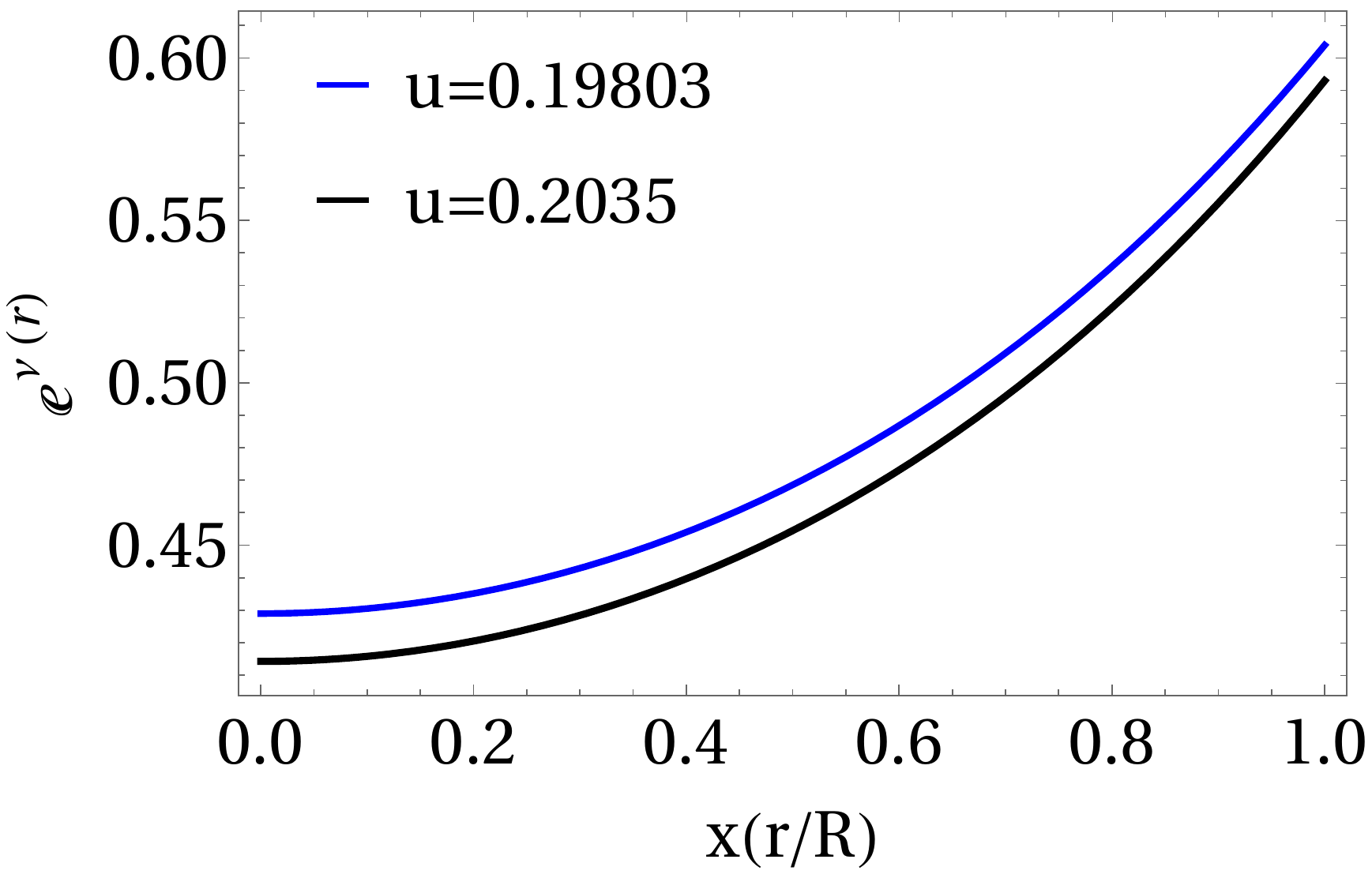}
        \label{mtemp3}}
        \subfigure[]{
        \includegraphics[scale=0.223]{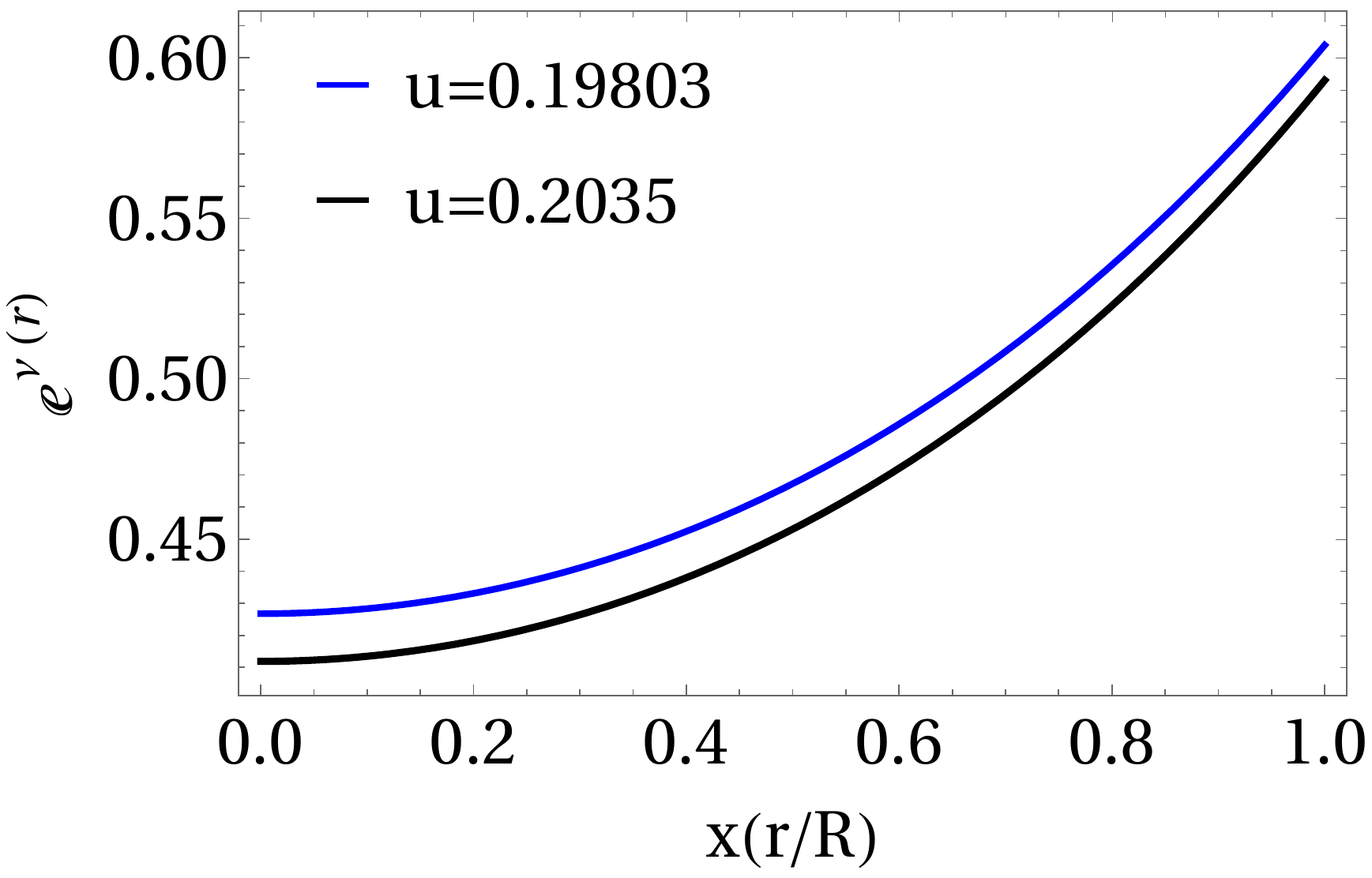}
        \label{mtemp4}}    
    \caption{\label{metrica-temporal} $e^{\nu}$ for Model 1 (a), Model 2 (b) Model 3 (c) and Model 4 (d)}
    \subfigure[]{
        \includegraphics[scale=0.223]{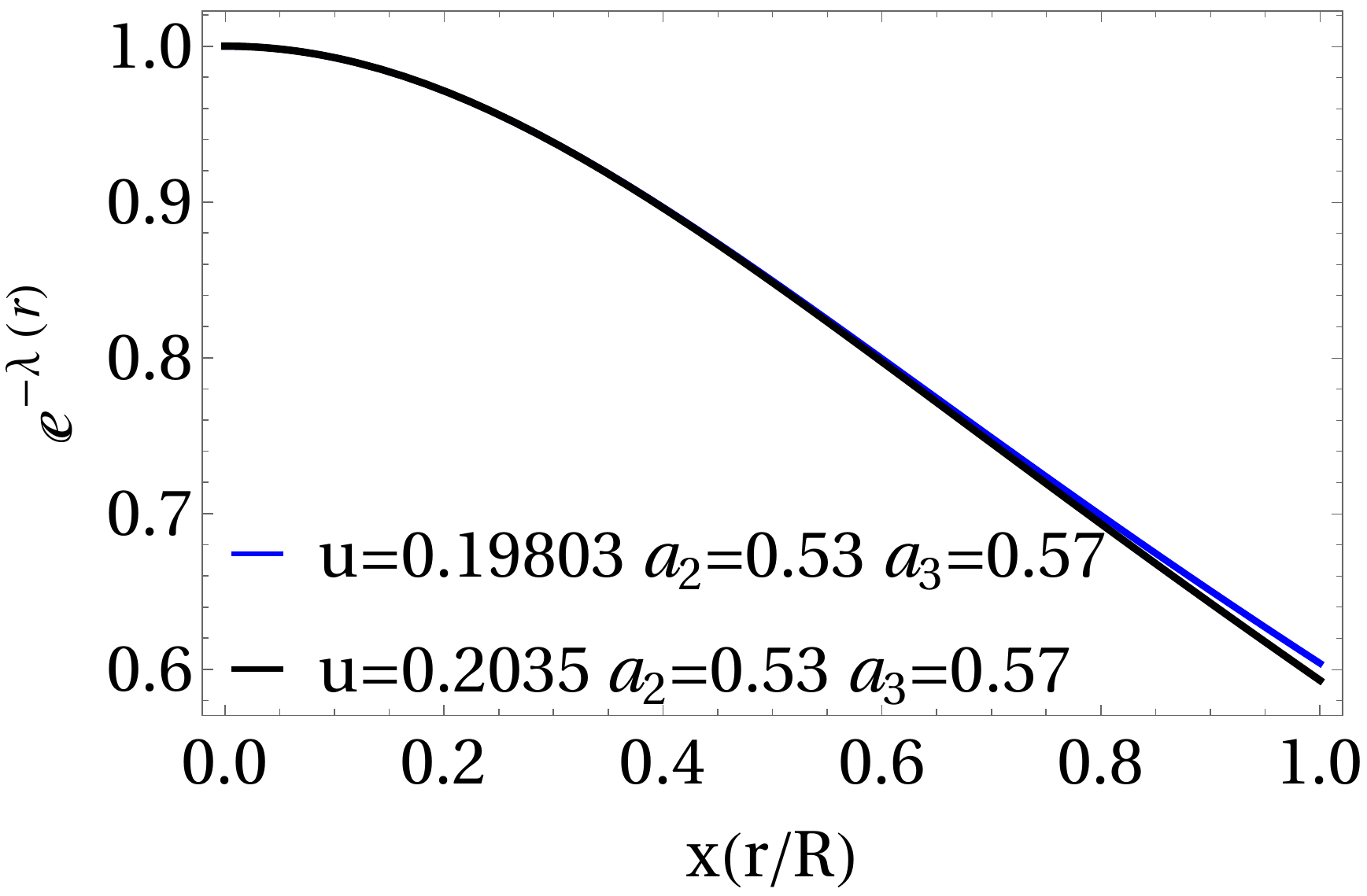}
        \label{mradial1}}
    \subfigure[]{
        \includegraphics[scale=0.223]{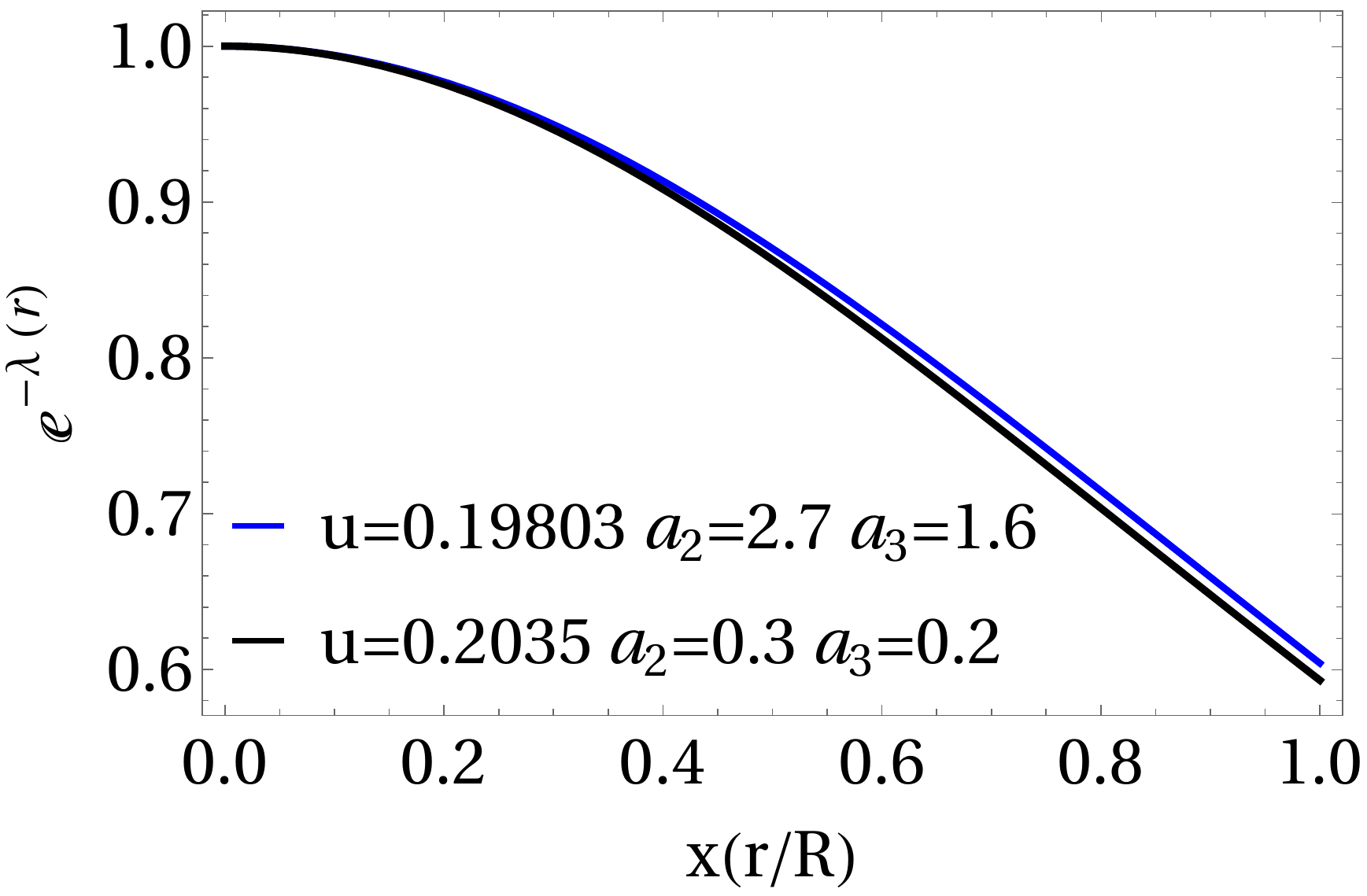}
        \label{mradial2}}
    \subfigure[]{
        \includegraphics[scale=0.223]{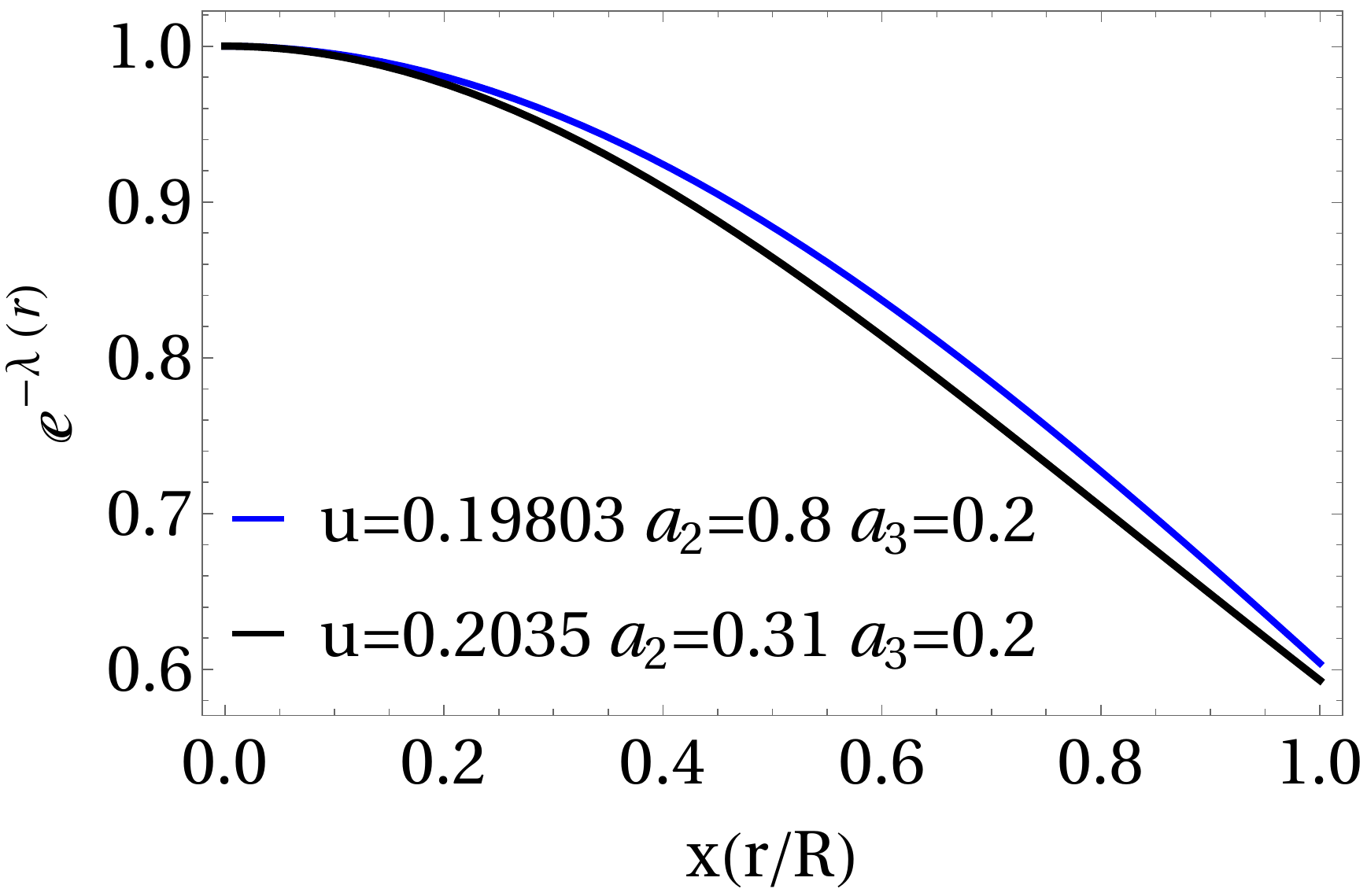}
        \label{mradial3}}
    \subfigure[]{
        \includegraphics[scale=0.223]{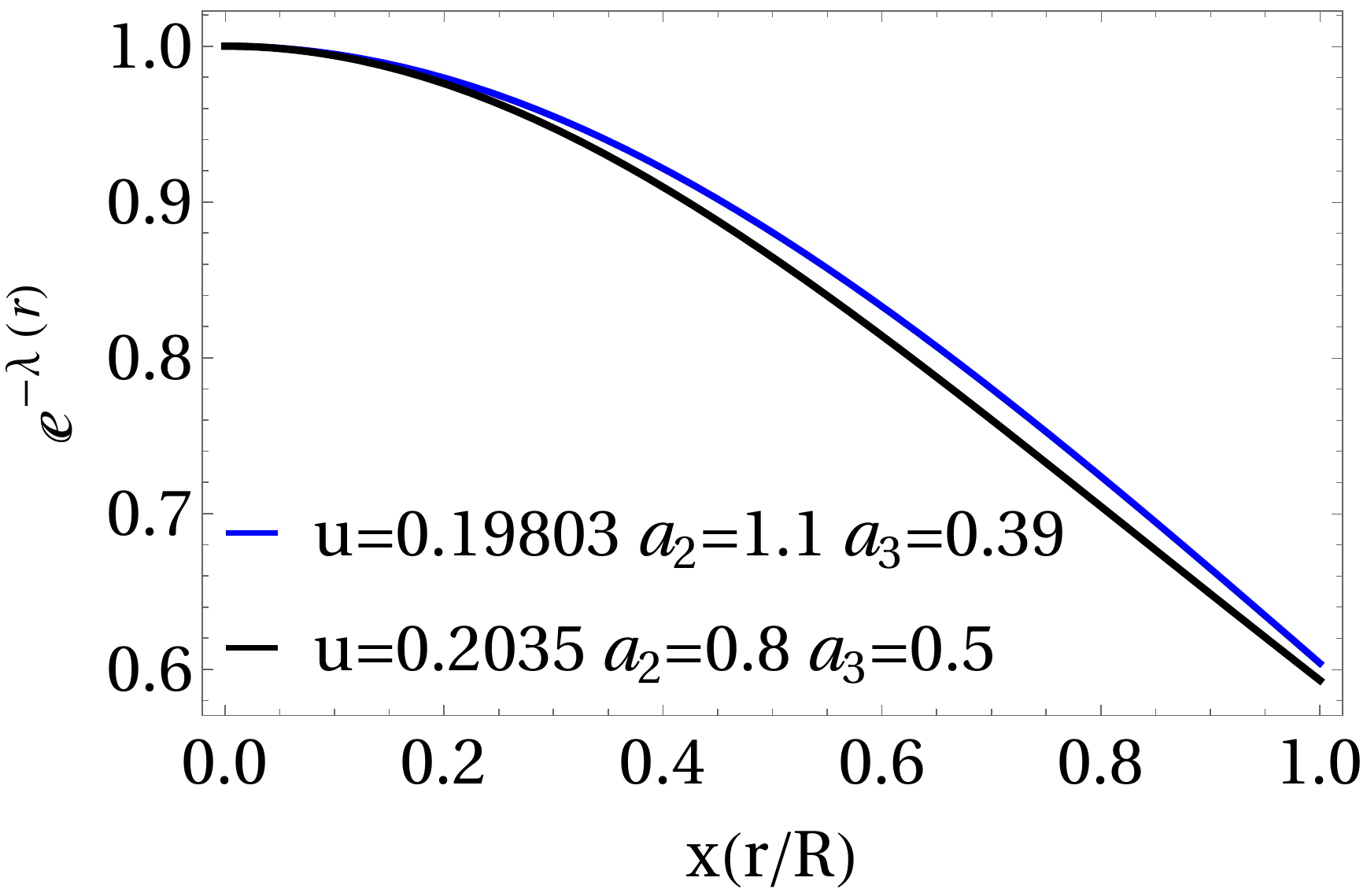}
        \label{mradial4}}
    \caption{\label{metrica-radial}$e^{-\lambda}$ for Model 1 (a), Model 2 (b) Model 3 (c) and Model 4 (d)}
    \label{metric4}
  \end{center}
\end{figure}

In figures \ref{metrica-temporal} and
\ref{metrica-radial}
we show the metric functions for the compactness parameter indicated in the legend. Note that on one hand $e^{\nu}$ is a monotonously increasing function with $e^{\nu(0)}=constant$. On the other hand $e^{-\lambda}$ is monotonously decreasing with $e^{-\lambda(0)}=1$, as expected.  

\subsection{Matter sector}
\begin{figure}[htb!]
  \begin{center}
   \subfigure[]{
        \includegraphics[scale=0.223]{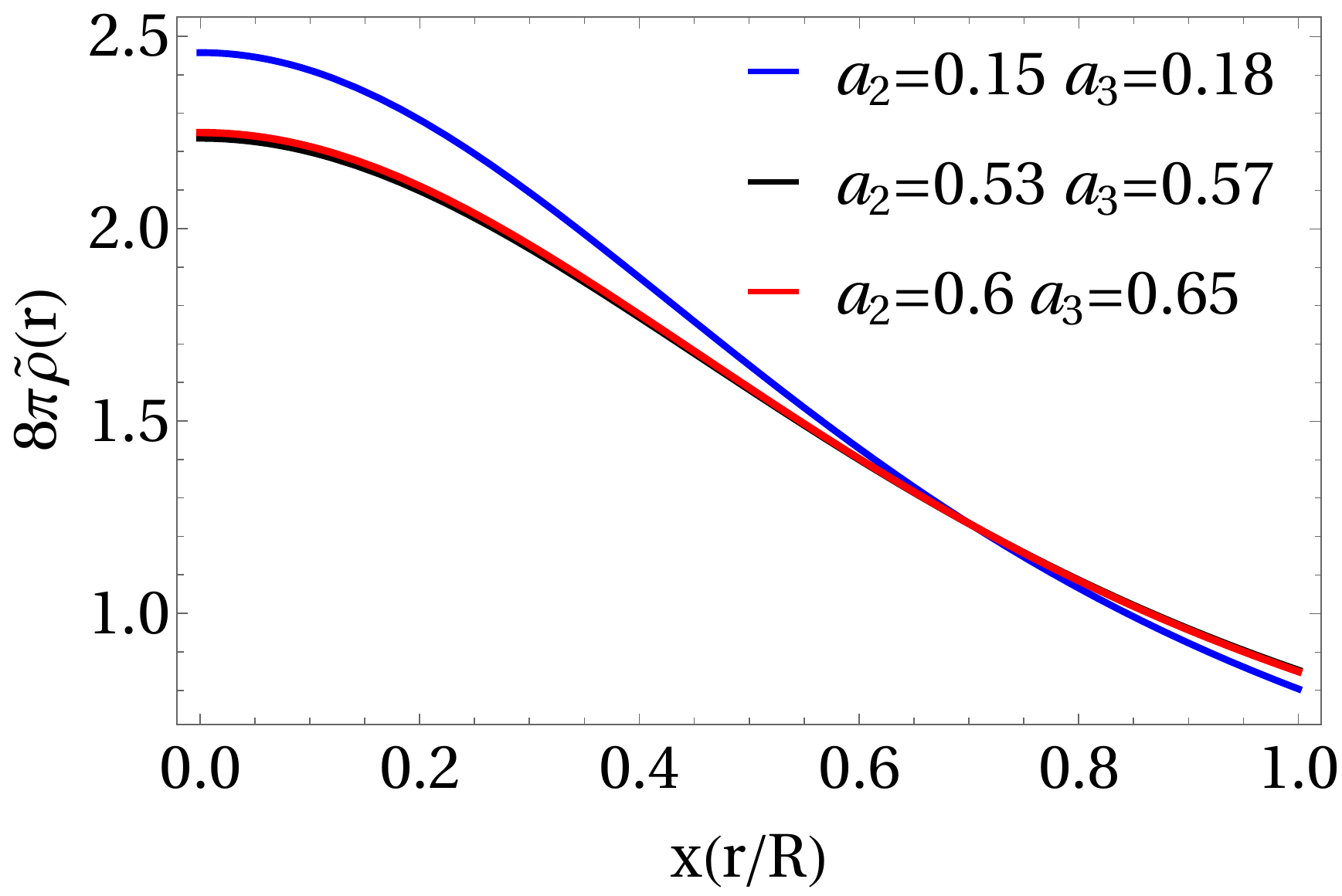}
        \label{density1a}}
    \subfigure[]{
        \includegraphics[scale=0.223]{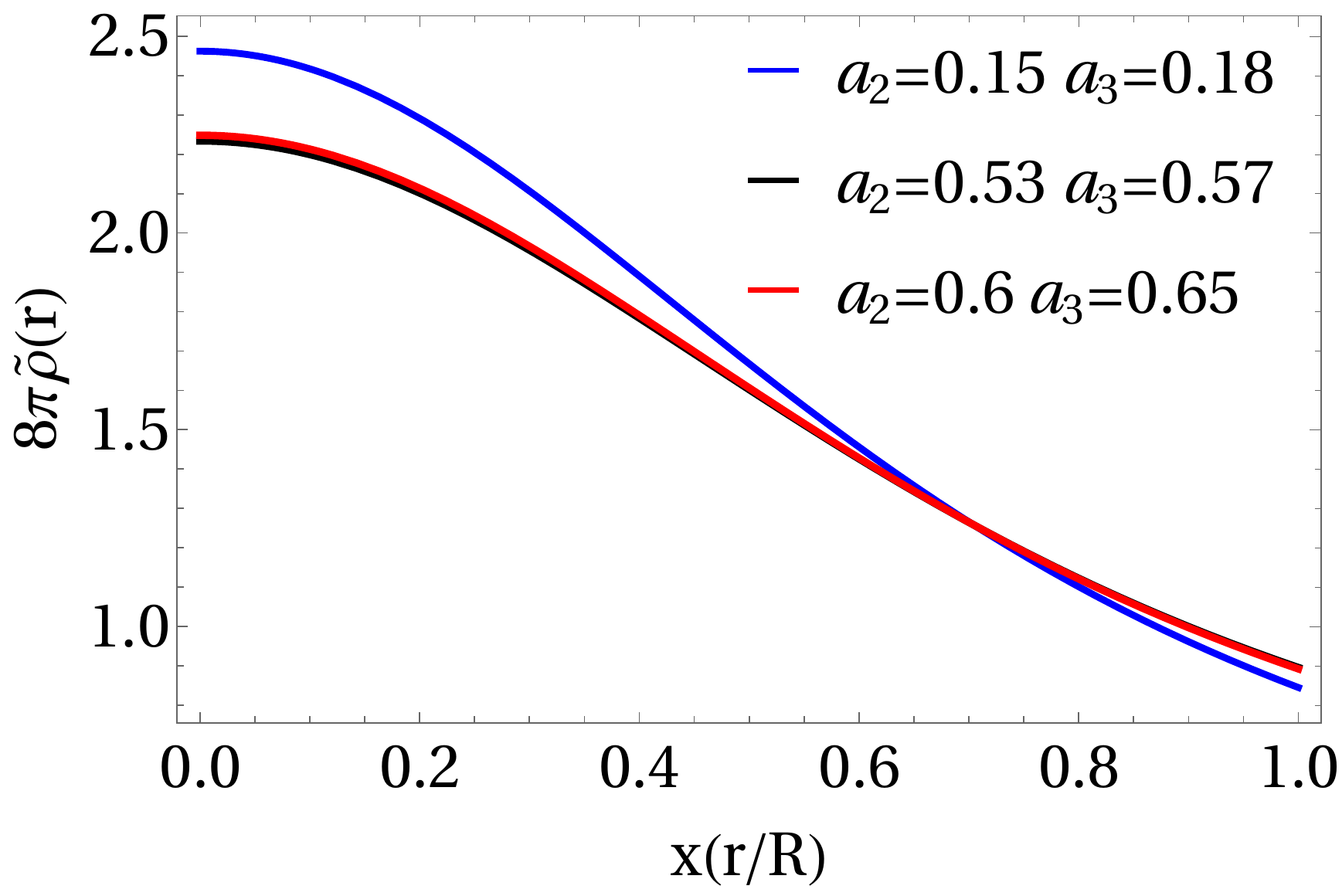}
        \label{density1b}}
    \subfigure[]{
        \includegraphics[scale=0.223]{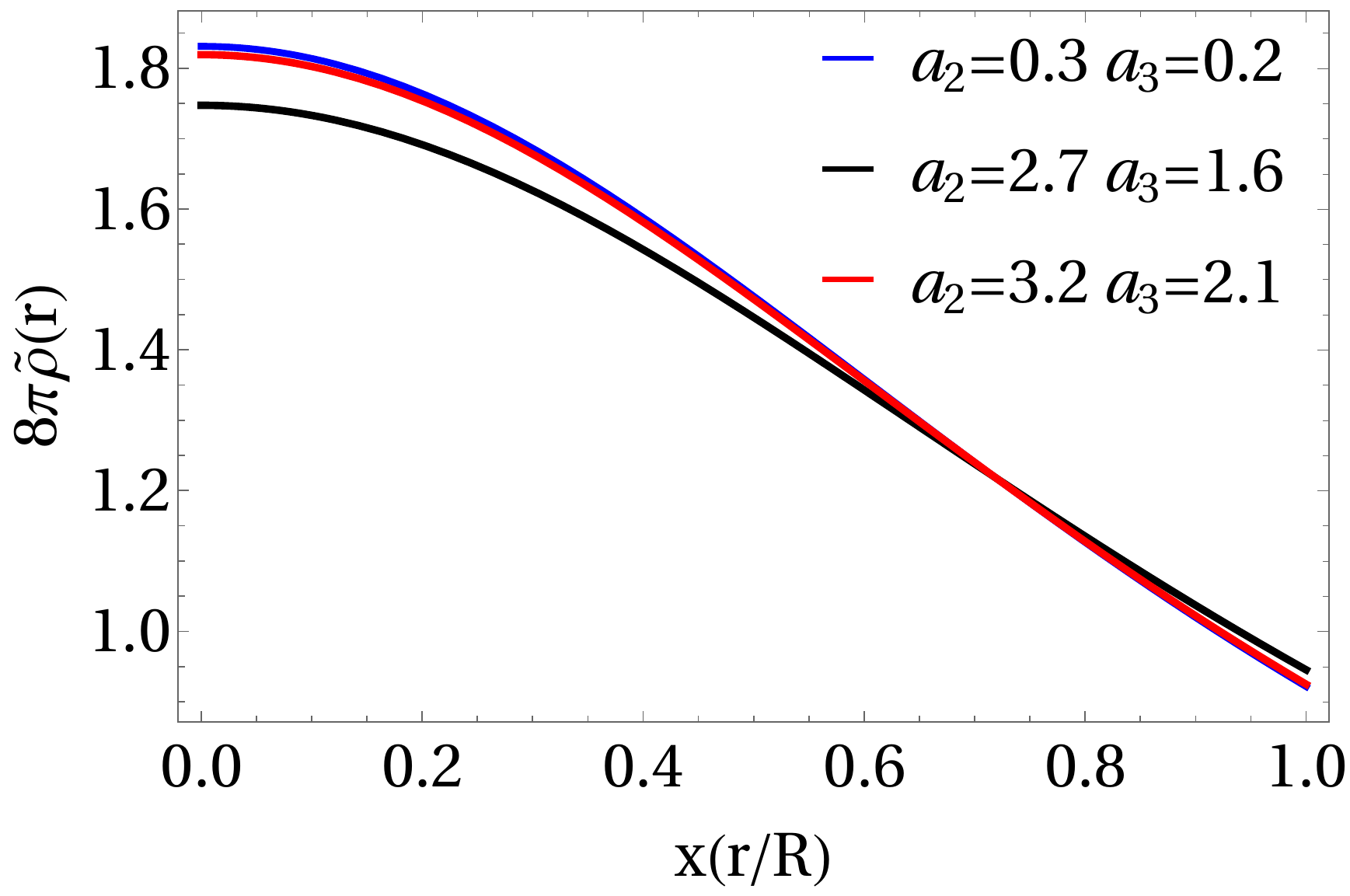}
        \label{density2a}}
    \subfigure[]{
        \includegraphics[scale=0.223]{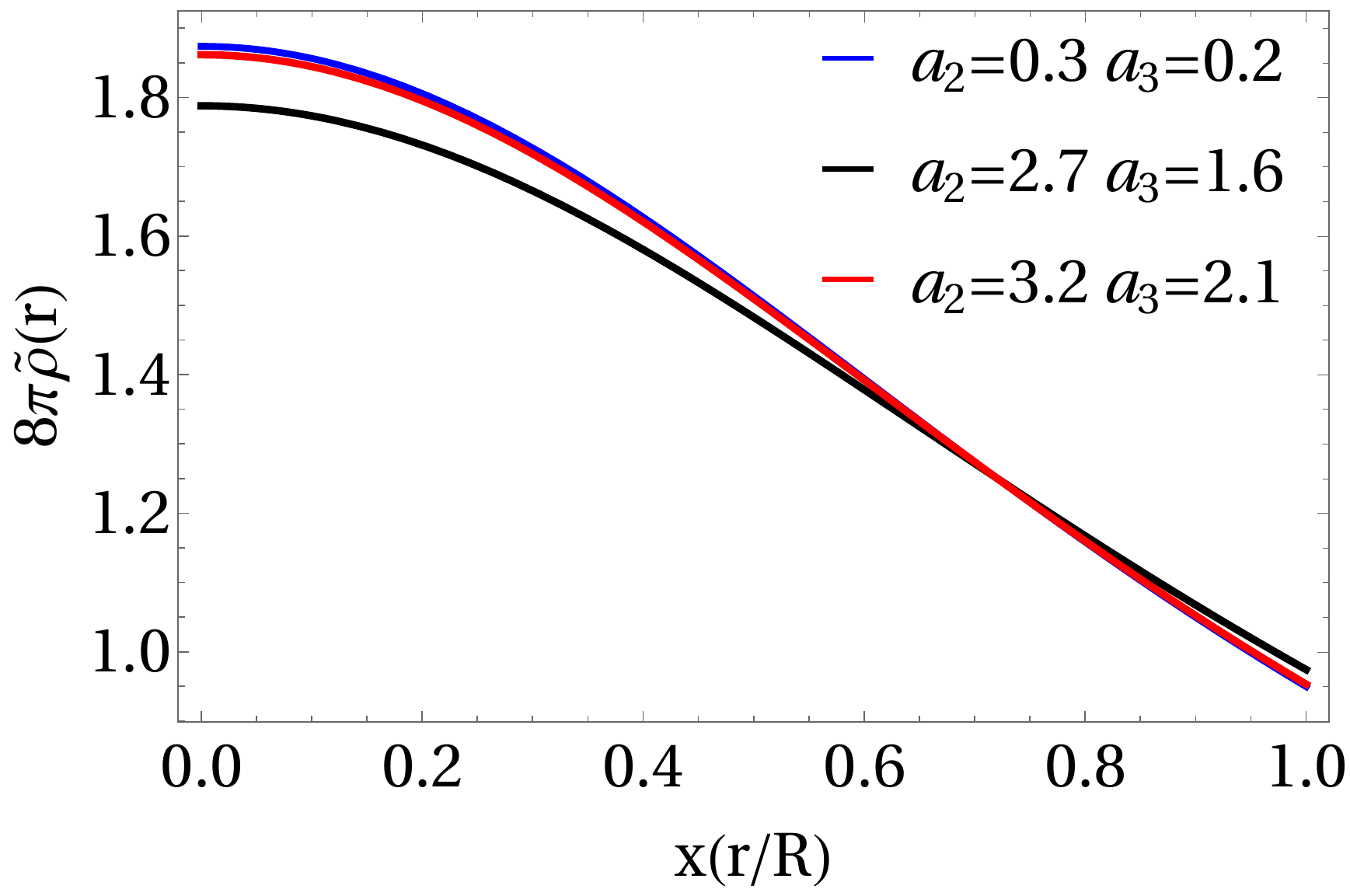}
        \label{density2b}}
    \subfigure[]{
        \includegraphics[scale=0.223]{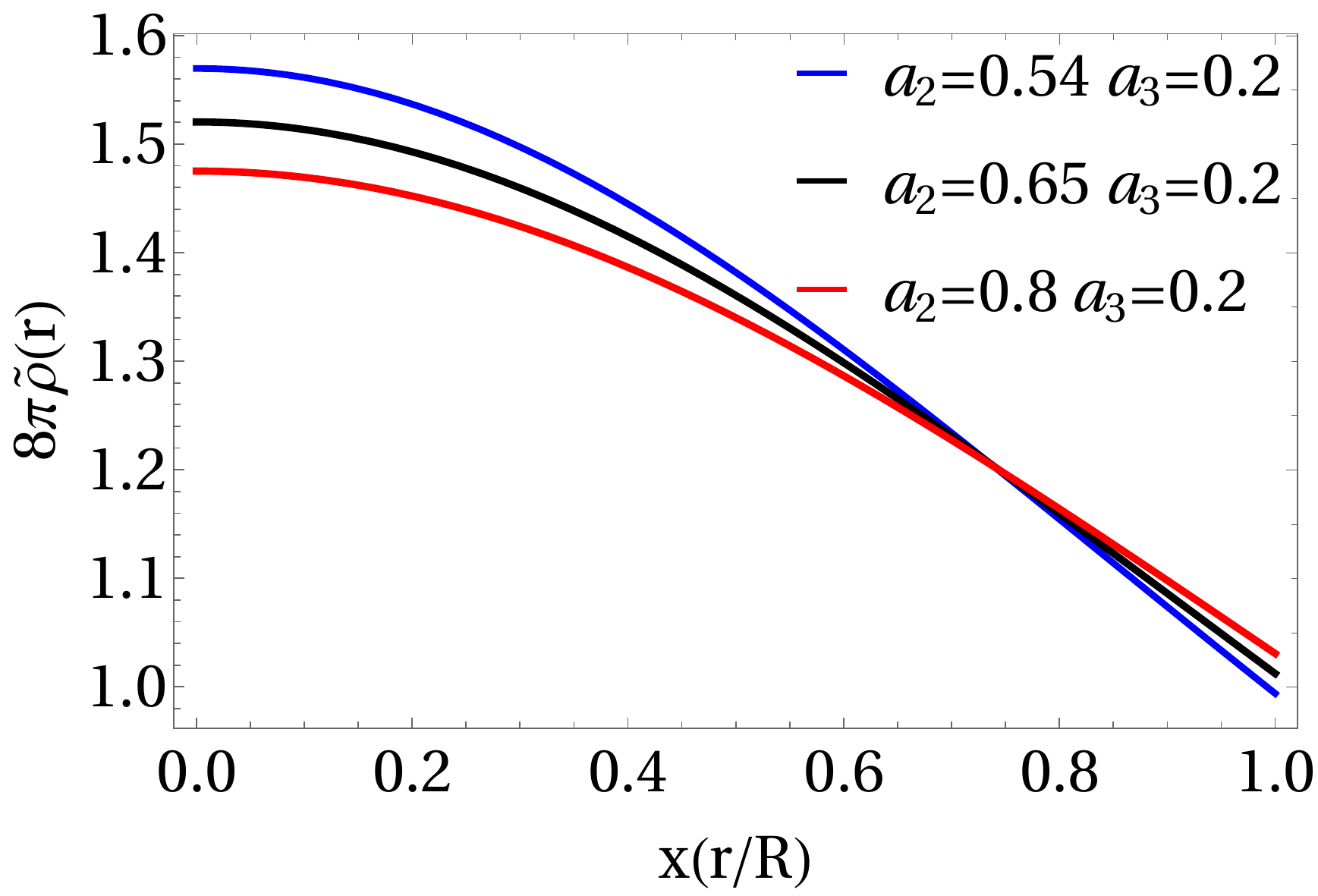}
        \label{density3a}}
    \subfigure[]{
        \includegraphics[scale=0.223]{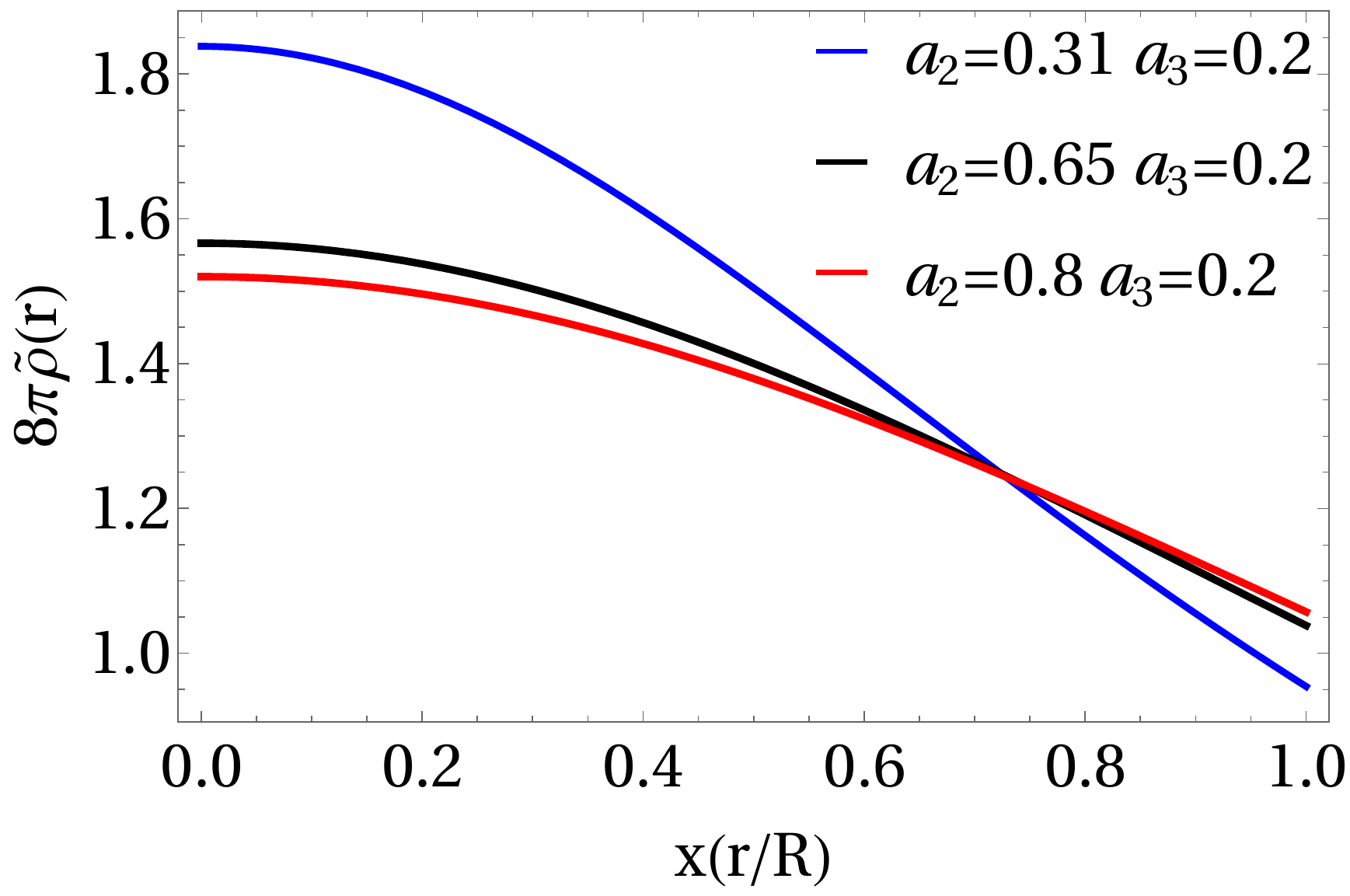}
        \label{density3b}}
        \subfigure[]{
        \includegraphics[scale=0.223]{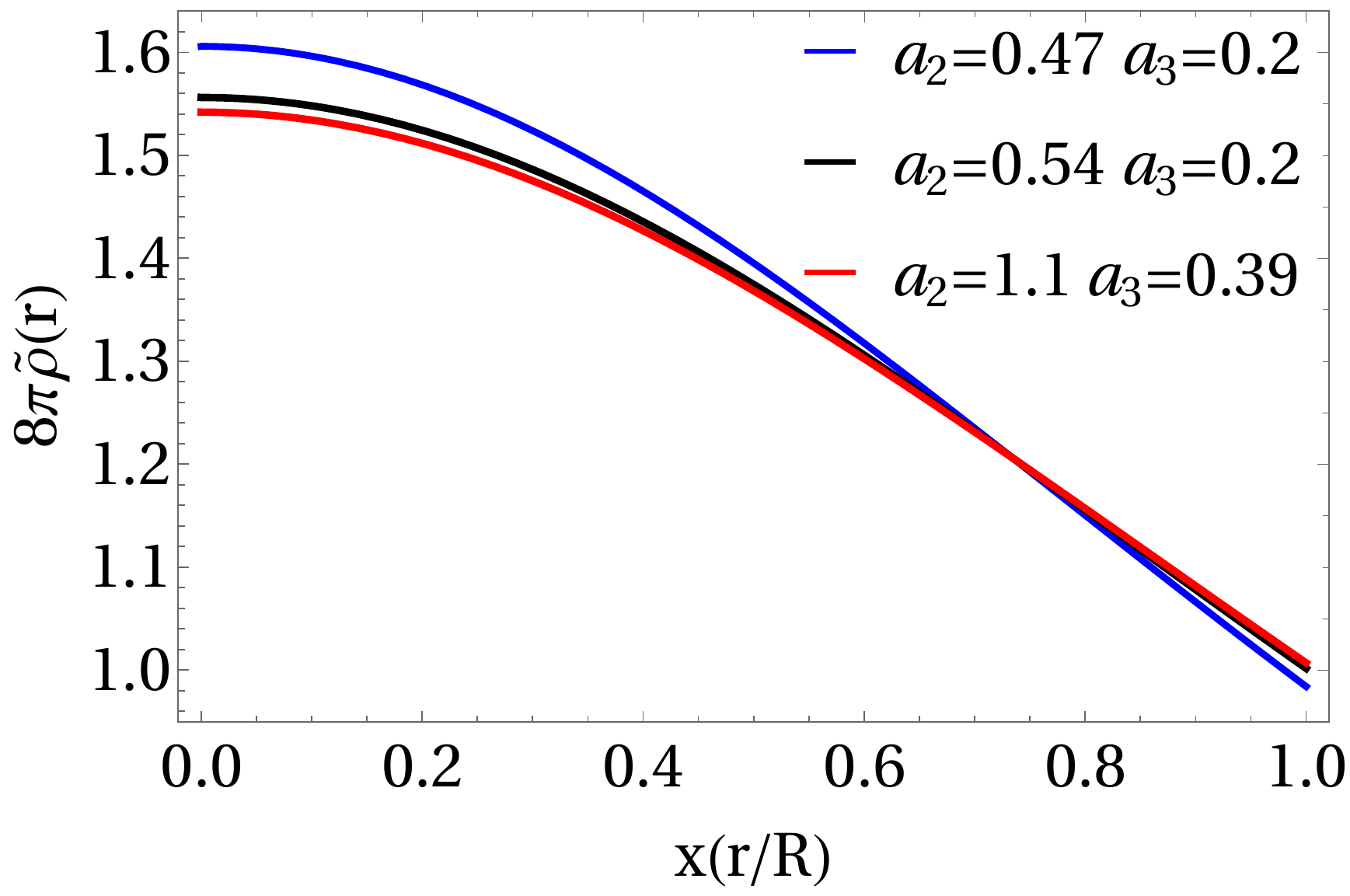}
        \label{density4a}}
    \subfigure[]{
        \includegraphics[scale=0.223]{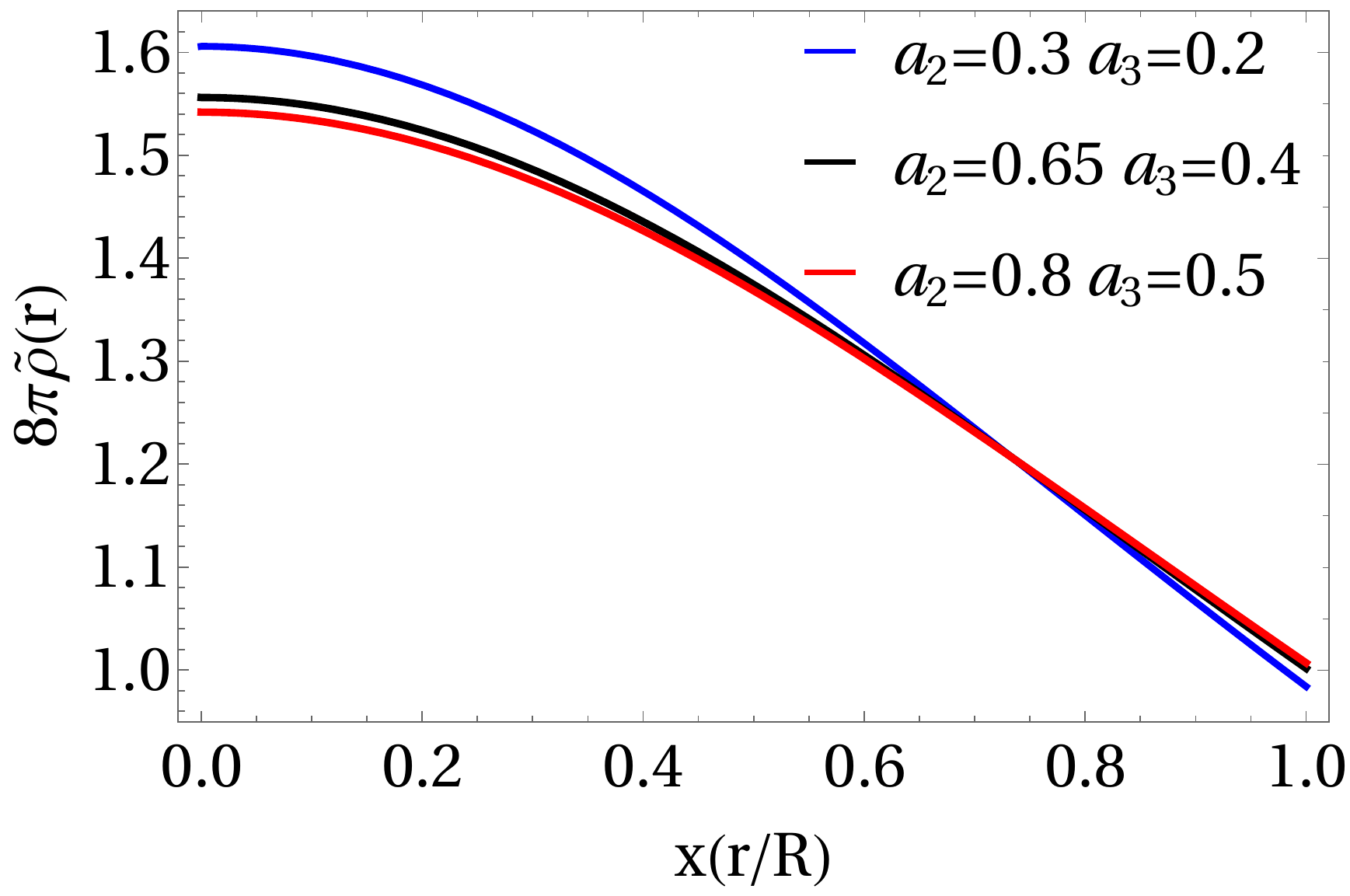}
        \label{density4b}}
    \caption{$\rho$ as a function of $r$ for Model 1: (a) u = 0.19803, (b)  u = 0.2035, Model 2: (c) u = 0.19803, (d) u = 0.2035,  Model 3: (e) u = 0.19803, (f) u = 0.2035,  Model 4: (g) u = 0.19803, (h)  u = 0.2035.} 
 \label{density}
  \end{center}
\end{figure}

\begin{figure}[htb!]
  \begin{center}
    \subfigure[]{
        \includegraphics[scale=0.223]{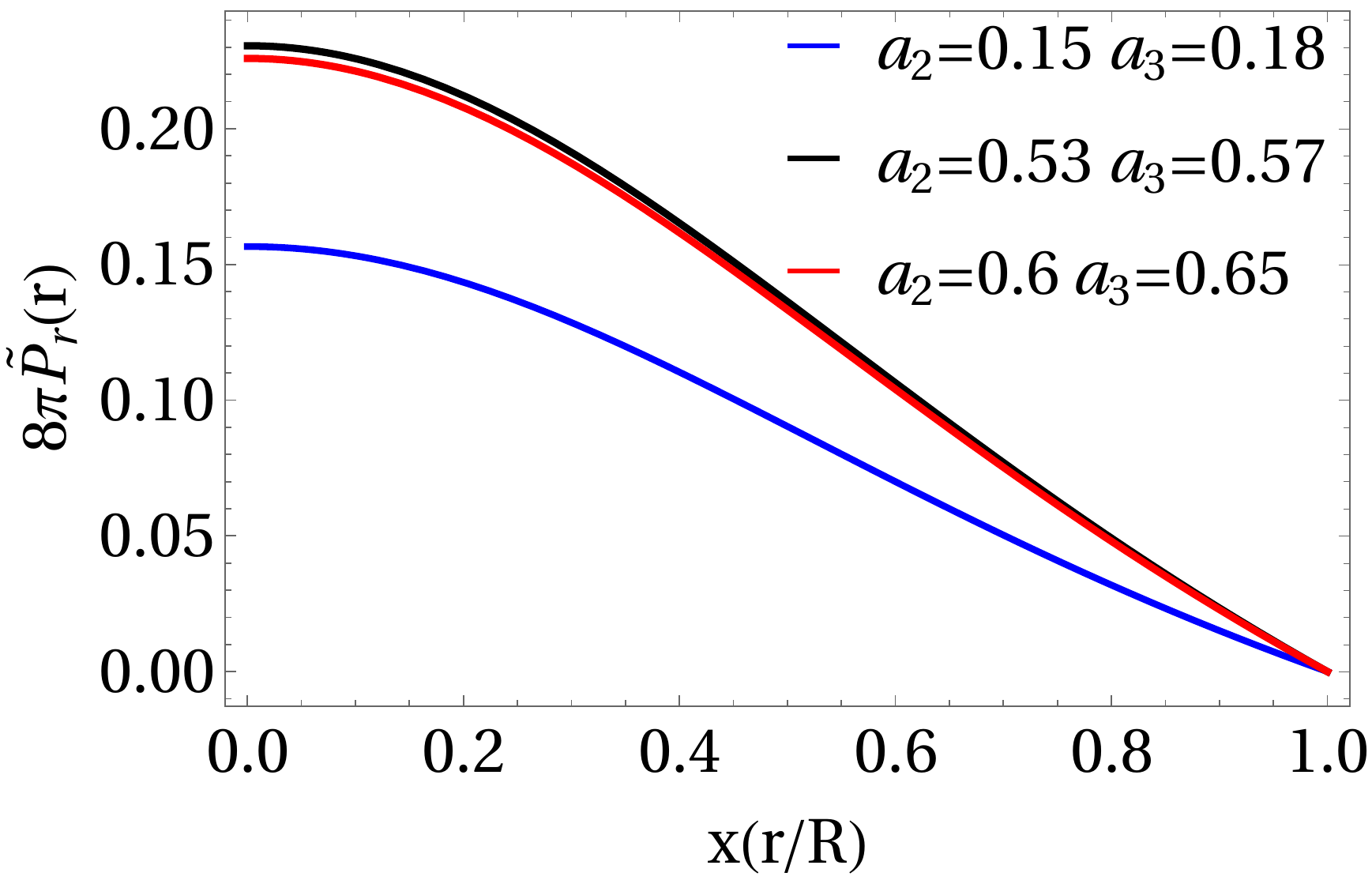}
        \label{rpressure1a}}
    \subfigure[]{
        \includegraphics[scale=0.223]{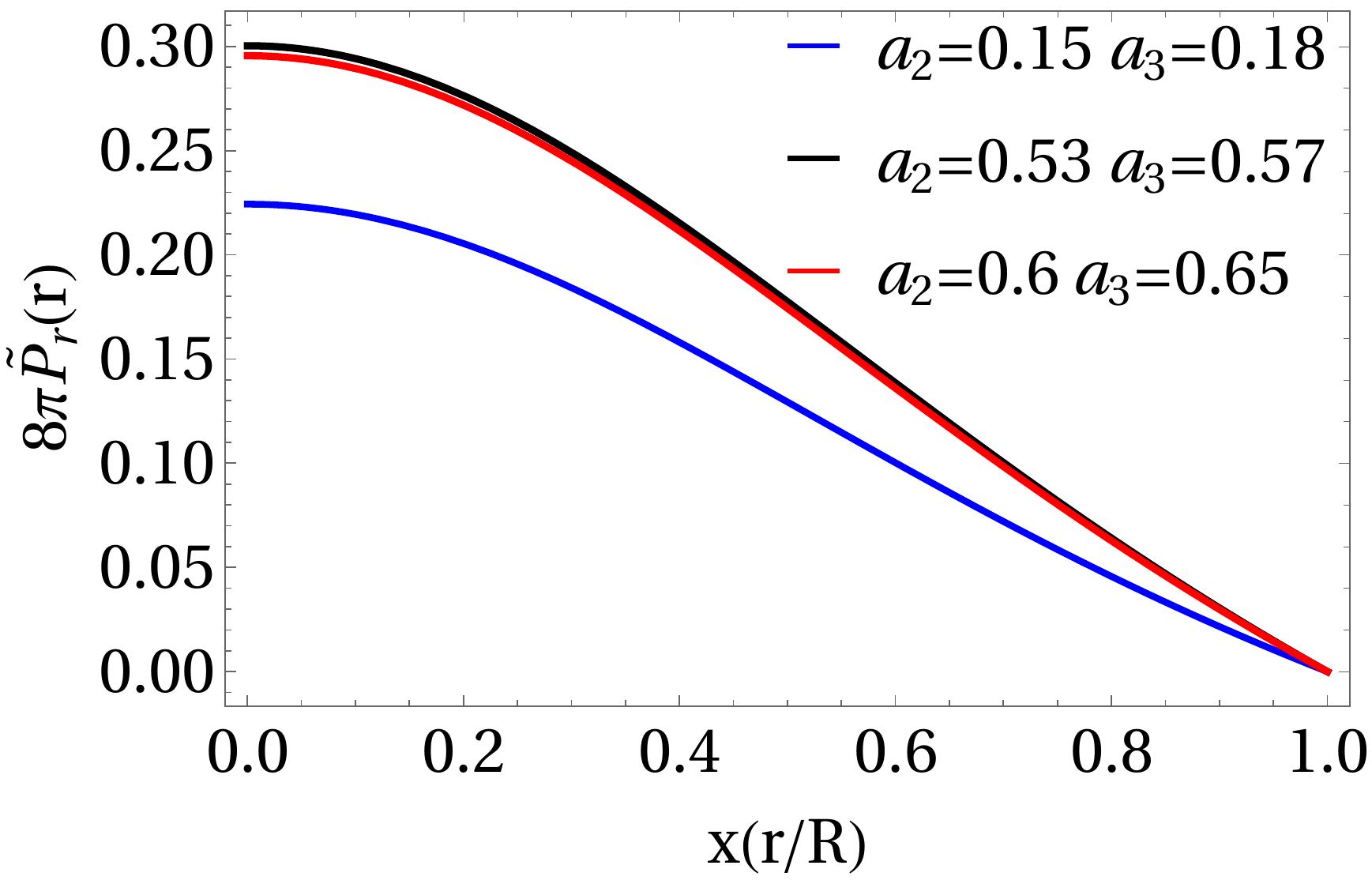}
        \label{rpressure1b}}
  \subfigure[]{
        \includegraphics[scale=0.223]{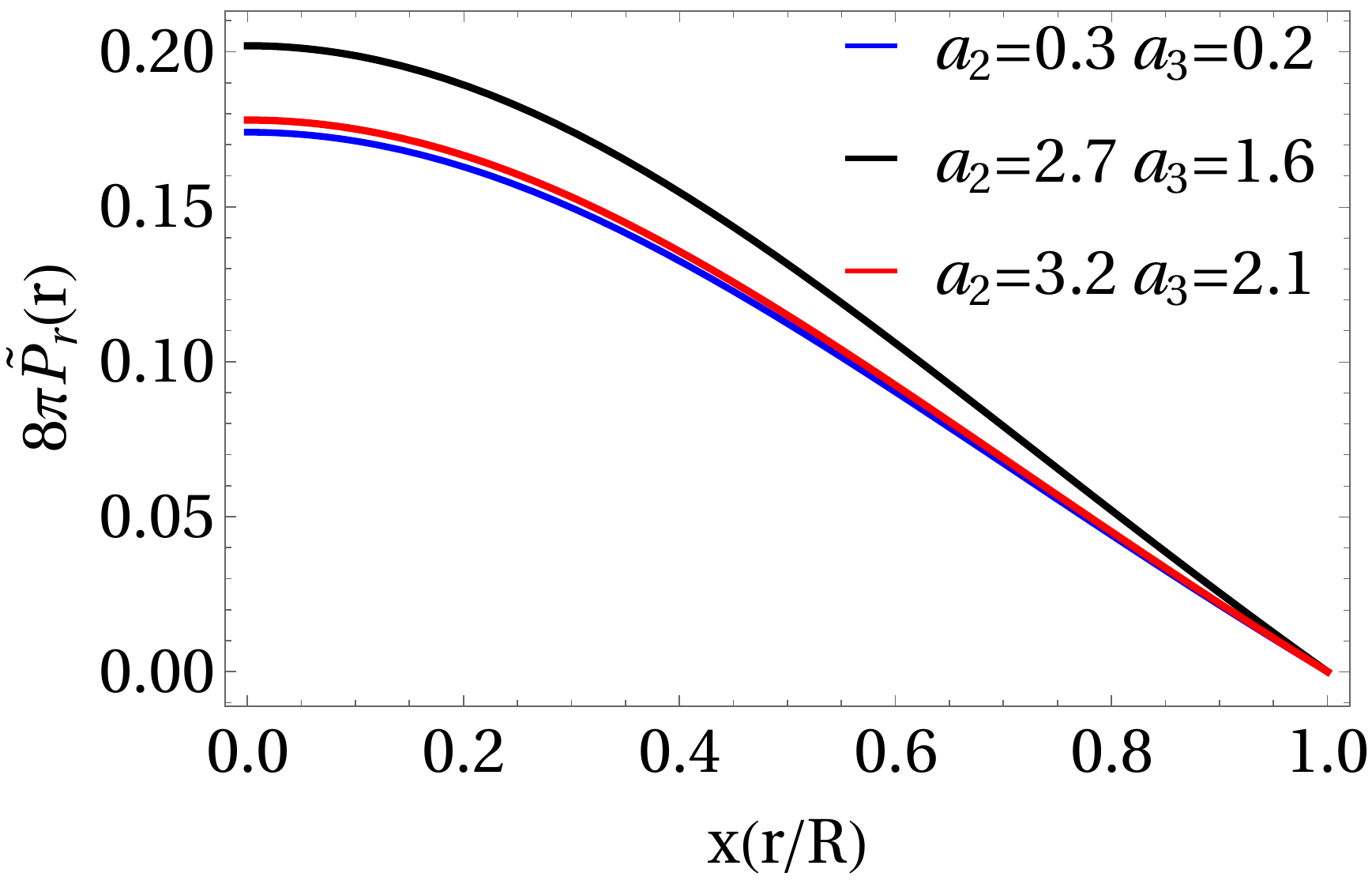}
        \label{rpressure2a}}
    \subfigure[]{
        \includegraphics[scale=0.223]{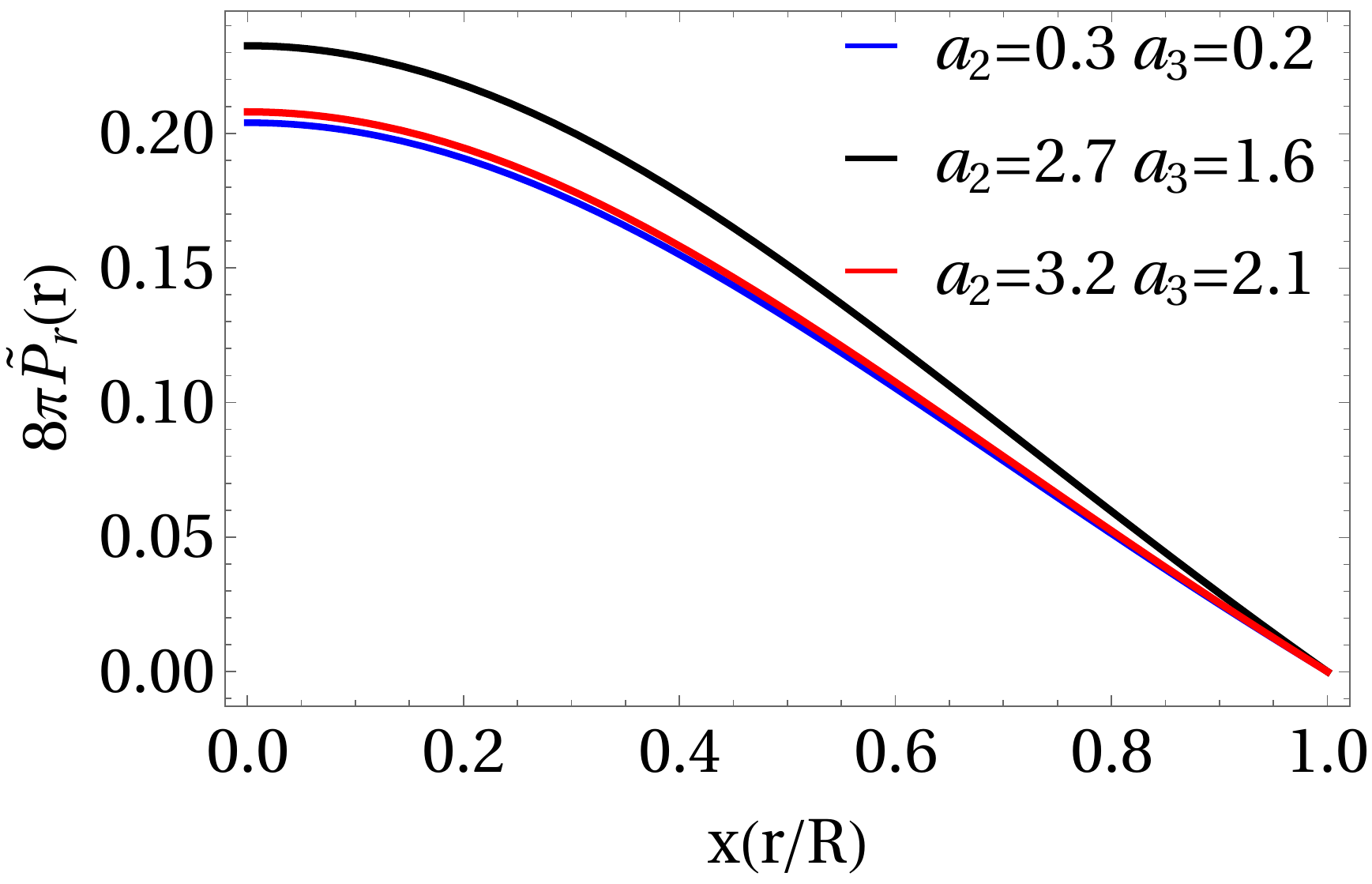}
        \label{rpressure2b}}
    \subfigure[]{
        \includegraphics[scale=0.223]{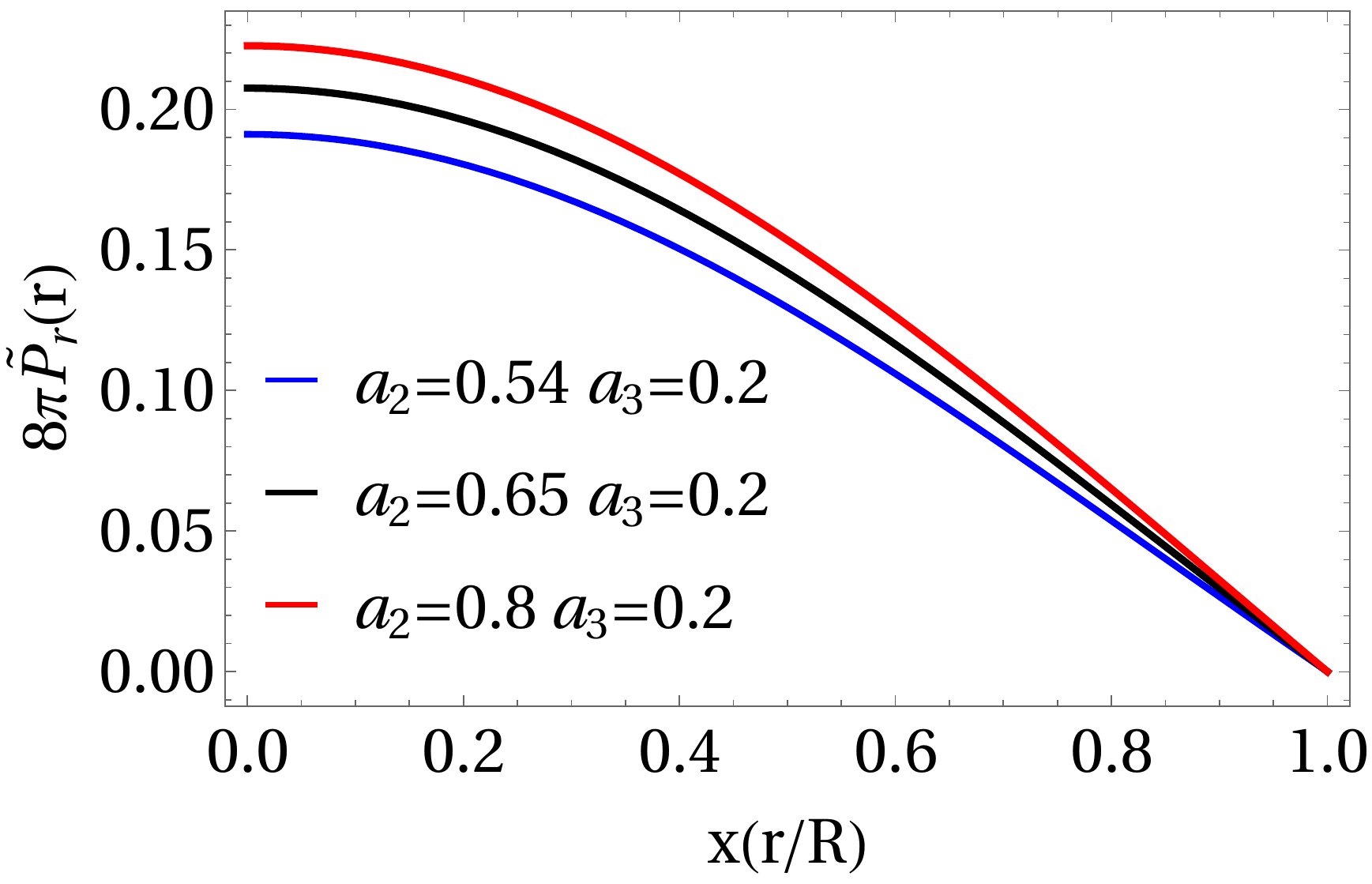}
        \label{rpressure3a}}
    \subfigure[]{
        \includegraphics[scale=0.223]{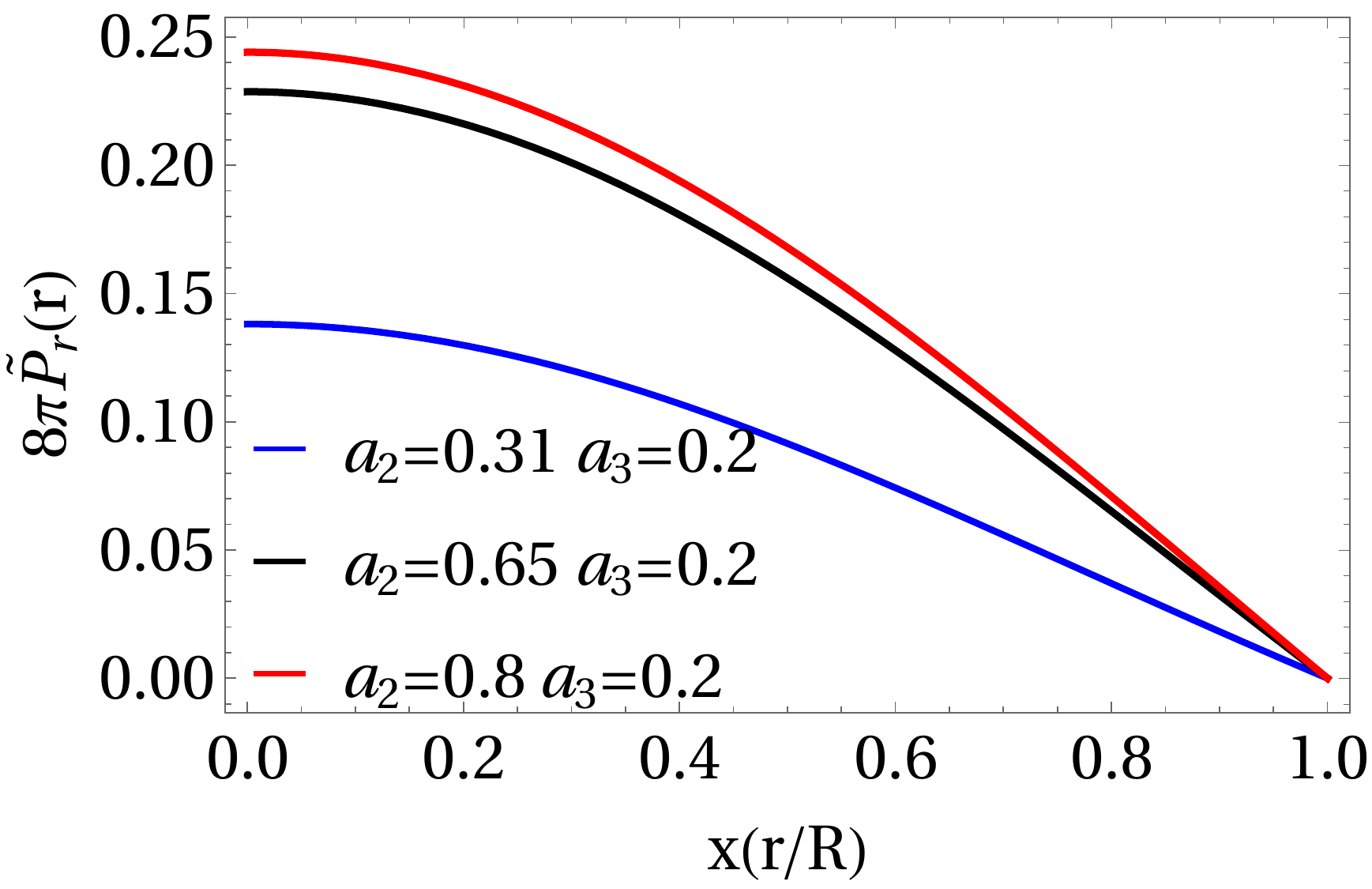}
        \label{rpressure3b}}
    \subfigure[]{
        \includegraphics[scale=0.223]{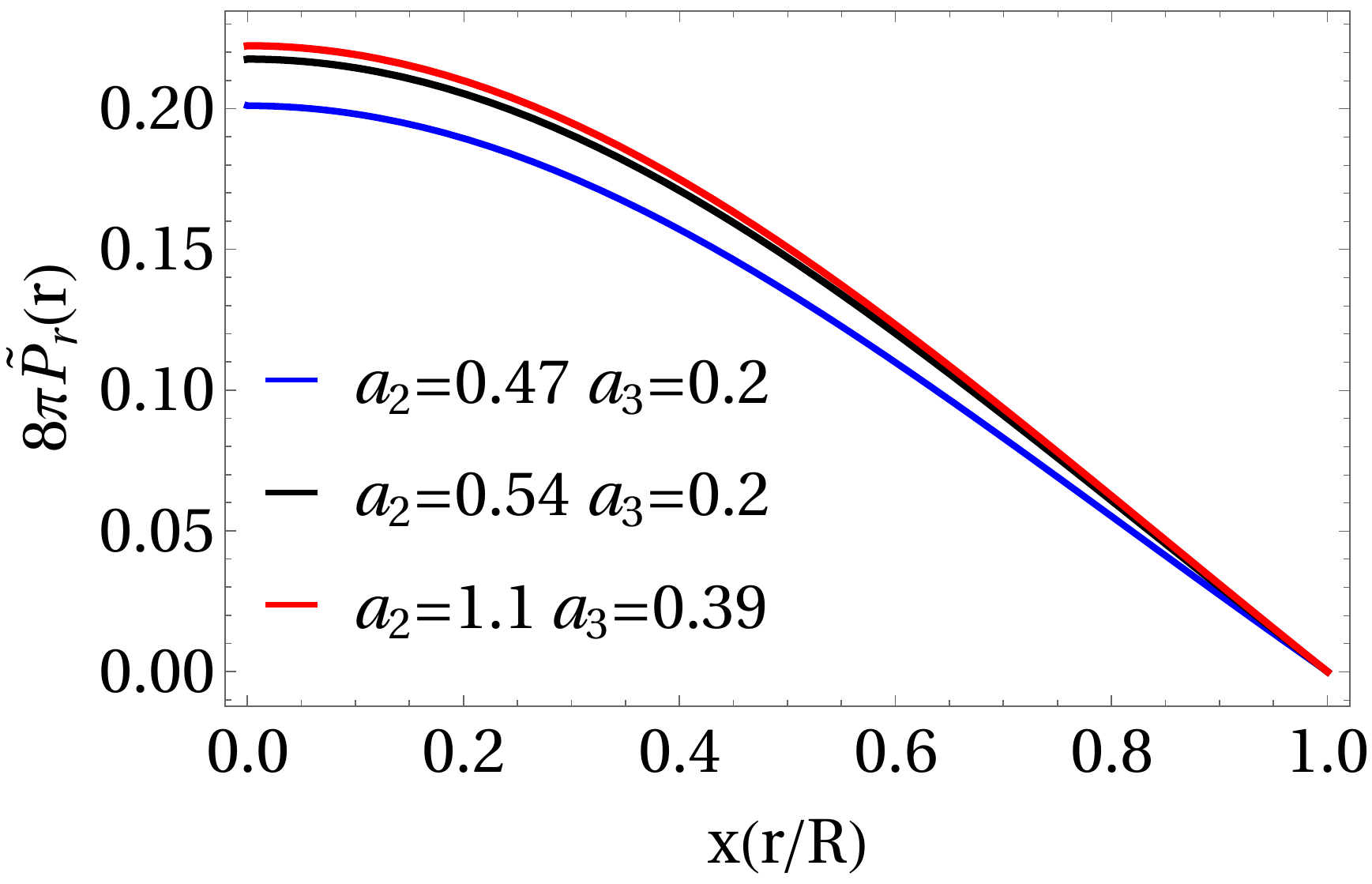}
        \label{rpressure4a}}
    \subfigure[]{
        \includegraphics[scale=0.223]{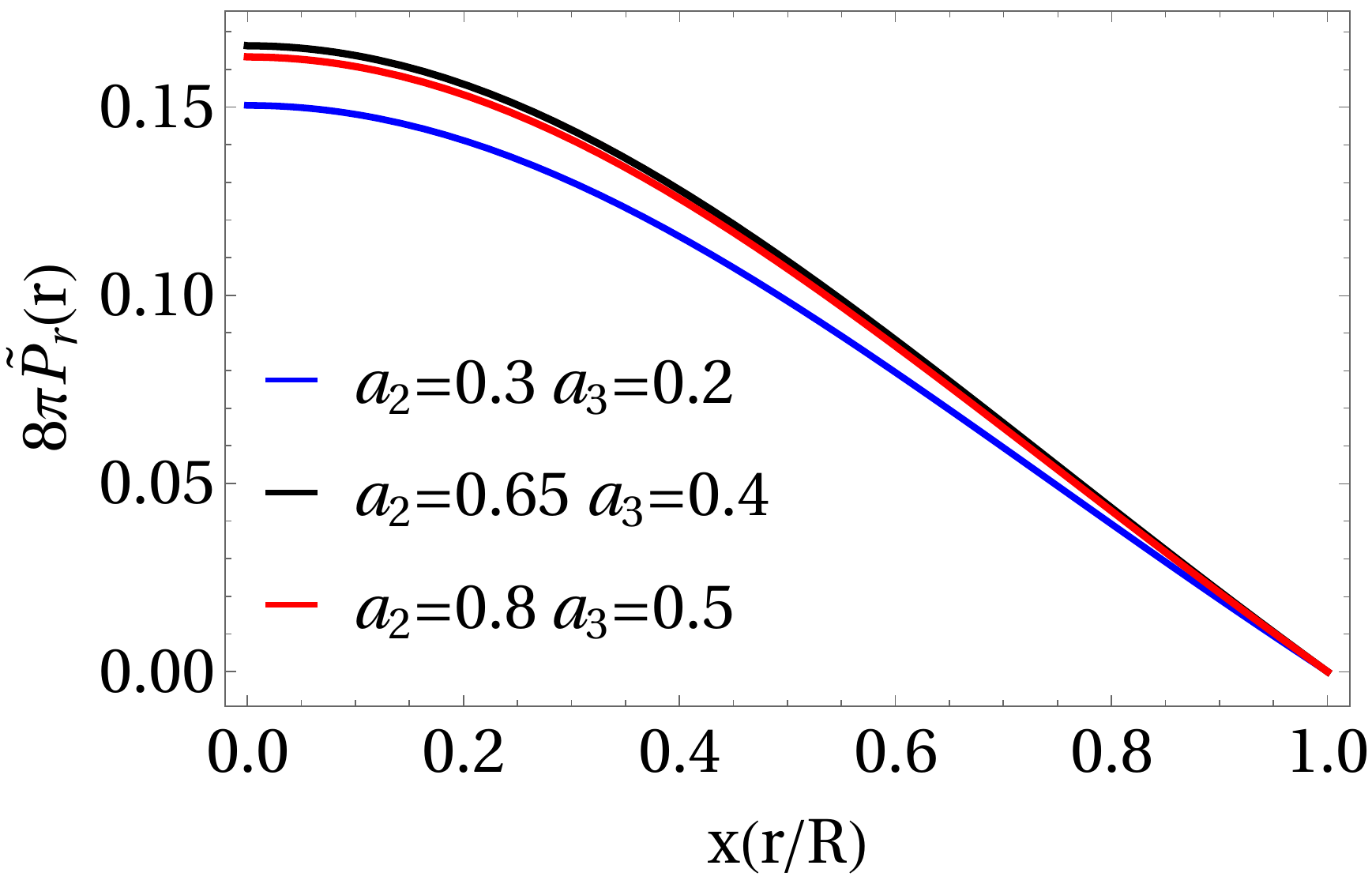}
        \label{rpressure4b}}    
        \caption{$\tilde{P}_r(r)$ as a function of $r$ for Model 1: (a) u = 0.19803, (b)  u = 0.2035,  Model 2: (c) u = 0.19803, (d) u = 0.2035,  Model 3: (e) u = 0.19803, (f) u = 0.2035,  Model 4: (g) u = 0.19803, (h)  u = 0.2035.}
        \label{rpressure}
  \end{center}
\end{figure}

\begin{figure}[htb!]
  \begin{center}
    \subfigure[]{
        \includegraphics[scale=0.223]{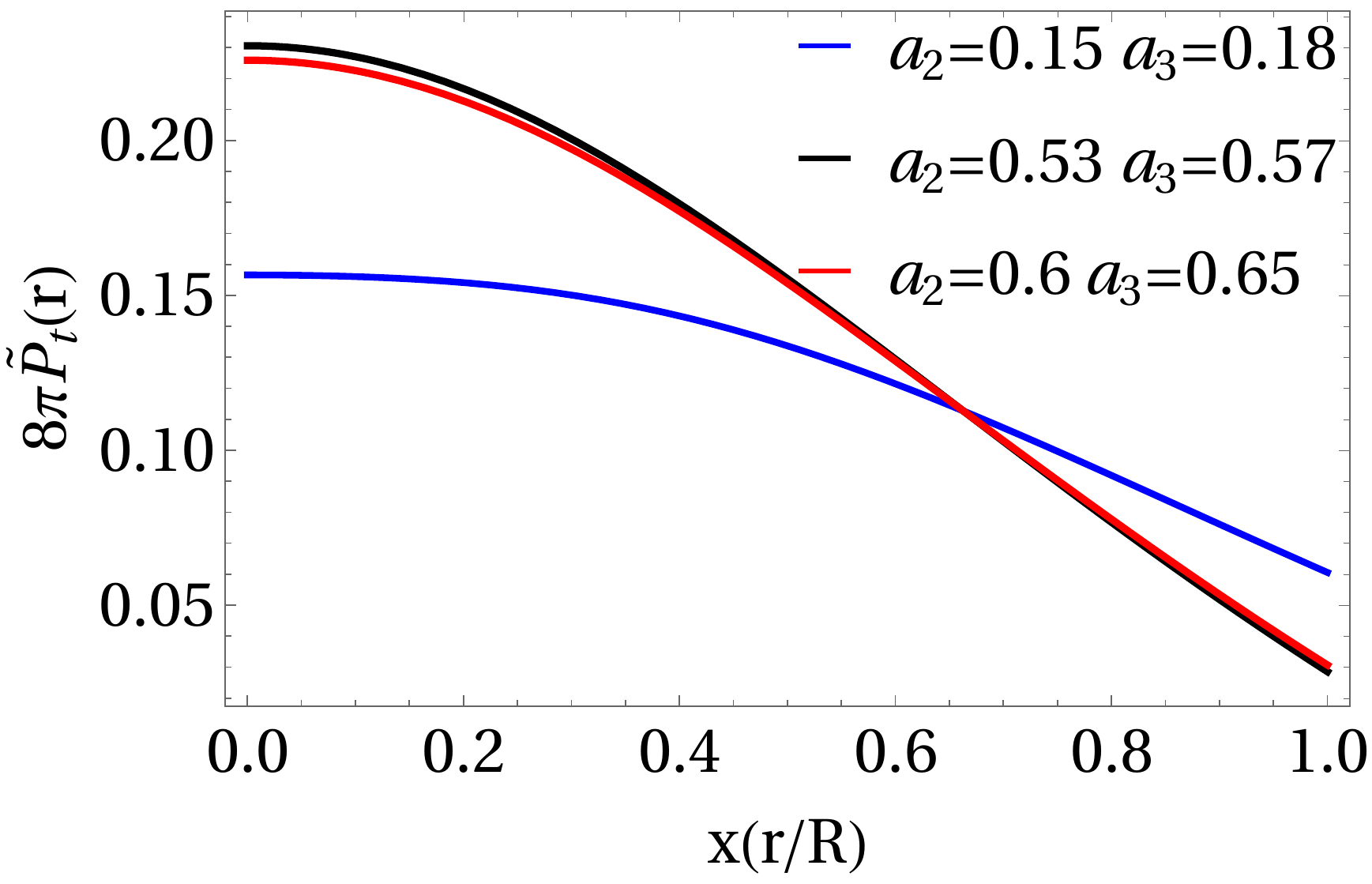}
        \label{tpressure1a}}
    \subfigure[]{
        \includegraphics[scale=0.223]{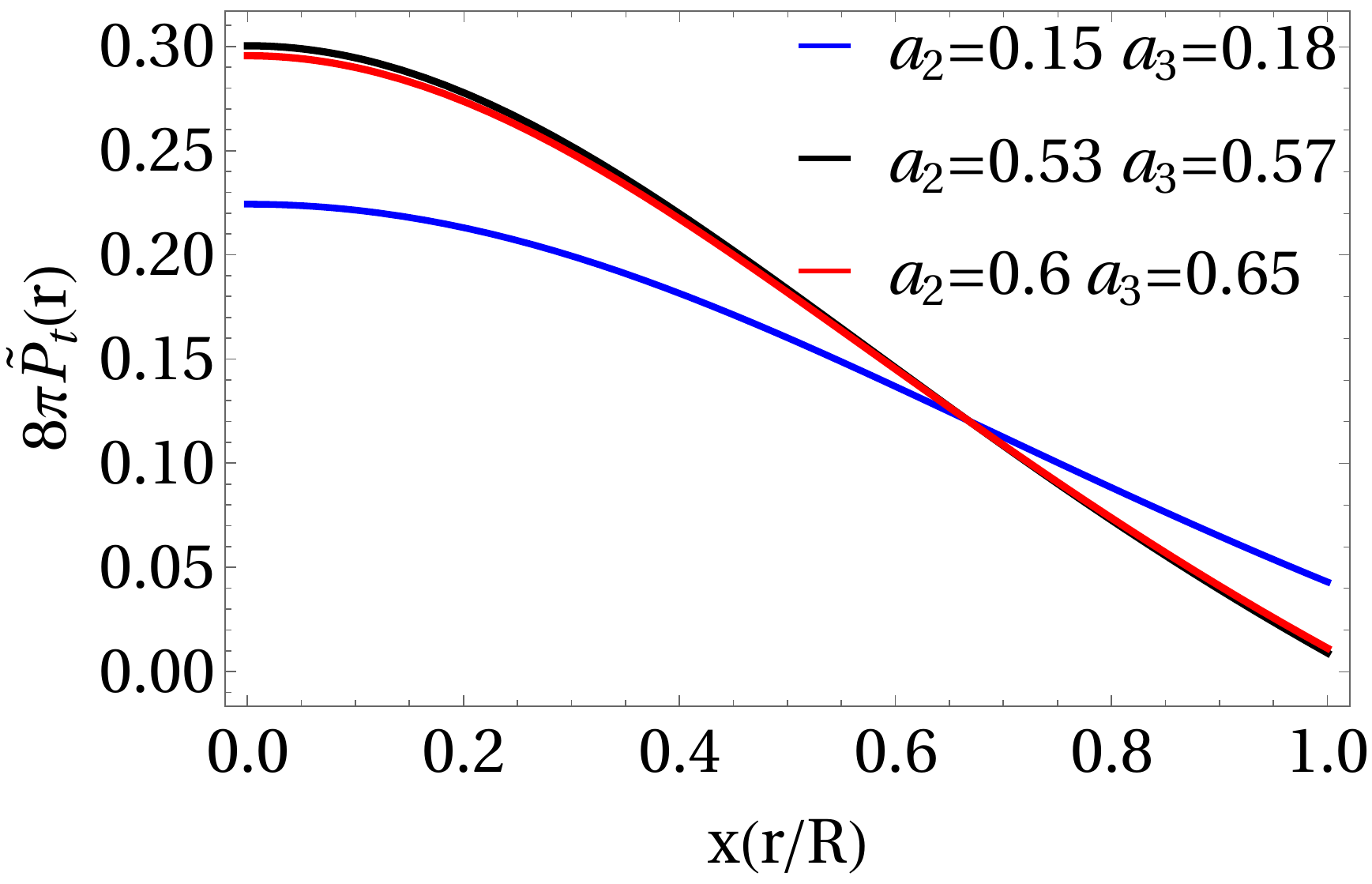}
        \label{tpressure1b}}
    \subfigure[]{
        \includegraphics[scale=0.223]{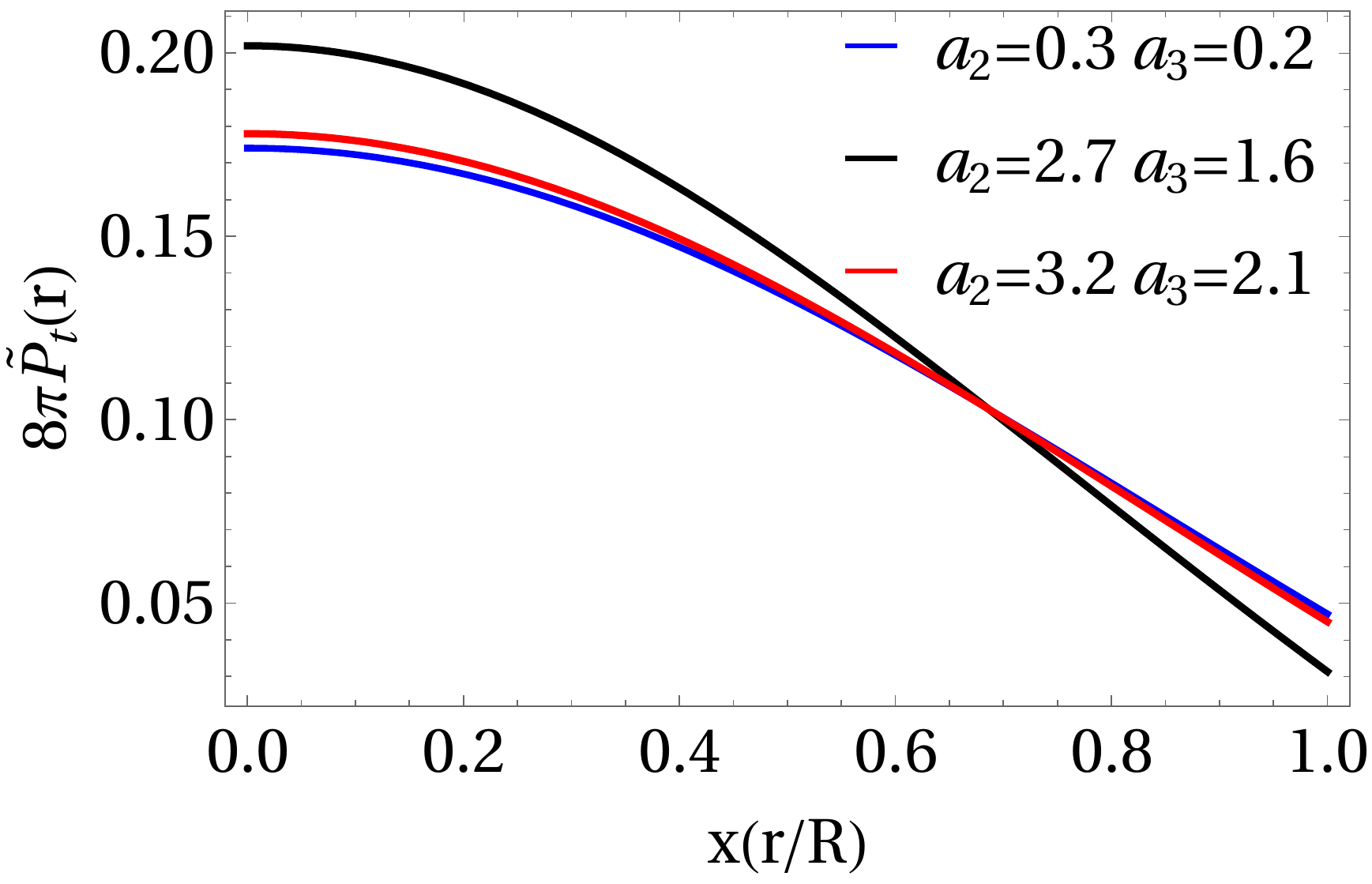}
        \label{tpressure2a}}
    \subfigure[]{
        \includegraphics[scale=0.223]{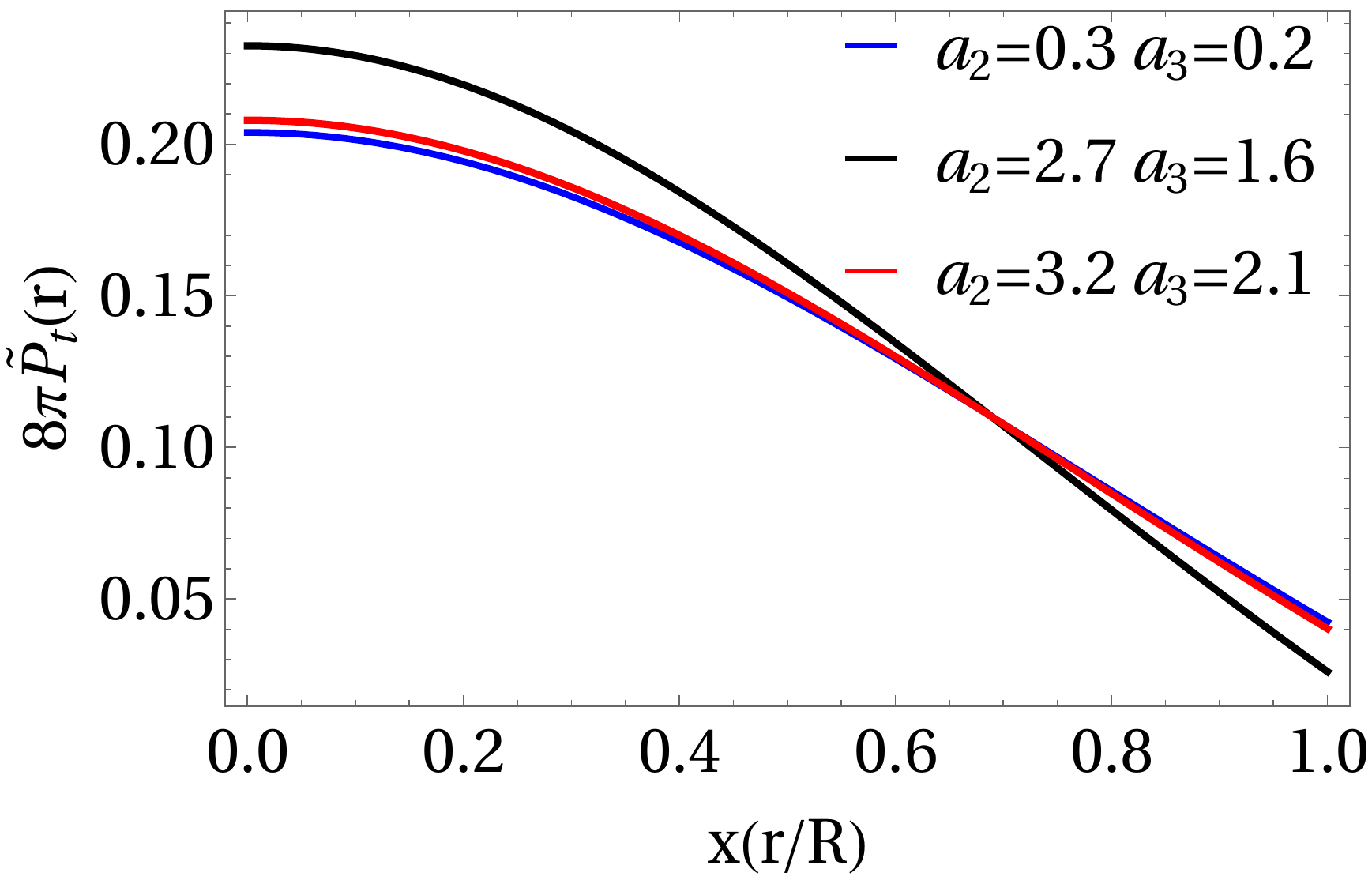}
        \label{tpressure2b}}
    \subfigure[]{
        \includegraphics[scale=0.223]{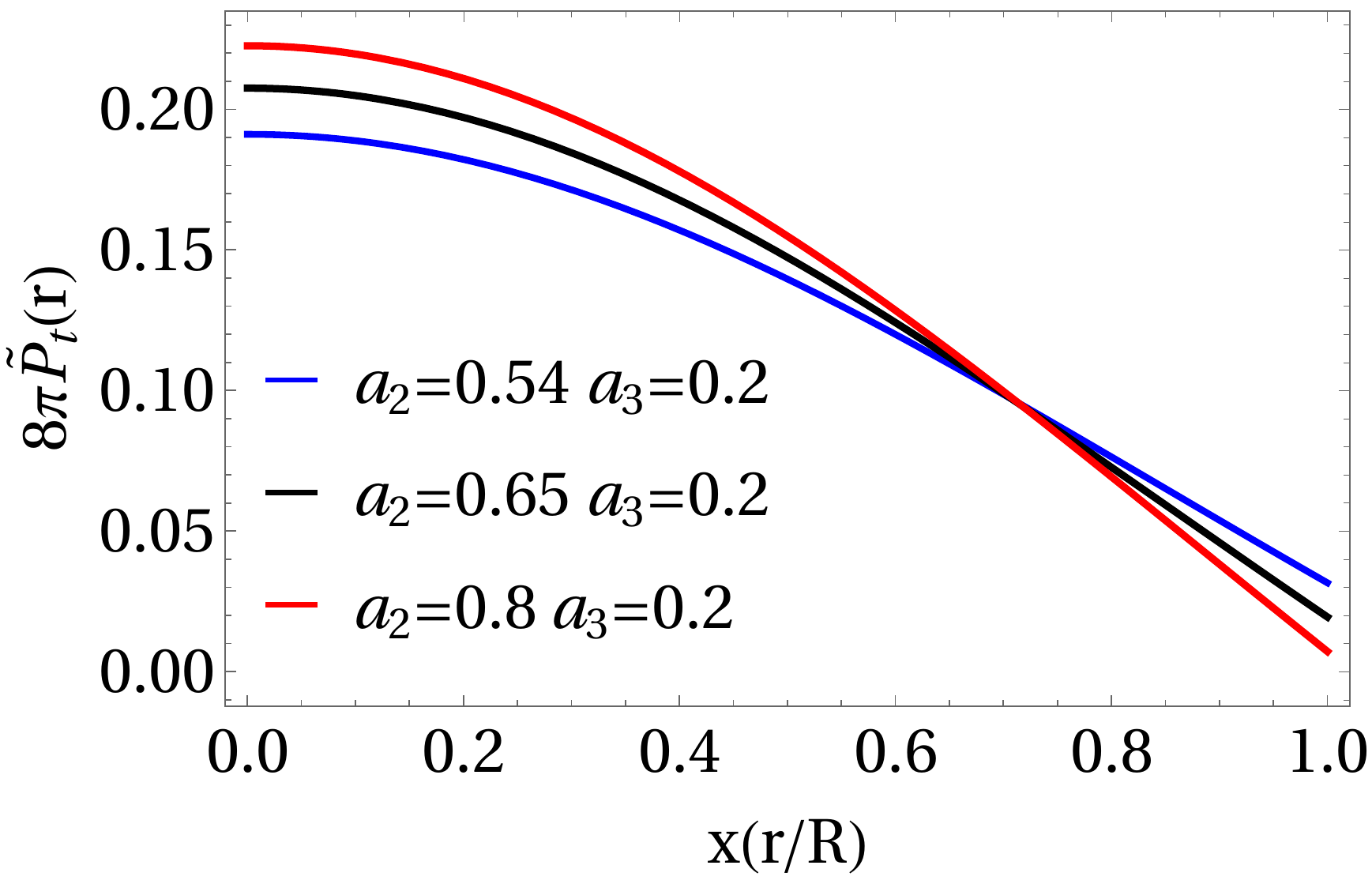}
        \label{tpressure3a}}
    \subfigure[]{
        \includegraphics[scale=0.223]{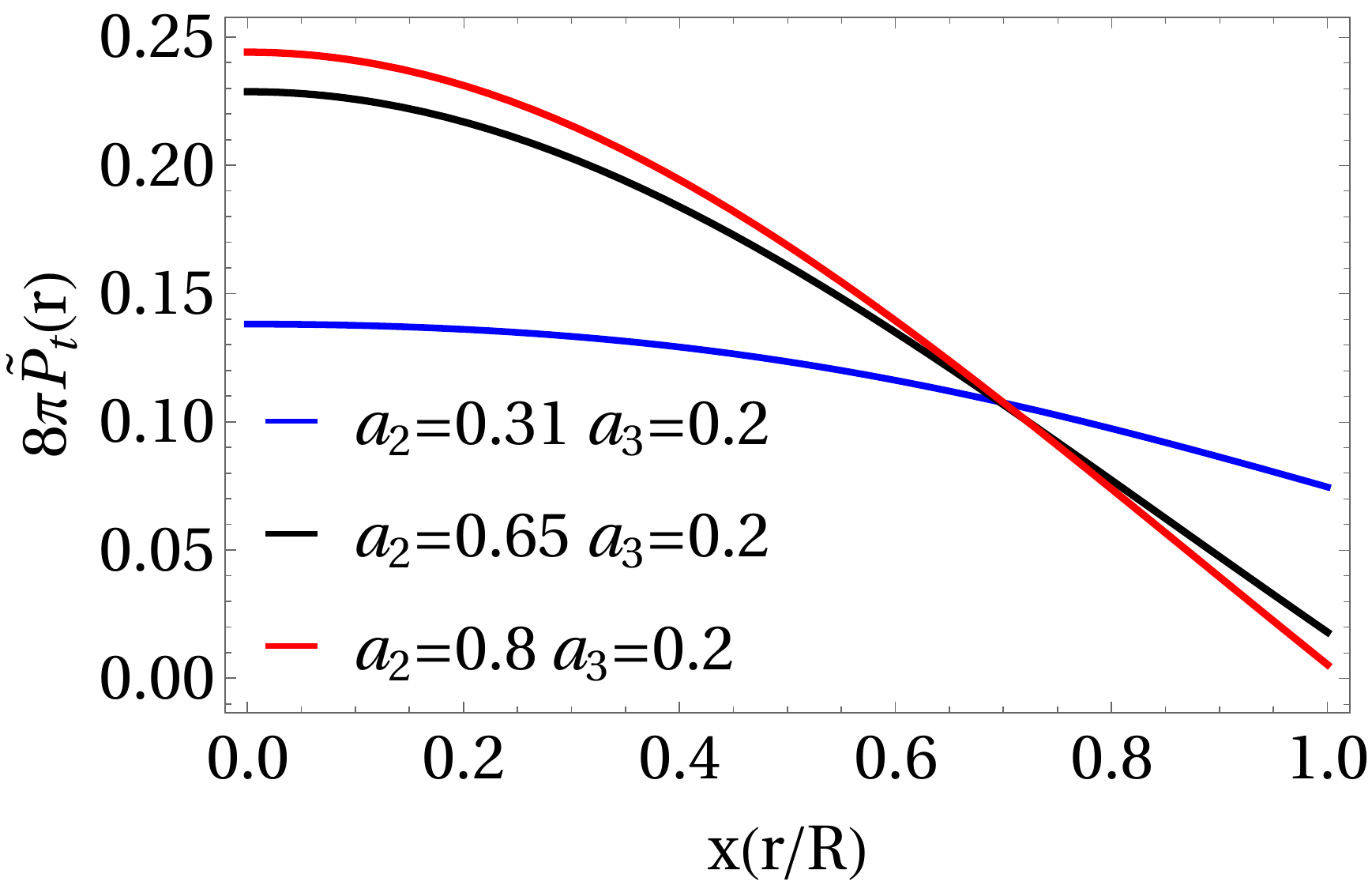}
        \label{tpressure3b}}
        \subfigure[]{
        \includegraphics[scale=0.223]{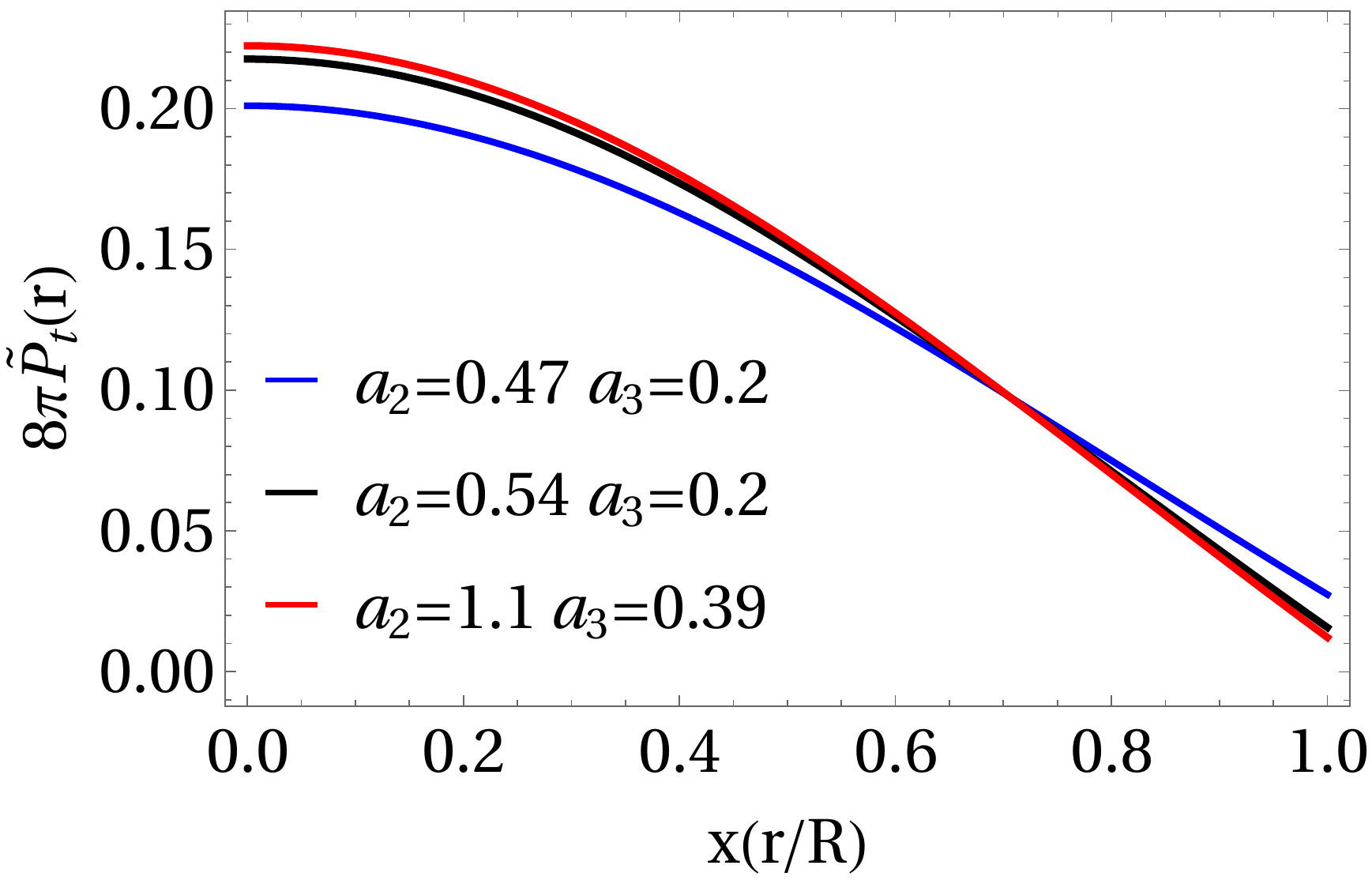}
        \label{tpressure4a}}
    \subfigure[]{
        \includegraphics[scale=0.223]{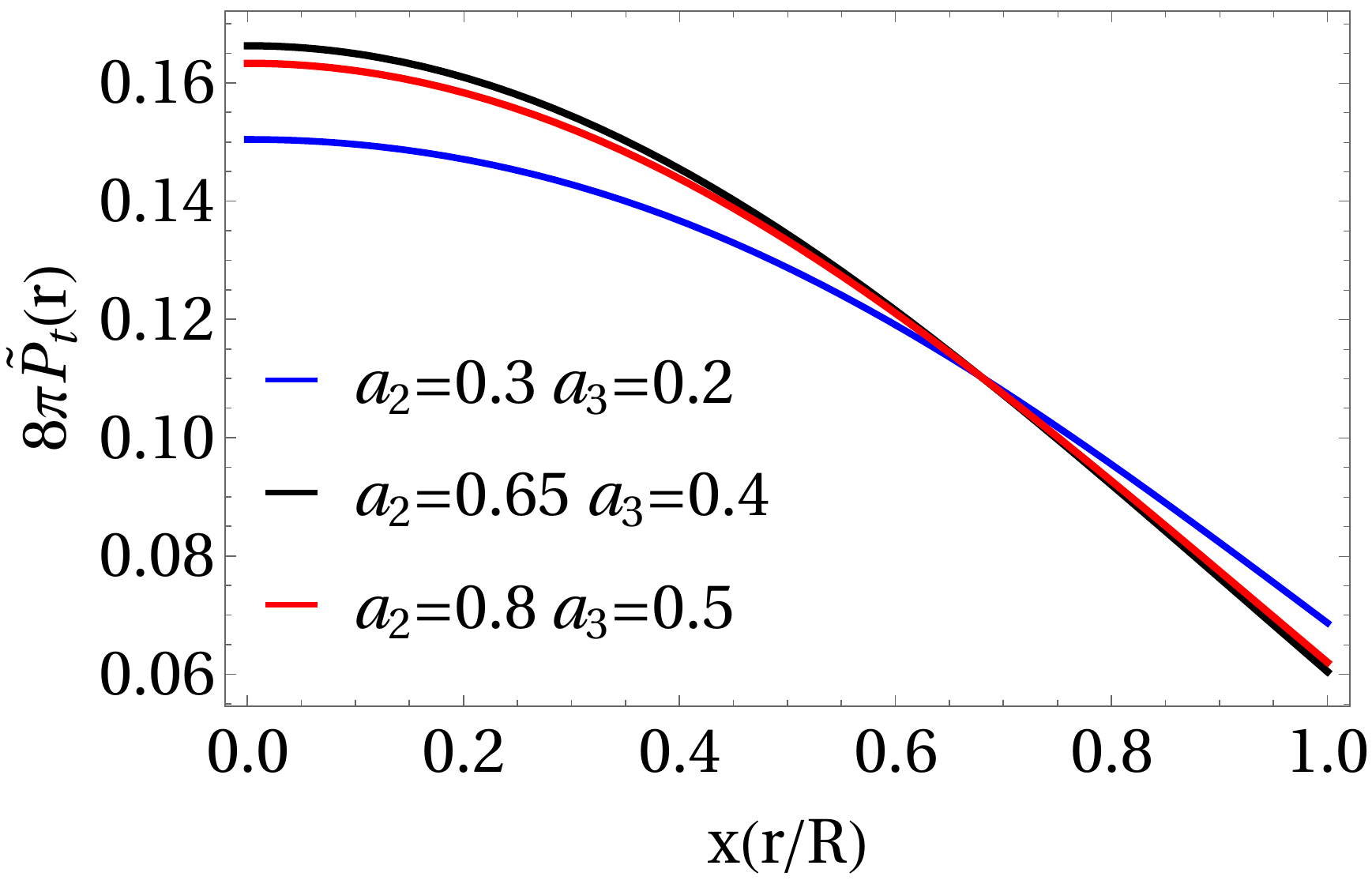}
        \label{tpressure4b}}  
    \caption{$\tilde{P}_t(r)$ as a function of $r$ for Model 1: (a) u = 0.19803, (b)  u = 0.2035,  Model 2: (c) u = 0.19803, (d) u = 0.2035,  Model 3: (e) u = 0.19803, (f) u = 0.2035,  Model 4: (g) u = 0.19803, (h)  u = 0.2035.}
    \label{tpressure}
  \end{center}
\end{figure}

\begin{figure}[ht!]
  \begin{center}
    \subfigure[]{
        \includegraphics[scale=0.223]{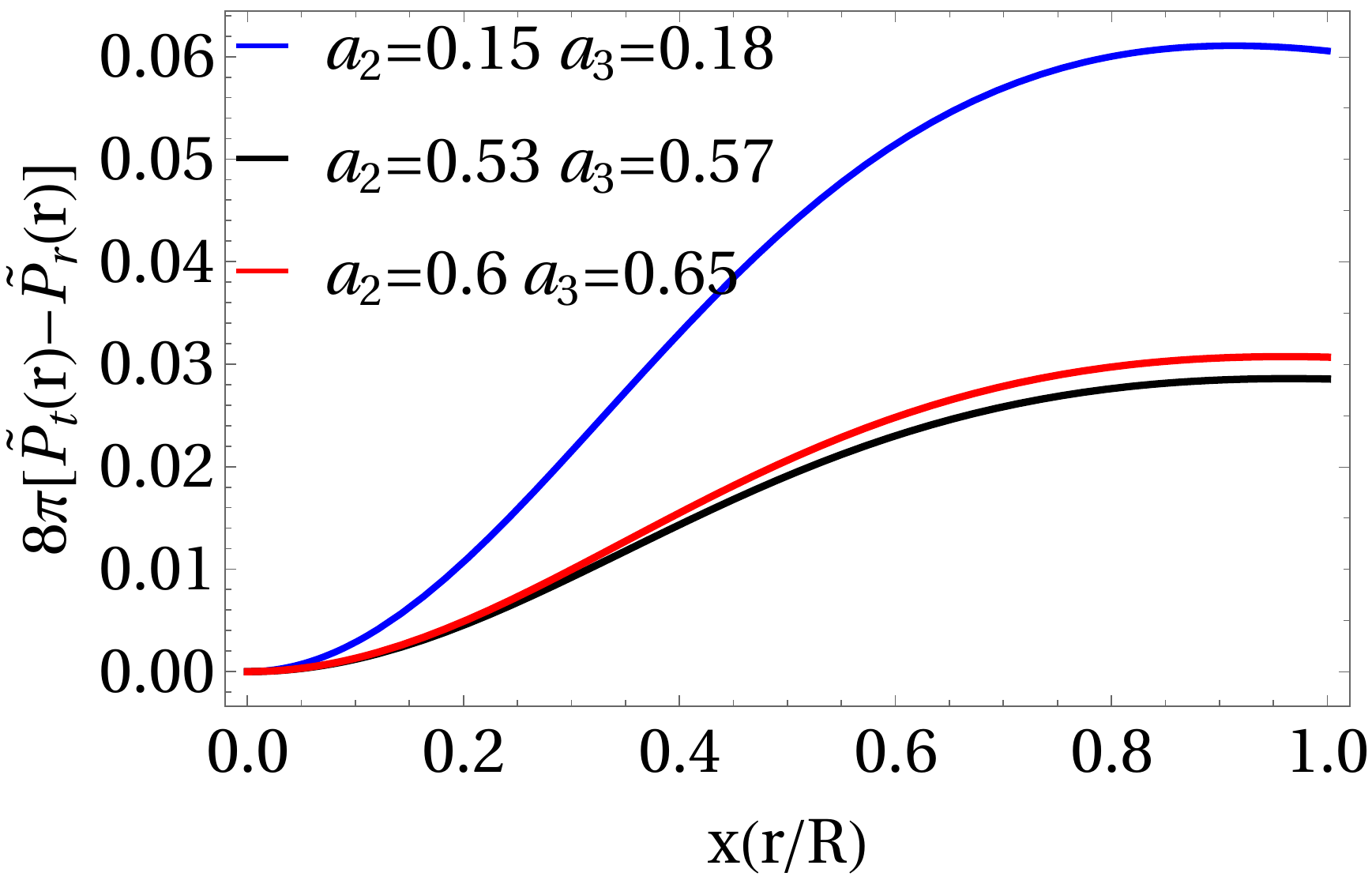}
        \label{delta1a}}
    \subfigure[]{
        \includegraphics[scale=0.223]{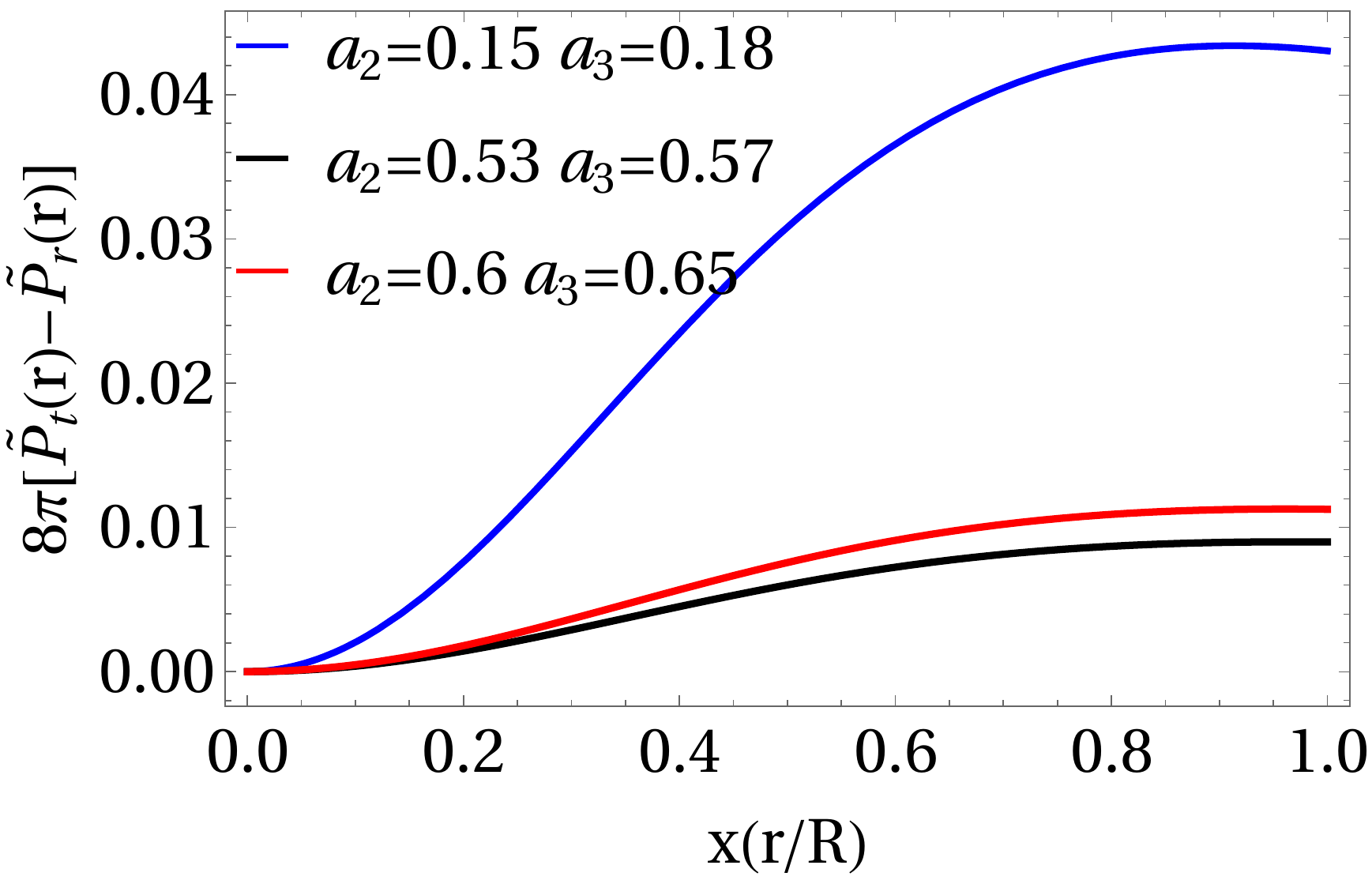}
        \label{delta1b}}
    \subfigure[]{
        \includegraphics[scale=0.223]{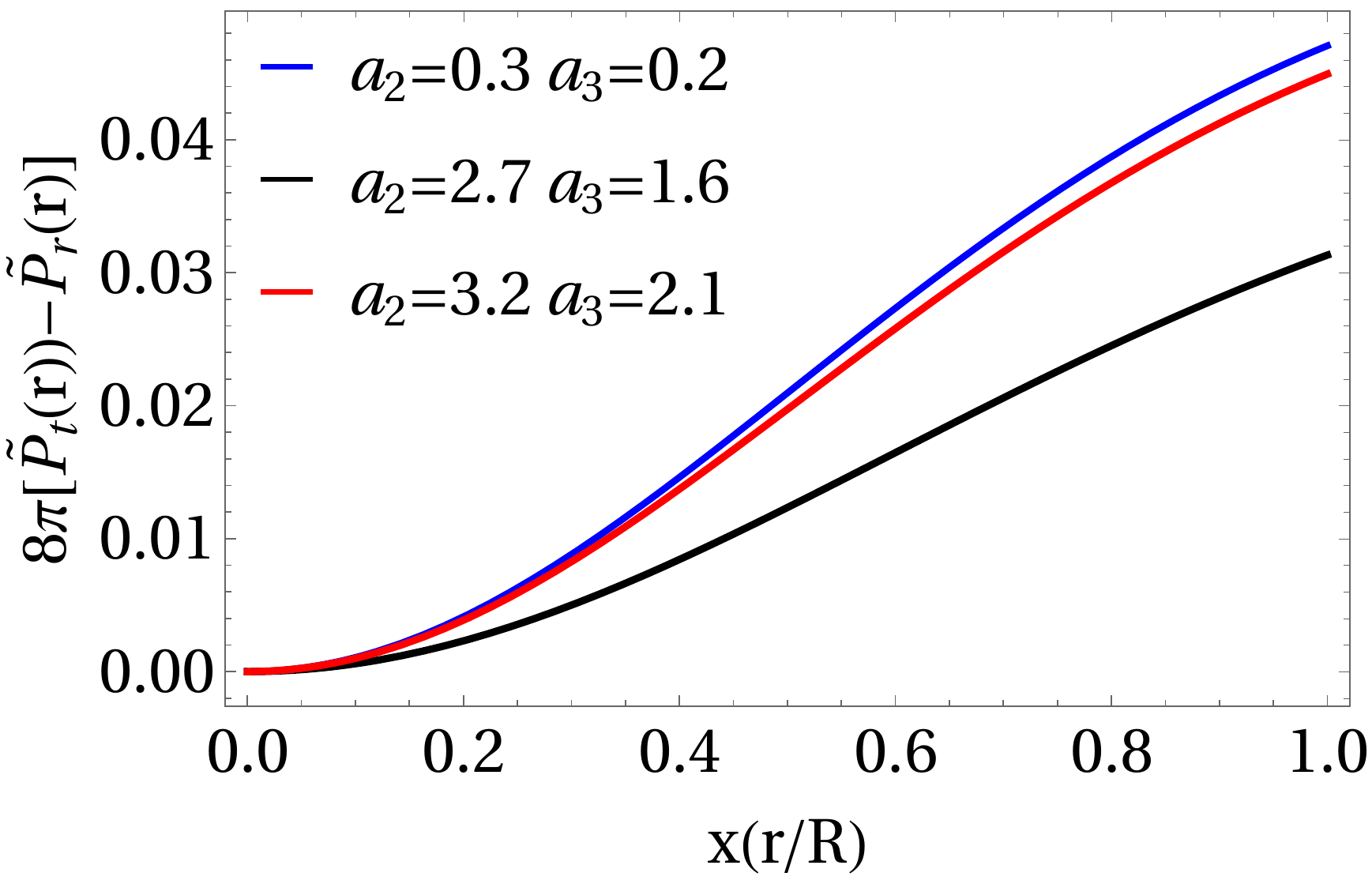}
        \label{delta2a}}
    \subfigure[]{
        \includegraphics[scale=0.223]{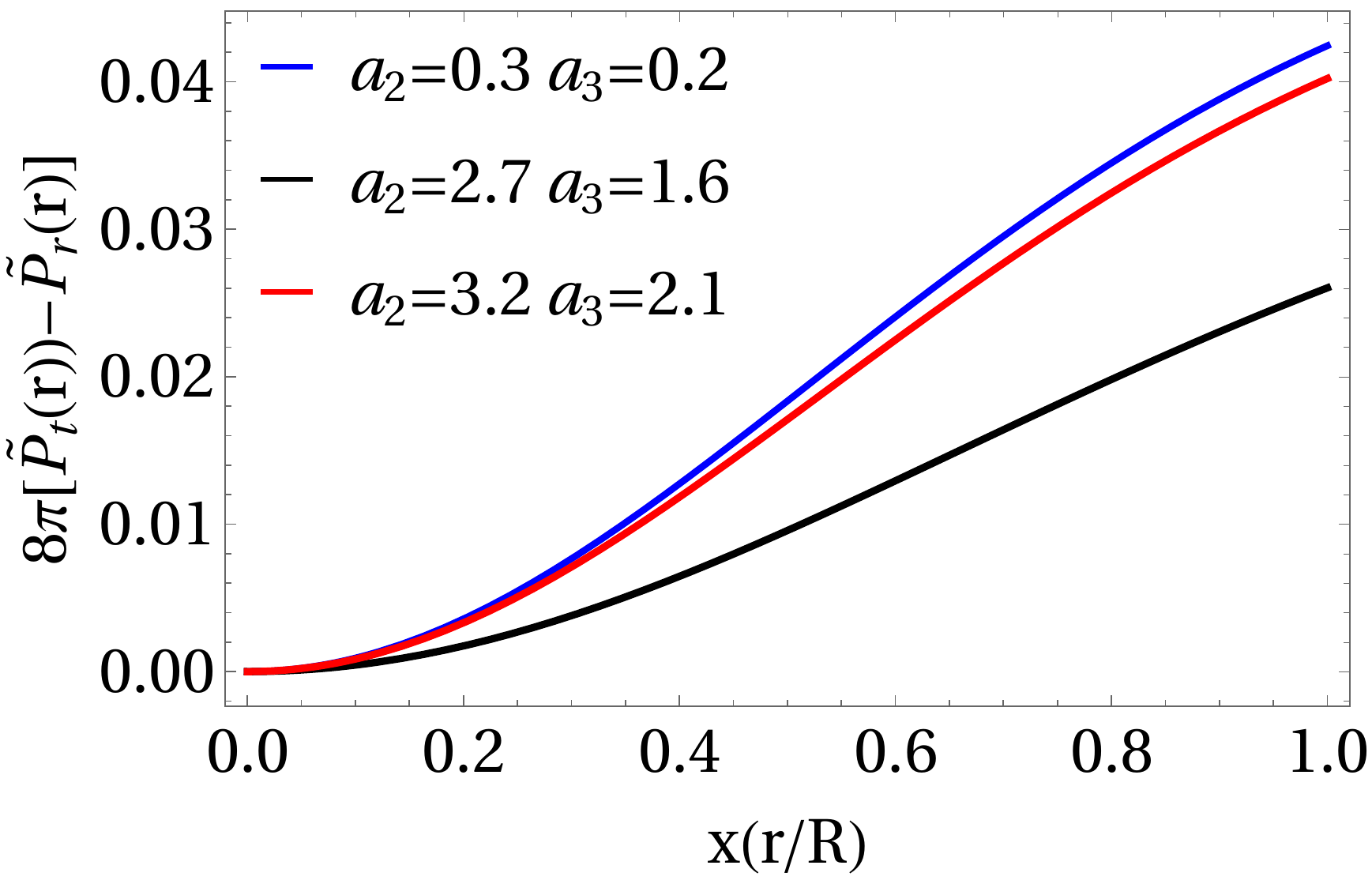}
        \label{delta2b}}
    \subfigure[]{
        \includegraphics[scale=0.223]{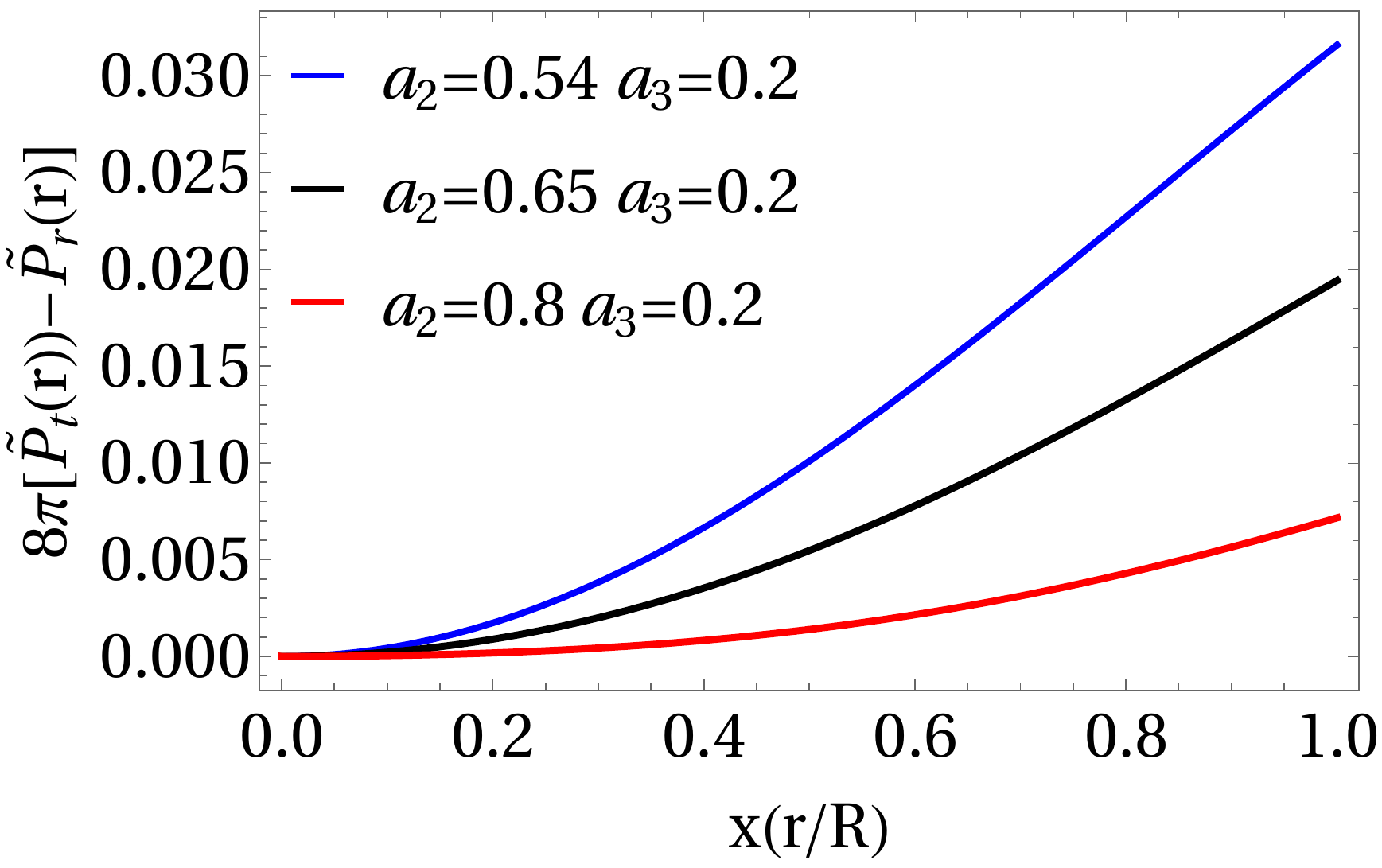}
        \label{delta3a}}
    \subfigure[]{
        \includegraphics[scale=0.223]{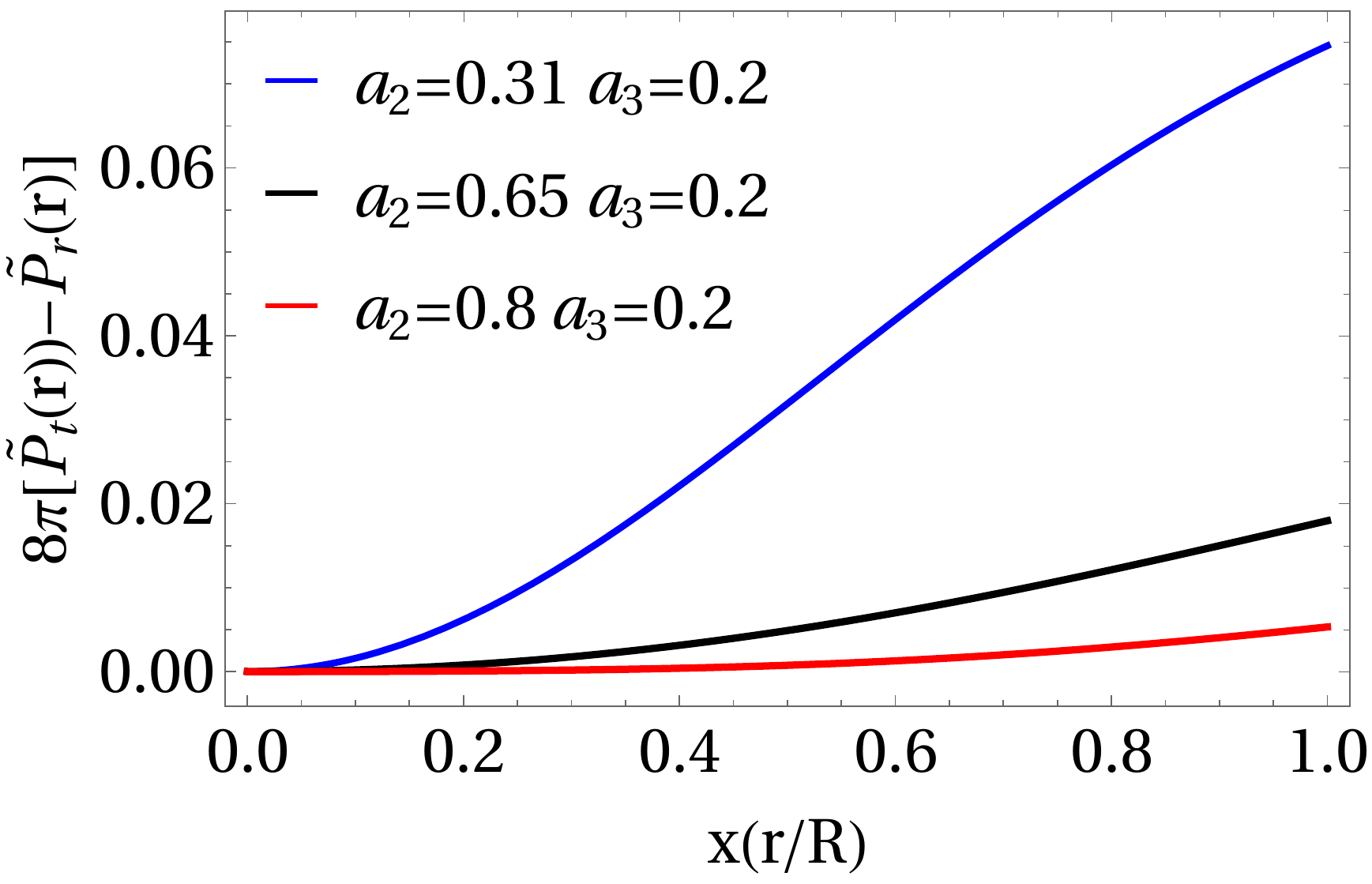}
        \label{delta3b}}
    \subfigure[]{
        \includegraphics[scale=0.223]{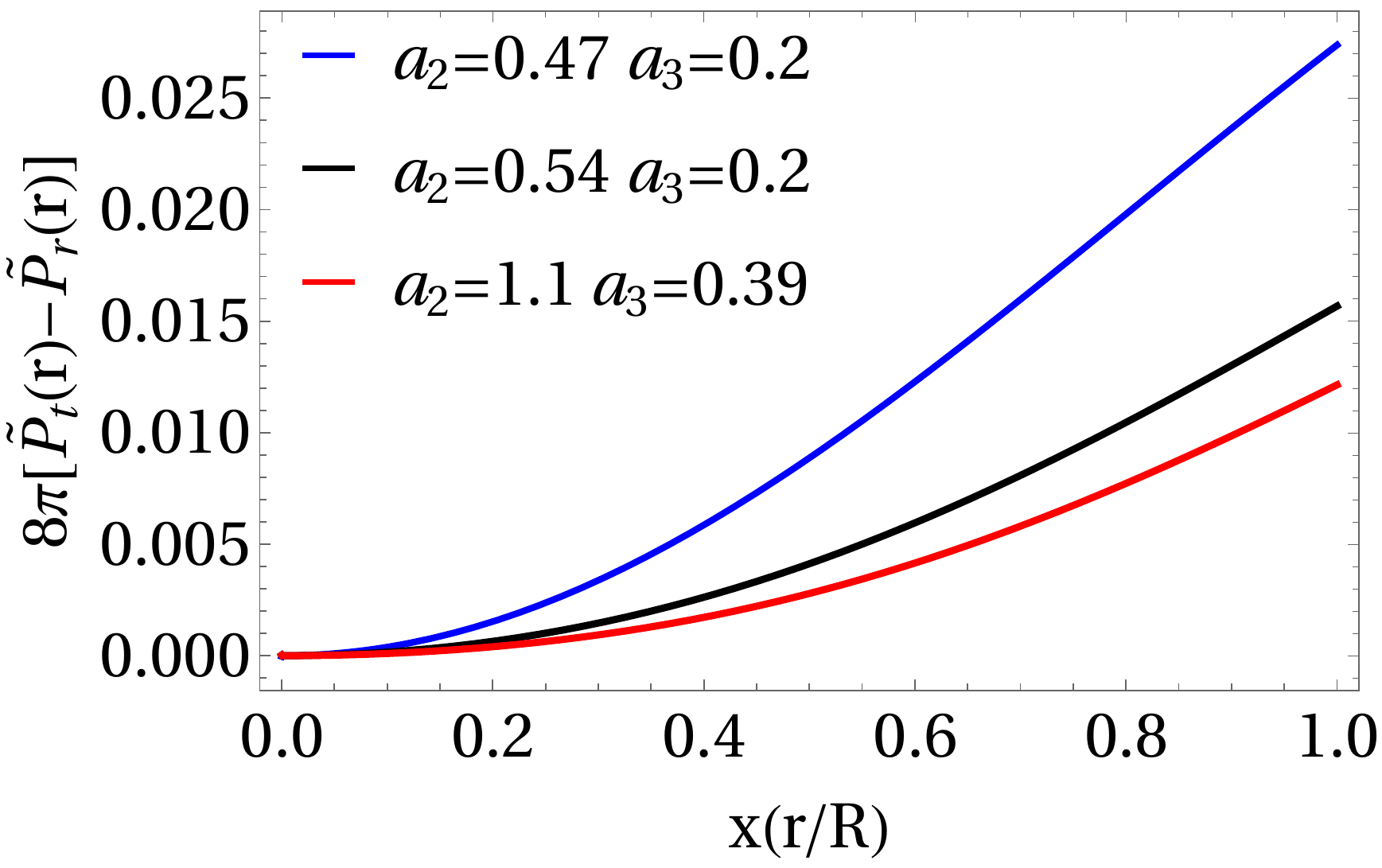}
        \label{delta4a}}
    \subfigure[]{
        \includegraphics[scale=0.223]{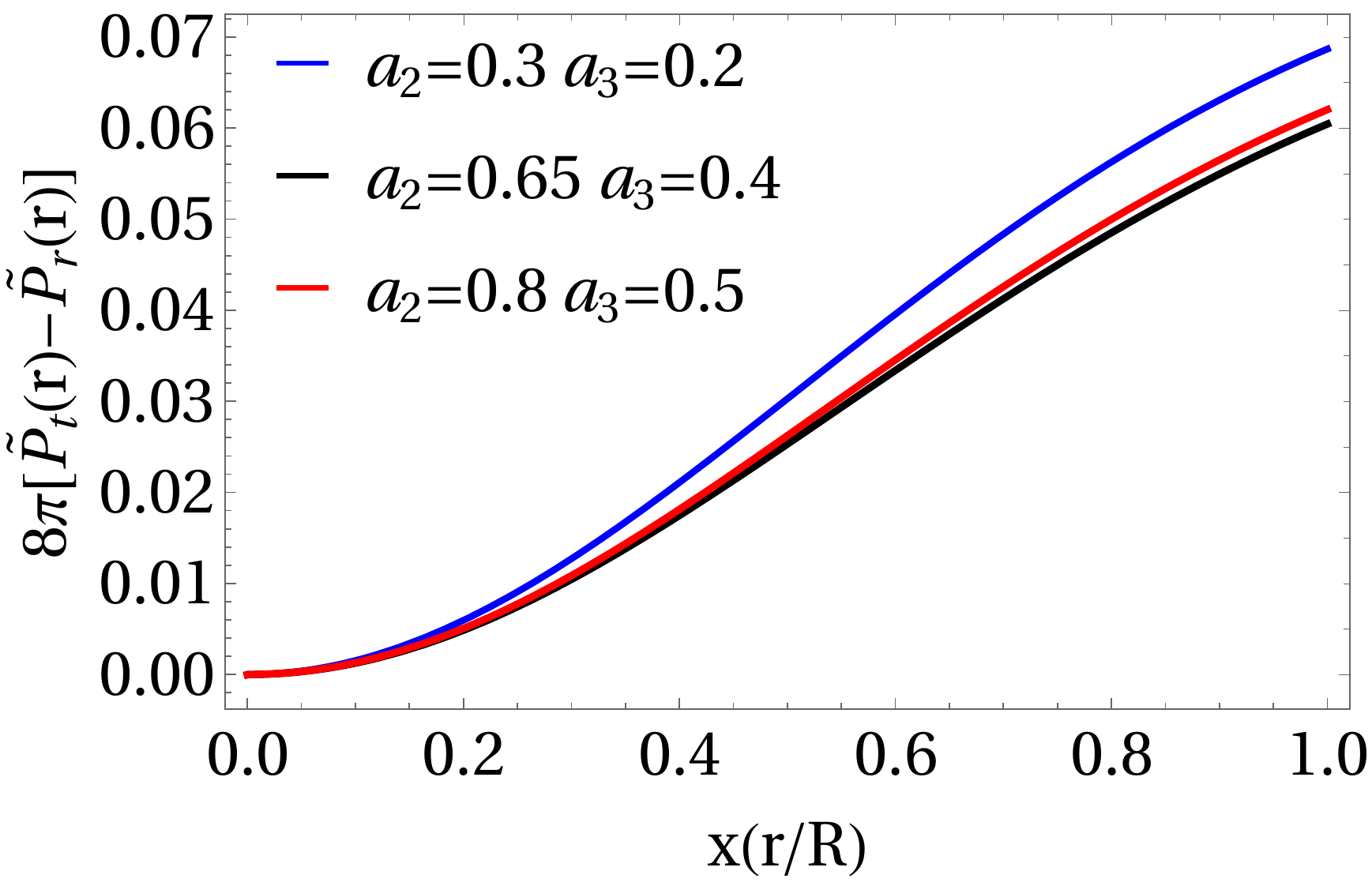}
        \label{delta4b}}                
    \caption{$\tilde{P}_t(r)-\tilde{P}_r(r)$ as a function of $r$ for Model 1: (a) u = 0.19803, (b)  u = 0.2035,  Model 2: (c) u = 0.19803, (d) u = 0.2035,  Model 3: (e) u = 0.19803, (f) u = 0.2035,  Model 4: (g) u = 0.19803, (h)  u = 0.2035.}
    \label{delta}
  \end{center}
\end{figure}

In figures \ref{density}, \ref{rpressure} and \ref{tpressure} we show the profile of $\tilde{\rho}$, $\tilde{P}_{r}$ and $\tilde{P}_{t}$ as a function of the radial coordinates for the values of the parameters in the legend. 
Note that all the quantities fulfill the physical requirements for all the parameters involved, namely $\tilde{\rho}$, $\tilde{P}_{r}$ and $\tilde{P}_{t}$ are finite at the center and decrease monotonously toward the surface. Besides, $\tilde{P}_{t}(0)=\tilde{P}_{r}(0)$ and $\tilde{P}_{t}(r)>\tilde{P}_{r}(r)$ for all $r>0$ as expected (see Fig. \ref{delta})

\subsection{Energy conditions and causality}
A suitable stellar model must satisfies the dominant energy condition (DEC) in order to avoid violation of causality. The DEC requires
\begin{eqnarray}
\tilde{\rho}-\tilde{P}_{r} \geq 0\label{decinequality1} \\
\tilde{\rho}-\tilde{P}_{t} \geq 0\label{decinequality2}.
\end{eqnarray}

\begin{figure}[ht!]
  \begin{center}
    \subfigure[]{
        \includegraphics[scale=0.223]{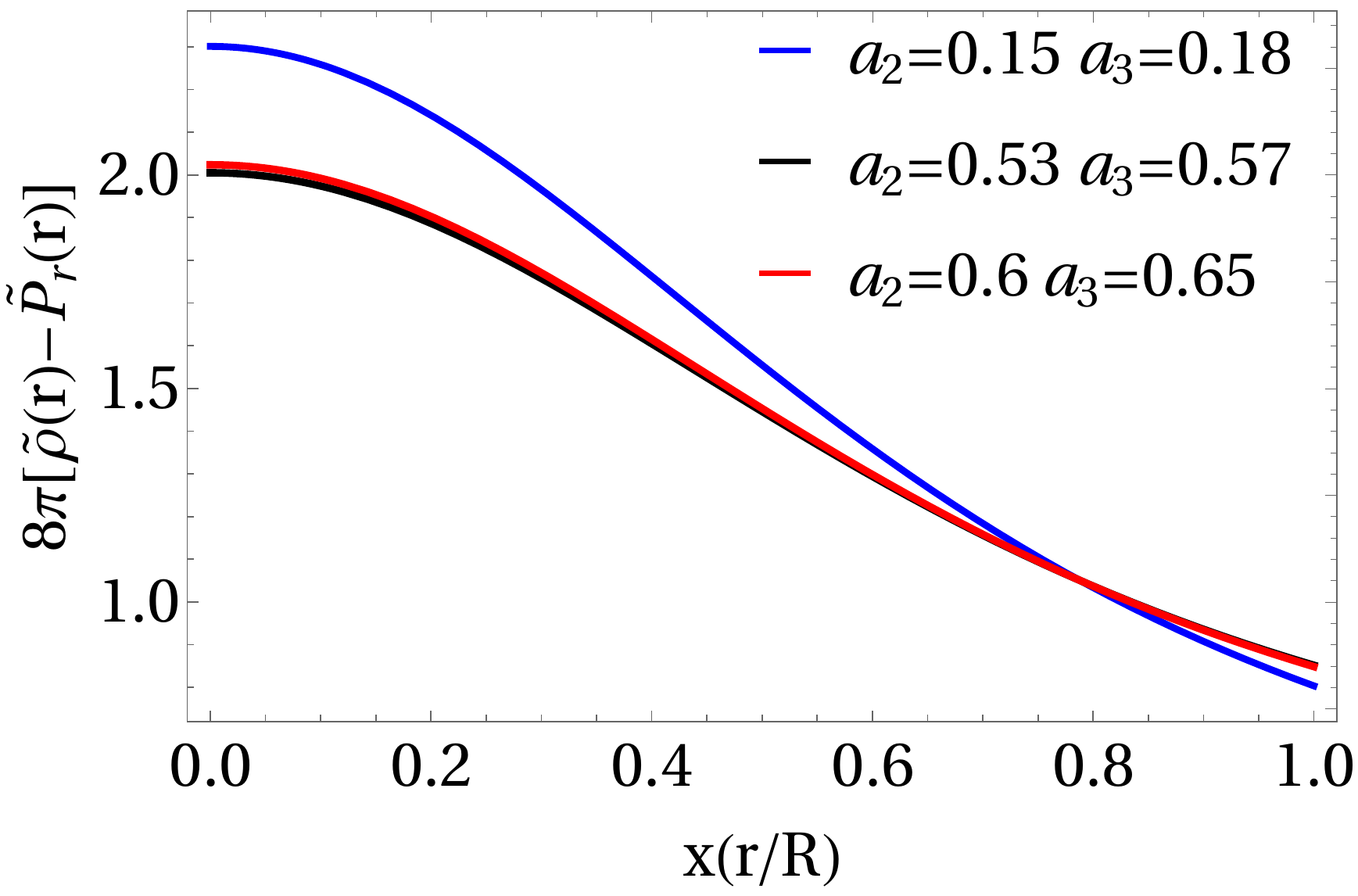}
        \label{decr1a}}
    \subfigure[]{
        \includegraphics[scale=0.223]{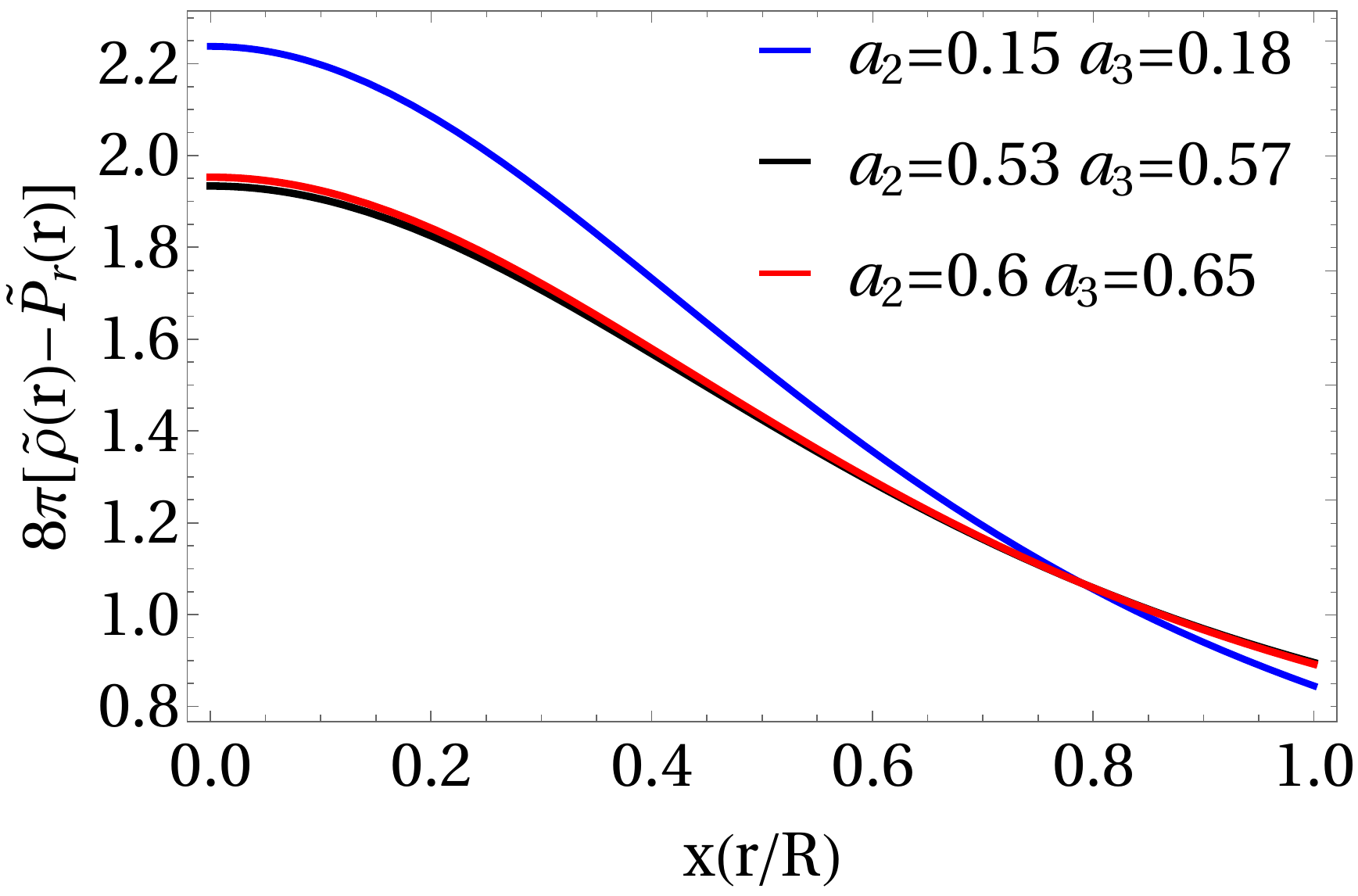}
        \label{decr1b}}
    \subfigure[]{
        \includegraphics[scale=0.223]{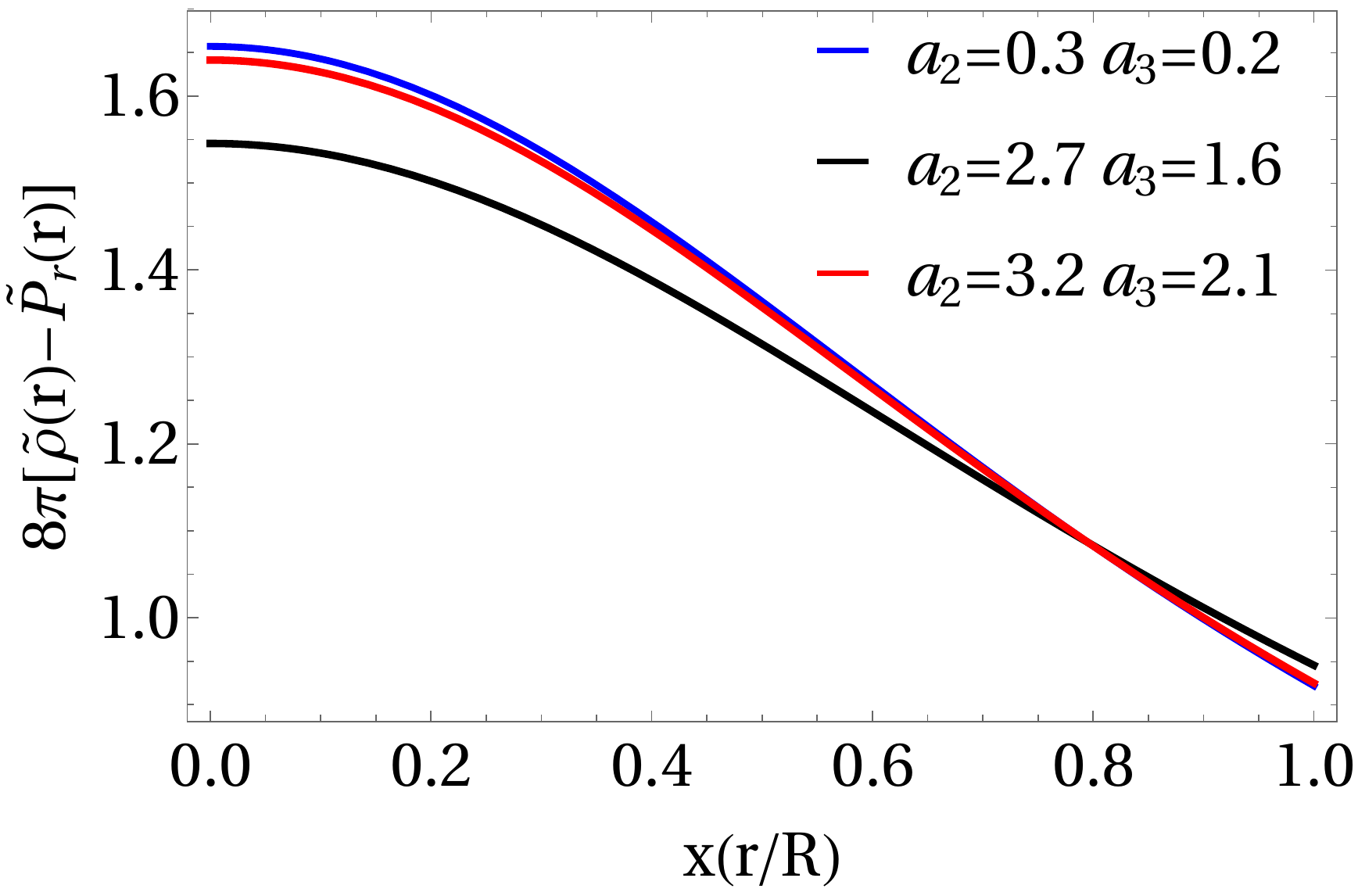}
        \label{decr2a}}
    \subfigure[]{
        \includegraphics[scale=0.223]{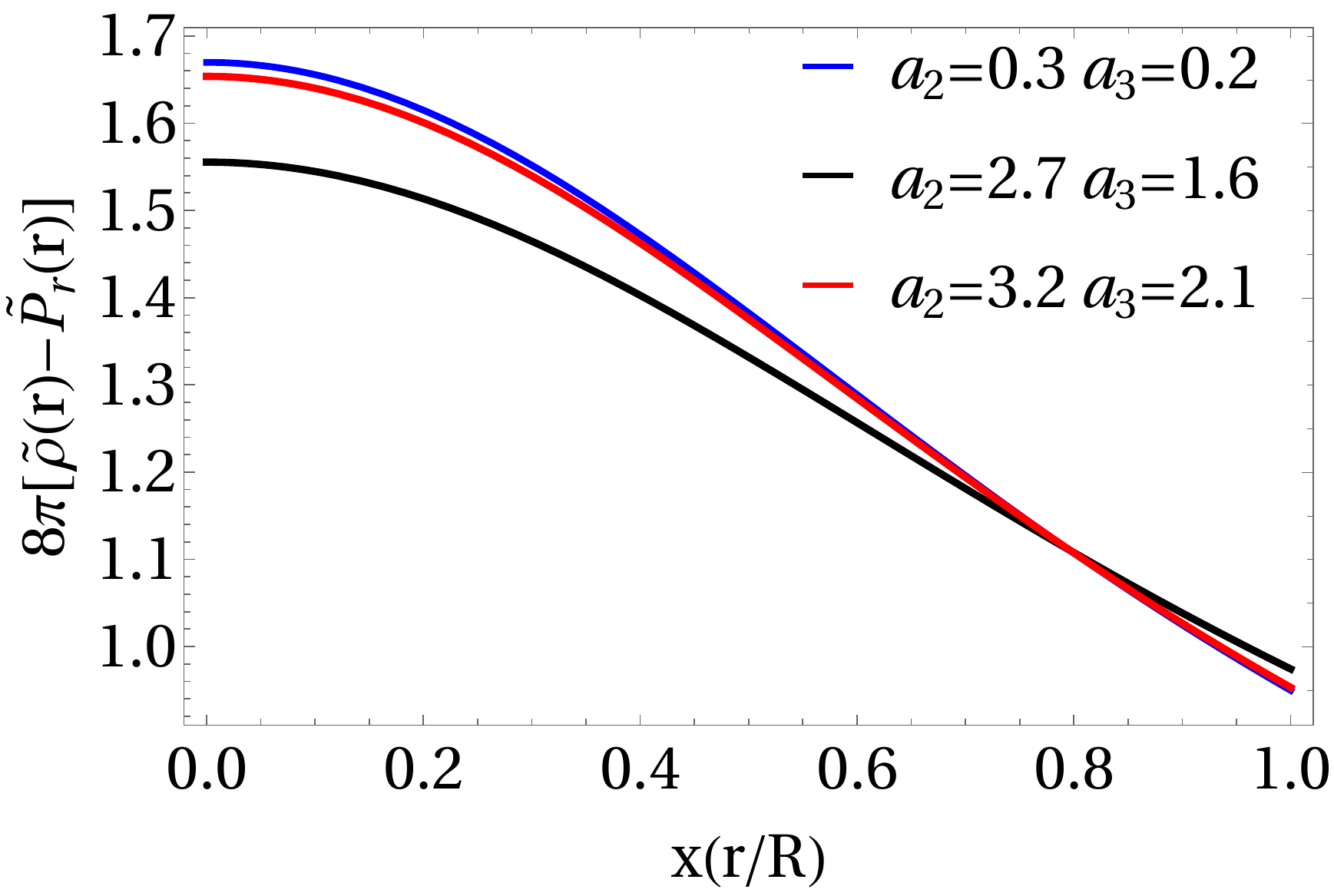}
        \label{decr2b}}
    \subfigure[]{
        \includegraphics[scale=0.223]{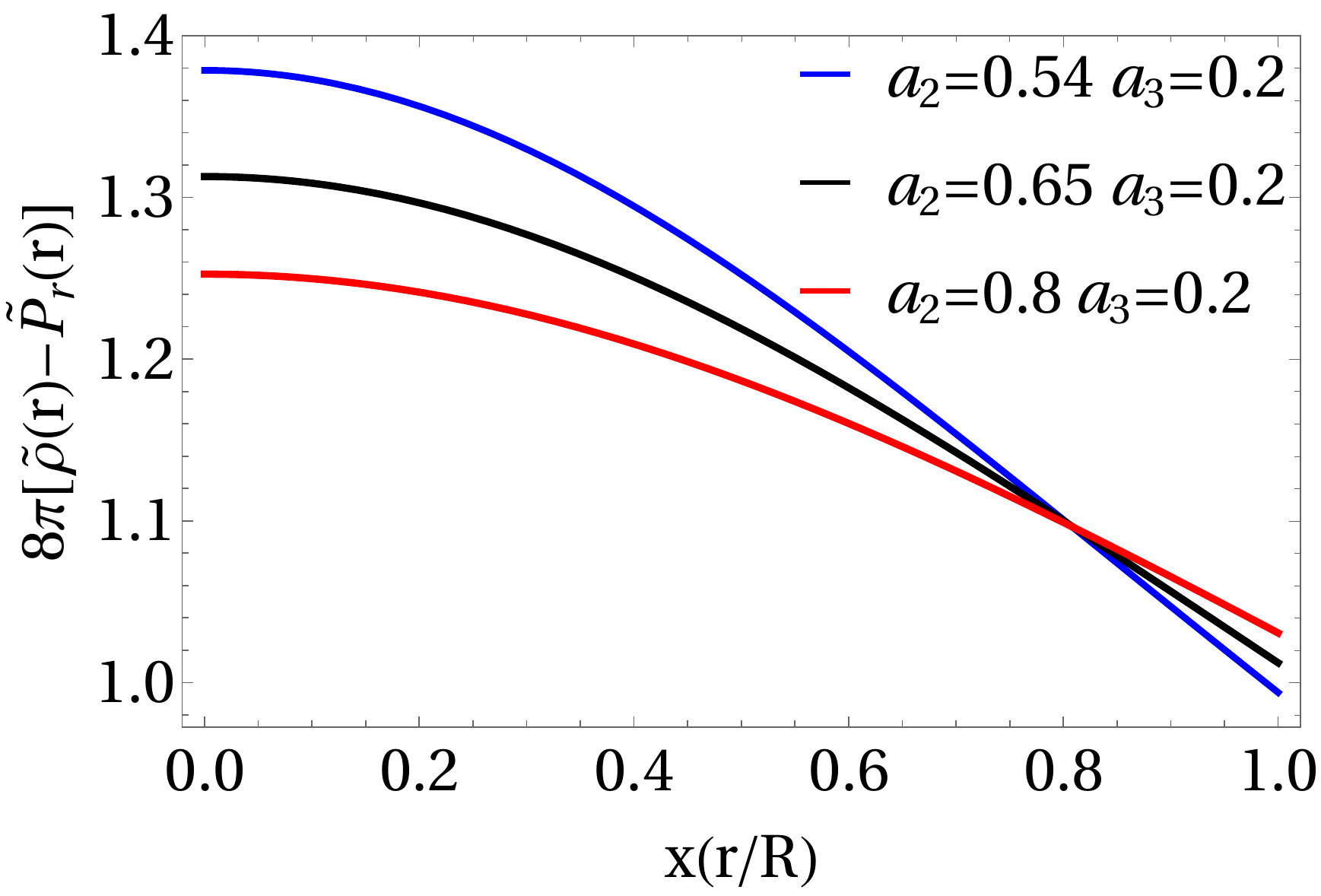}
        \label{decr3a}}
    \subfigure[]{
        \includegraphics[scale=0.223]{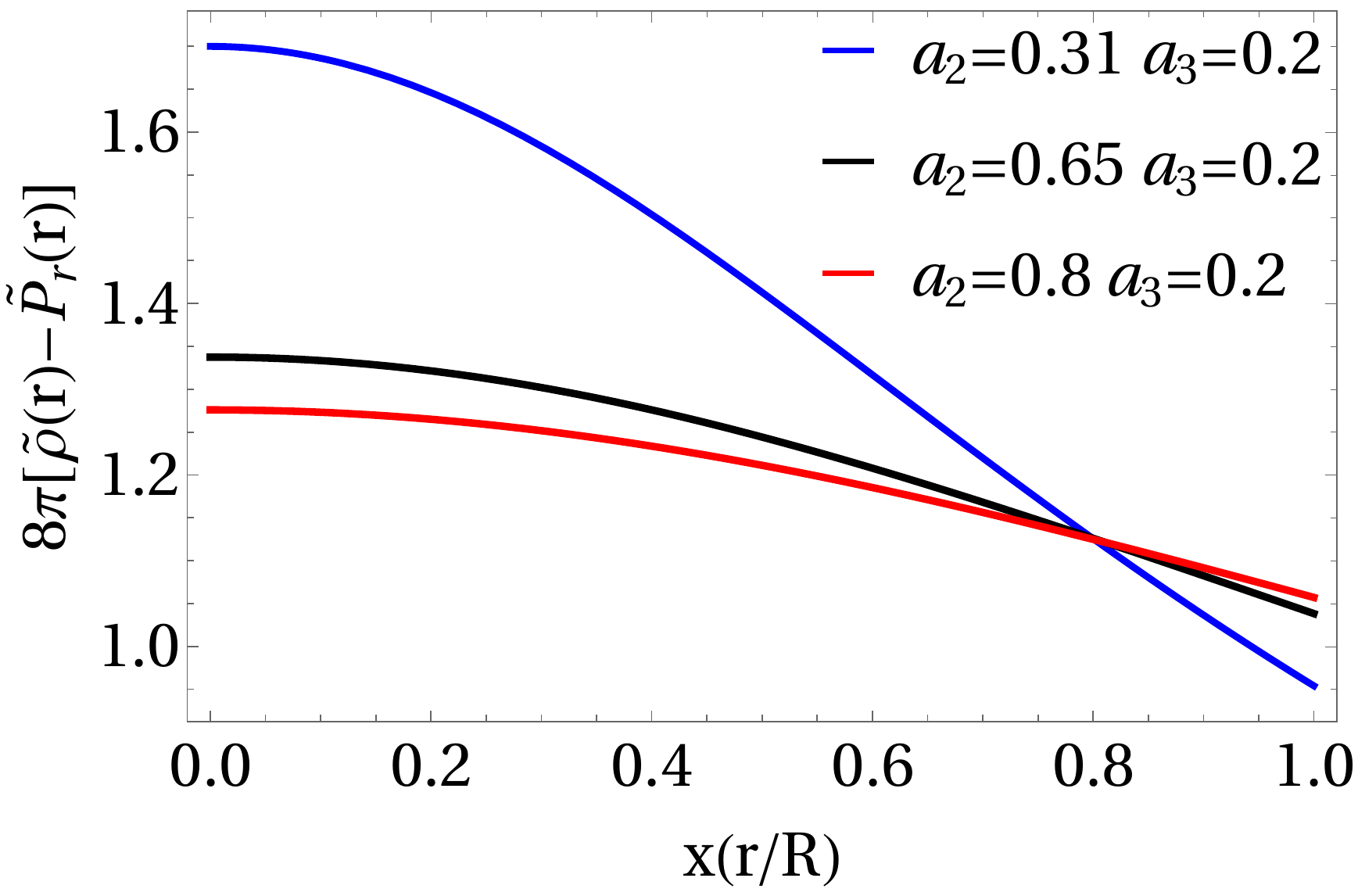}
        \label{decr3b}}
    \subfigure[]{
        \includegraphics[scale=0.223]{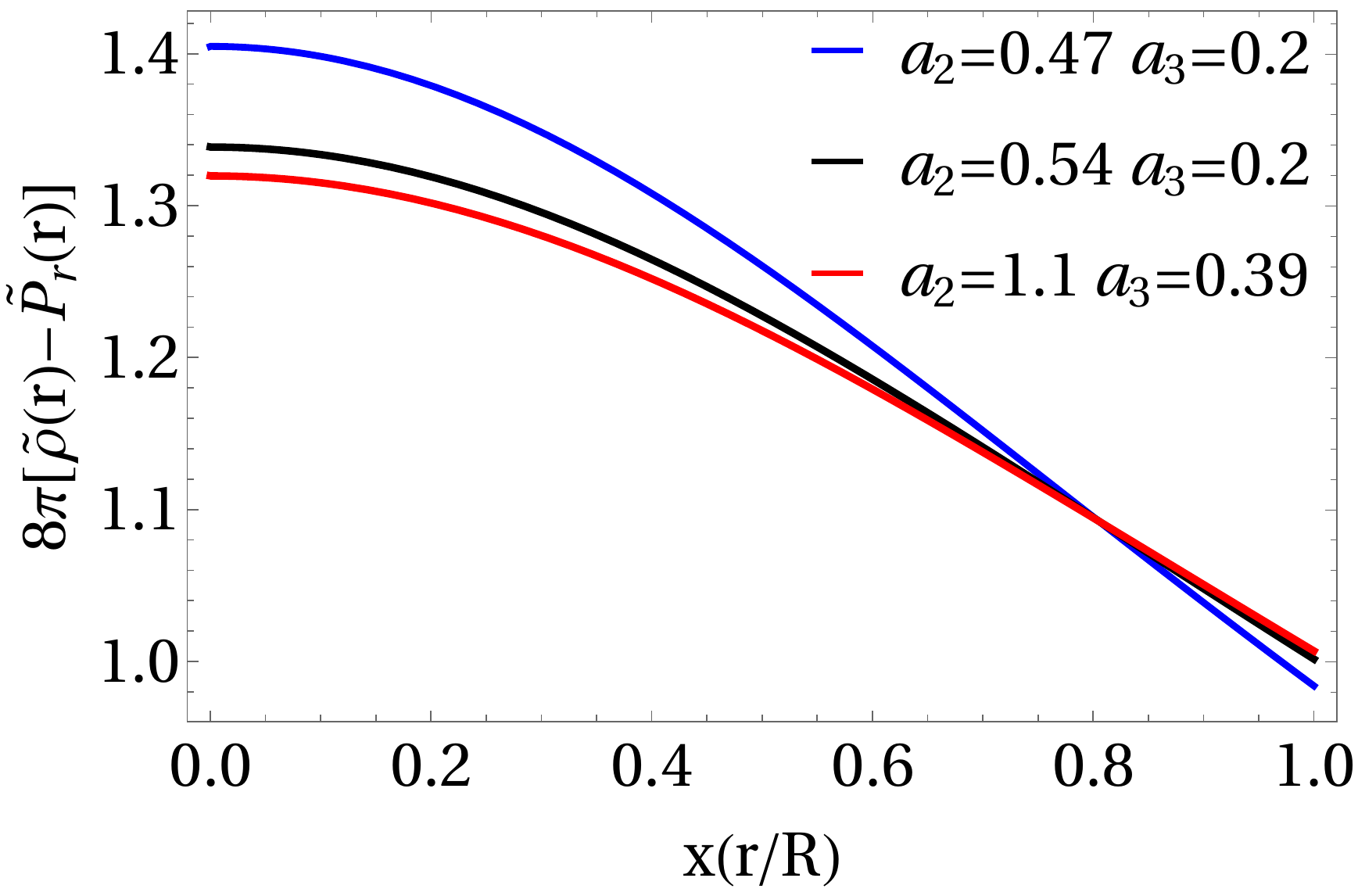}
        \label{decr4a}}
    \subfigure[]{
        \includegraphics[scale=0.223]{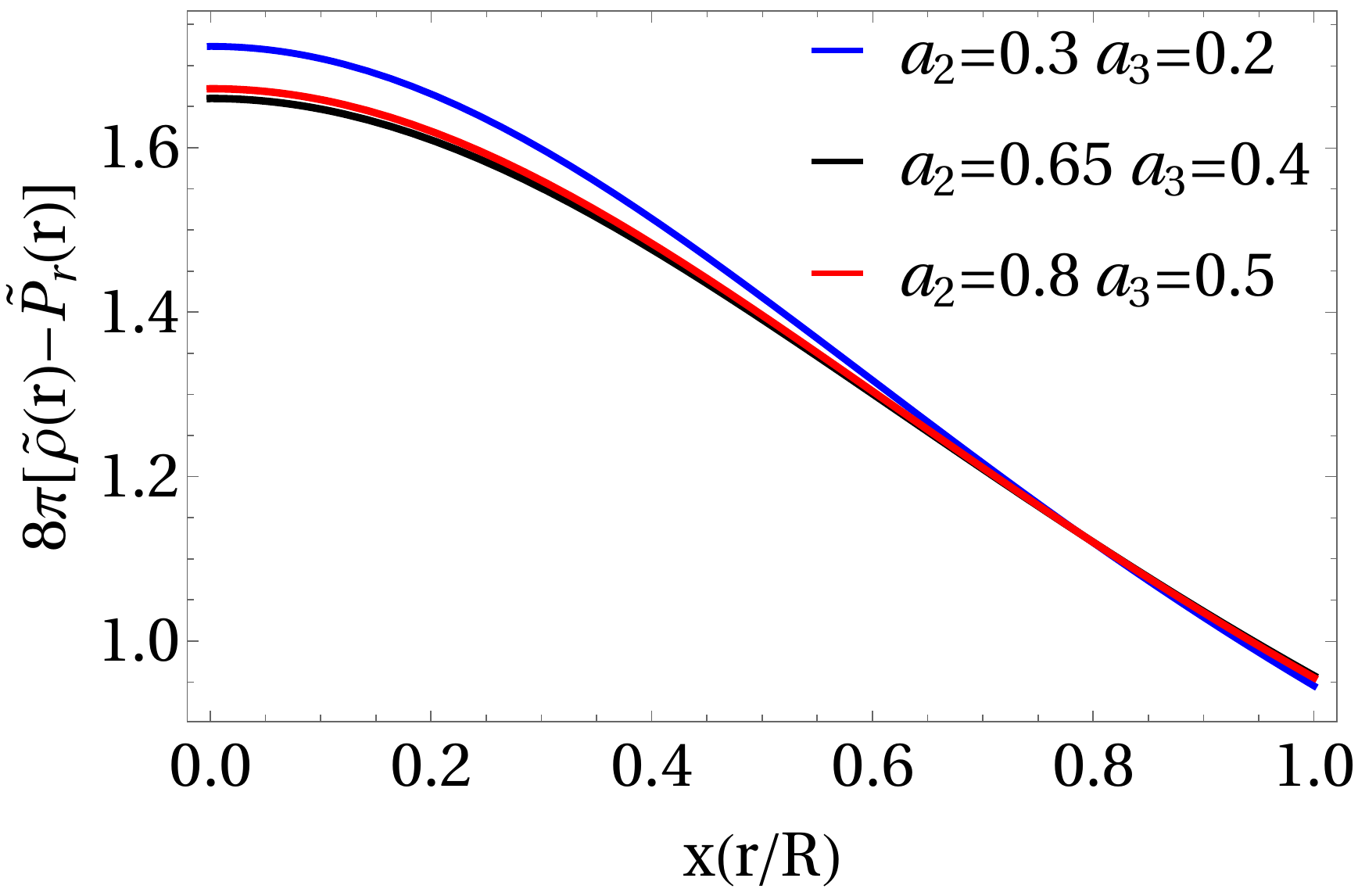}
        \label{decr4b}}                
    \caption{$\tilde{\rho}(r)-\tilde{P}_{r}(r)$ as a function of $r$ for} Model 1: (a) u = 0.19803, (b)  u = 0.2035,  Model 2: (c) u = 0.19803, (d) u = 0.2035,  Model 3: (e) u = 0.19803, (f)  u = 0.2035,  Model 4: (g) u = 0.19803, (h)  u = 0.2035.
    \label{dec1}
  \end{center}
\end{figure}

\begin{figure}[ht!]
  \begin{center}
    \subfigure[]{
        \includegraphics[scale=0.223]{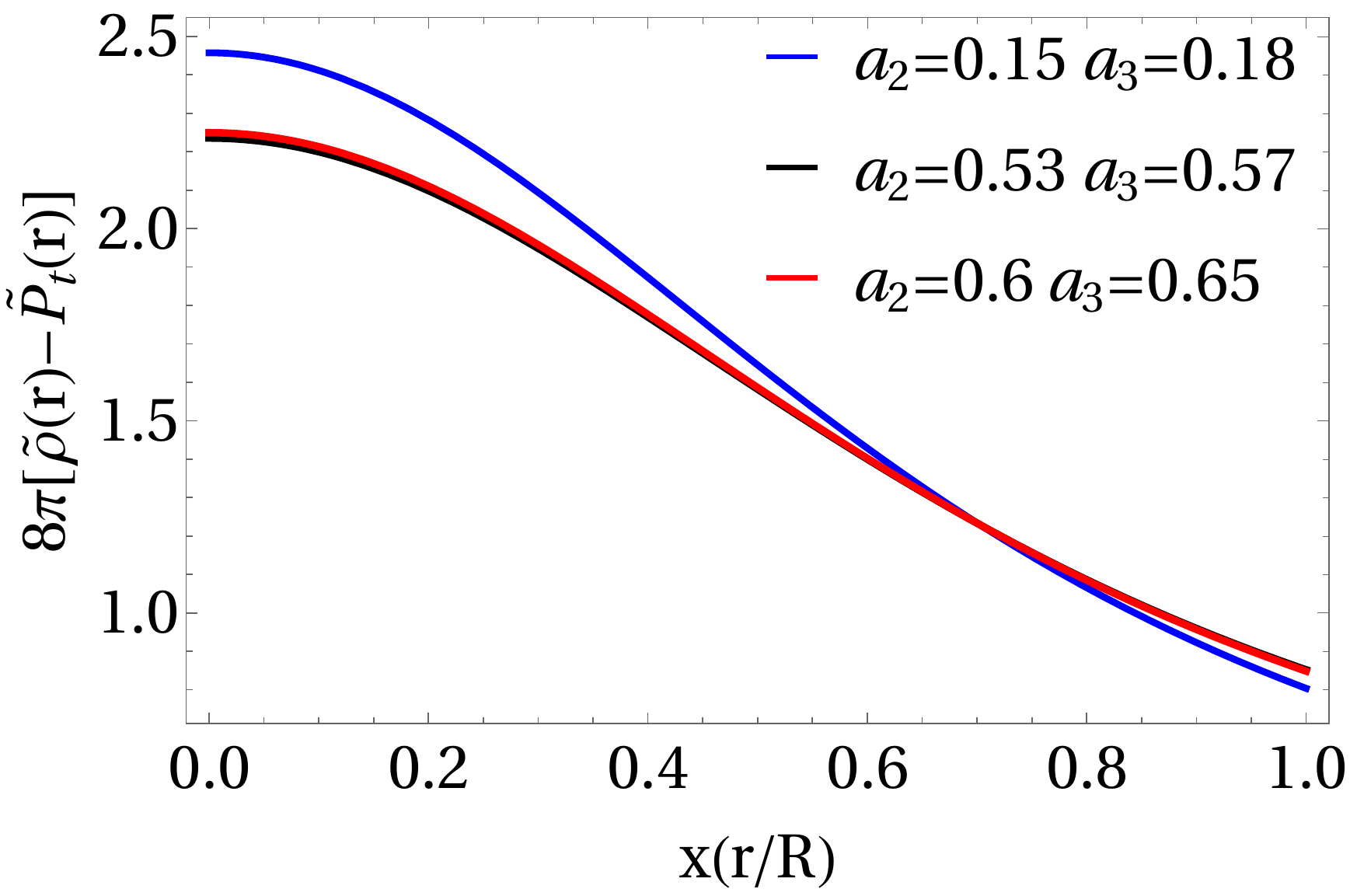}
        \label{dect1a}}
    \subfigure[]{
        \includegraphics[scale=0.223]{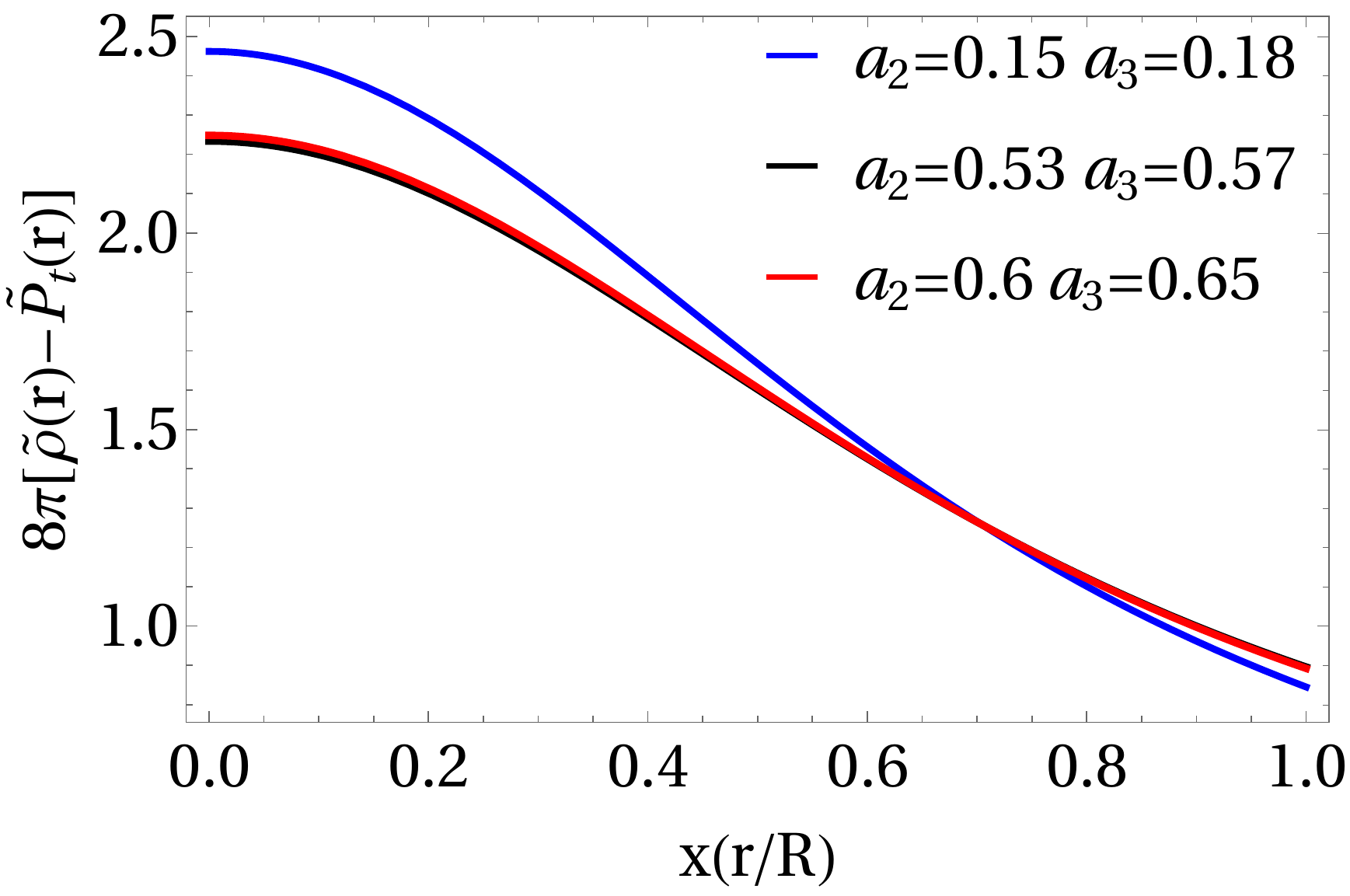}
        \label{dect1b}}
    \subfigure[]{
        \includegraphics[scale=0.223]{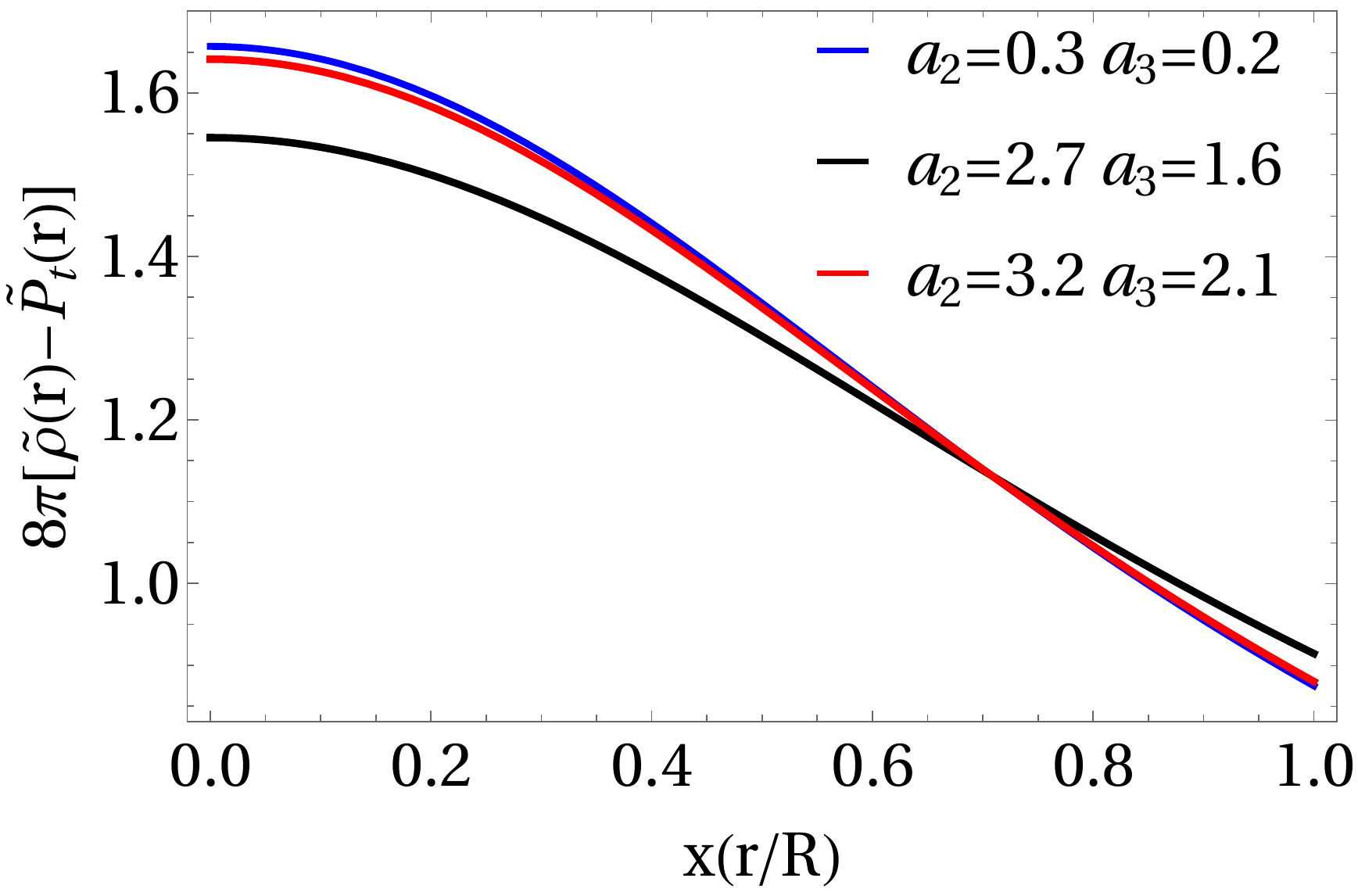}
        \label{dect2a}}
    \subfigure[]{
        \includegraphics[scale=0.223]{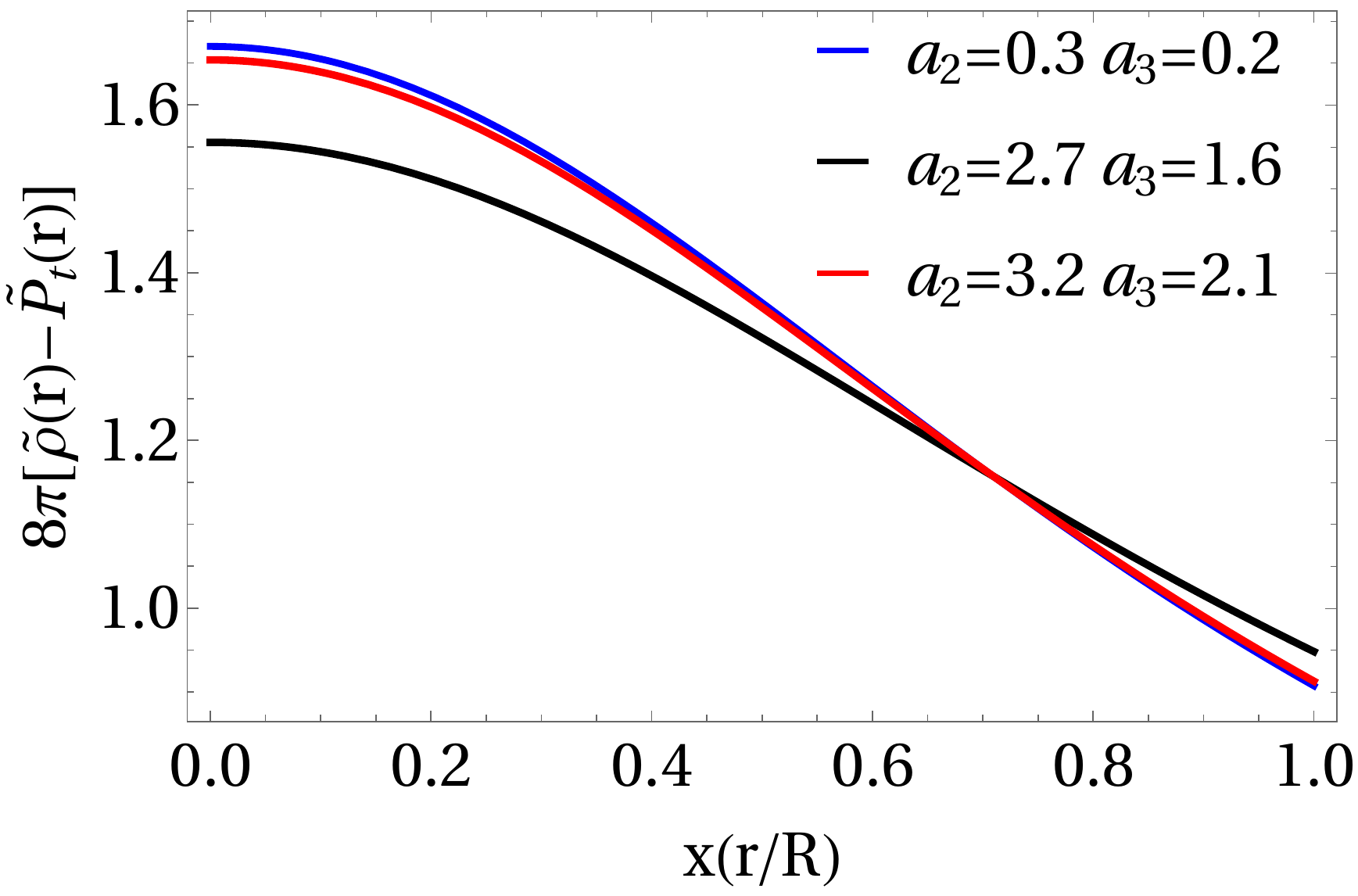}
        \label{dect2b}}
    \subfigure[]{
        \includegraphics[scale=0.223]{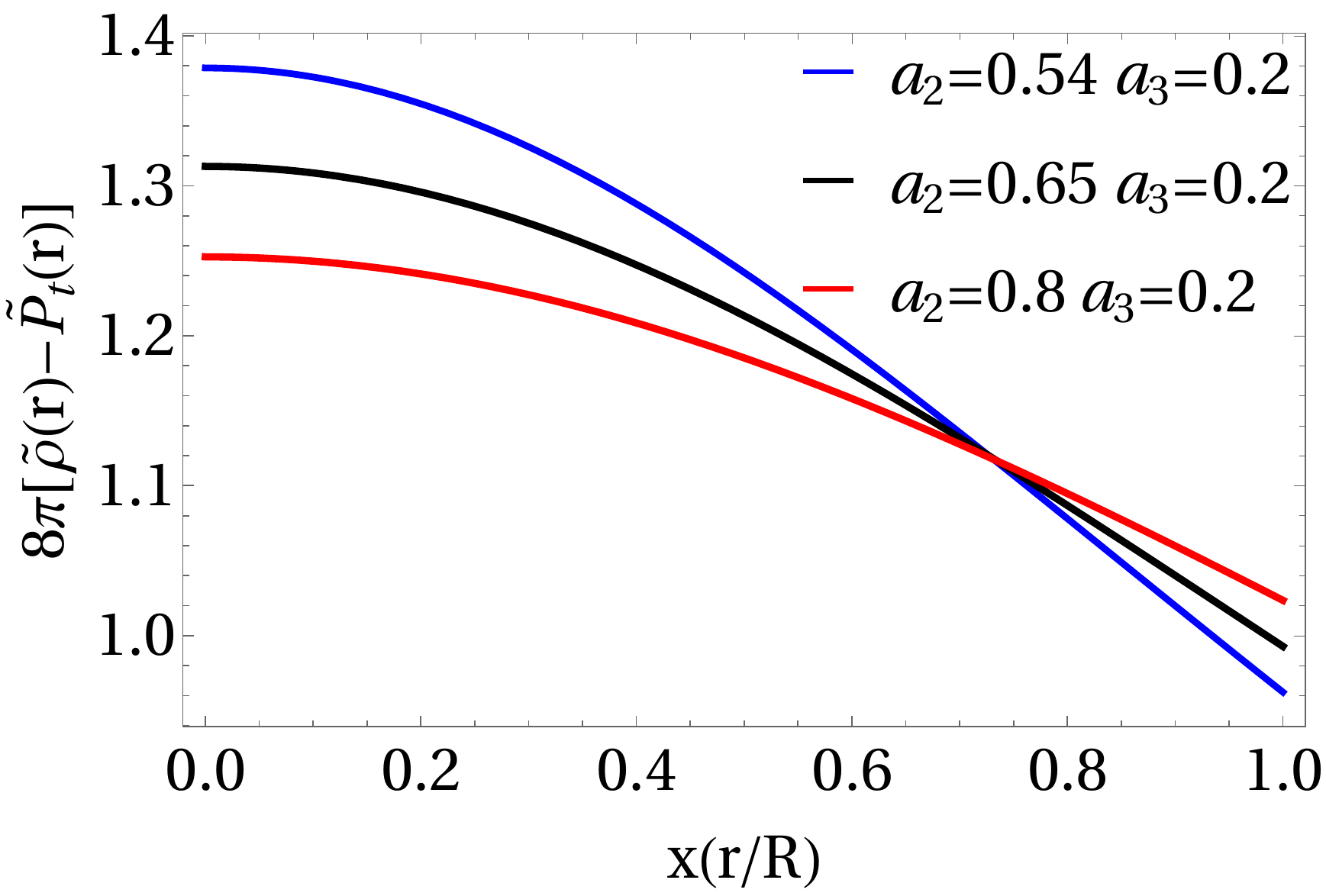}
        \label{dect3a}}
    \subfigure[]{
        \includegraphics[scale=0.223]{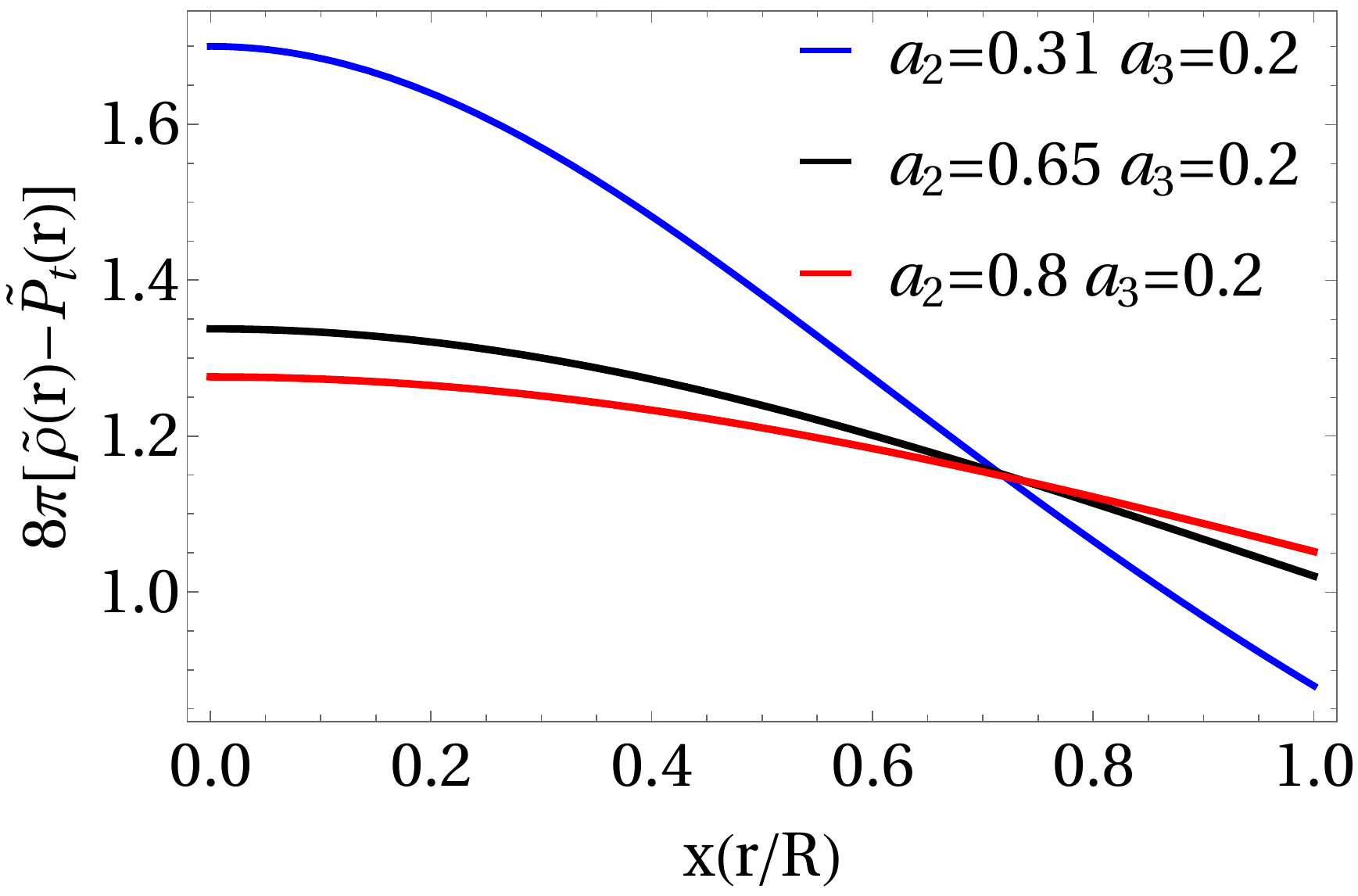}
        \label{dect3b}}
    \subfigure[]{
        \includegraphics[scale=0.223]{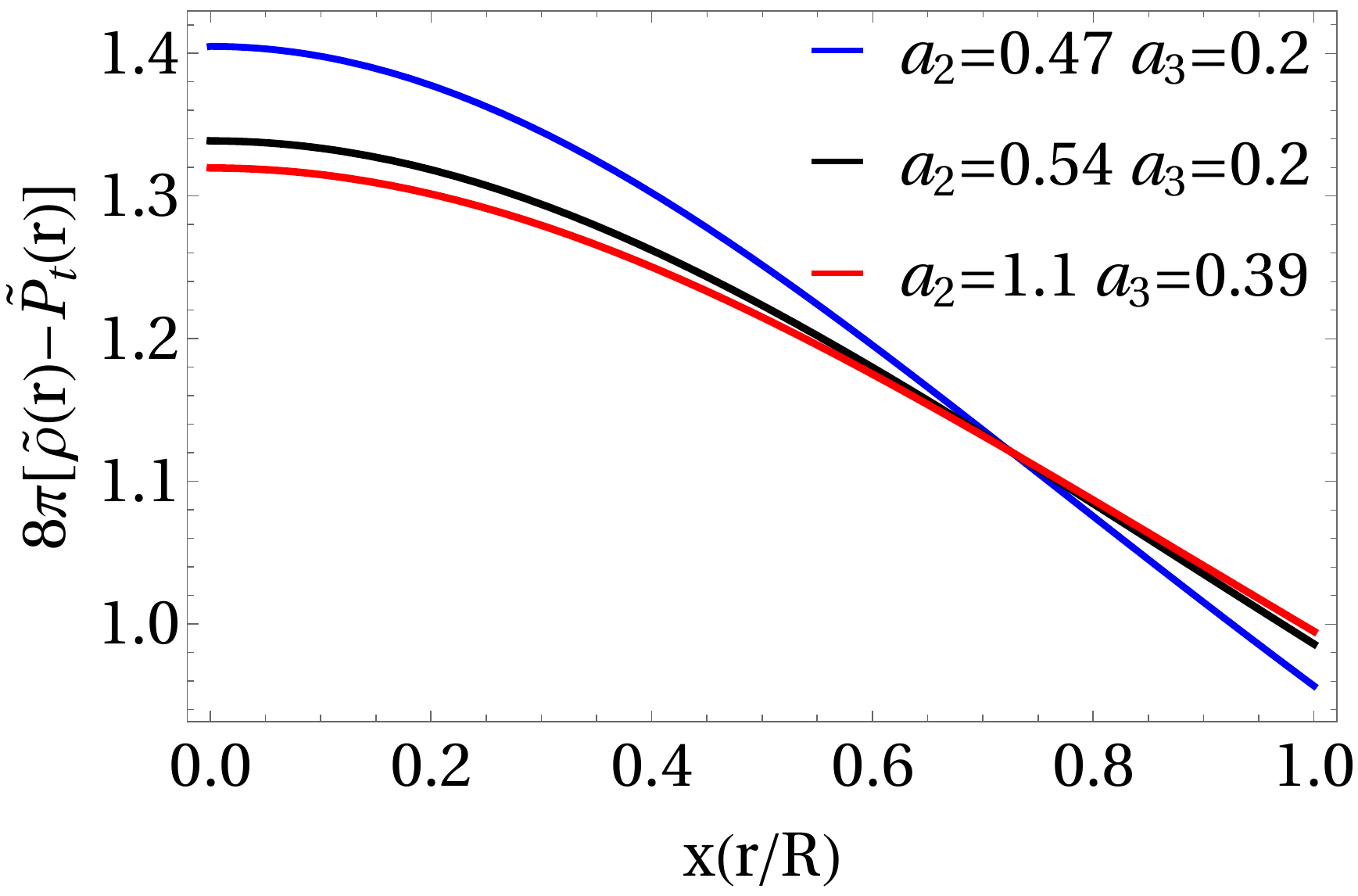}
        \label{dect4a}}
    \subfigure[]{
        \includegraphics[scale=0.223]{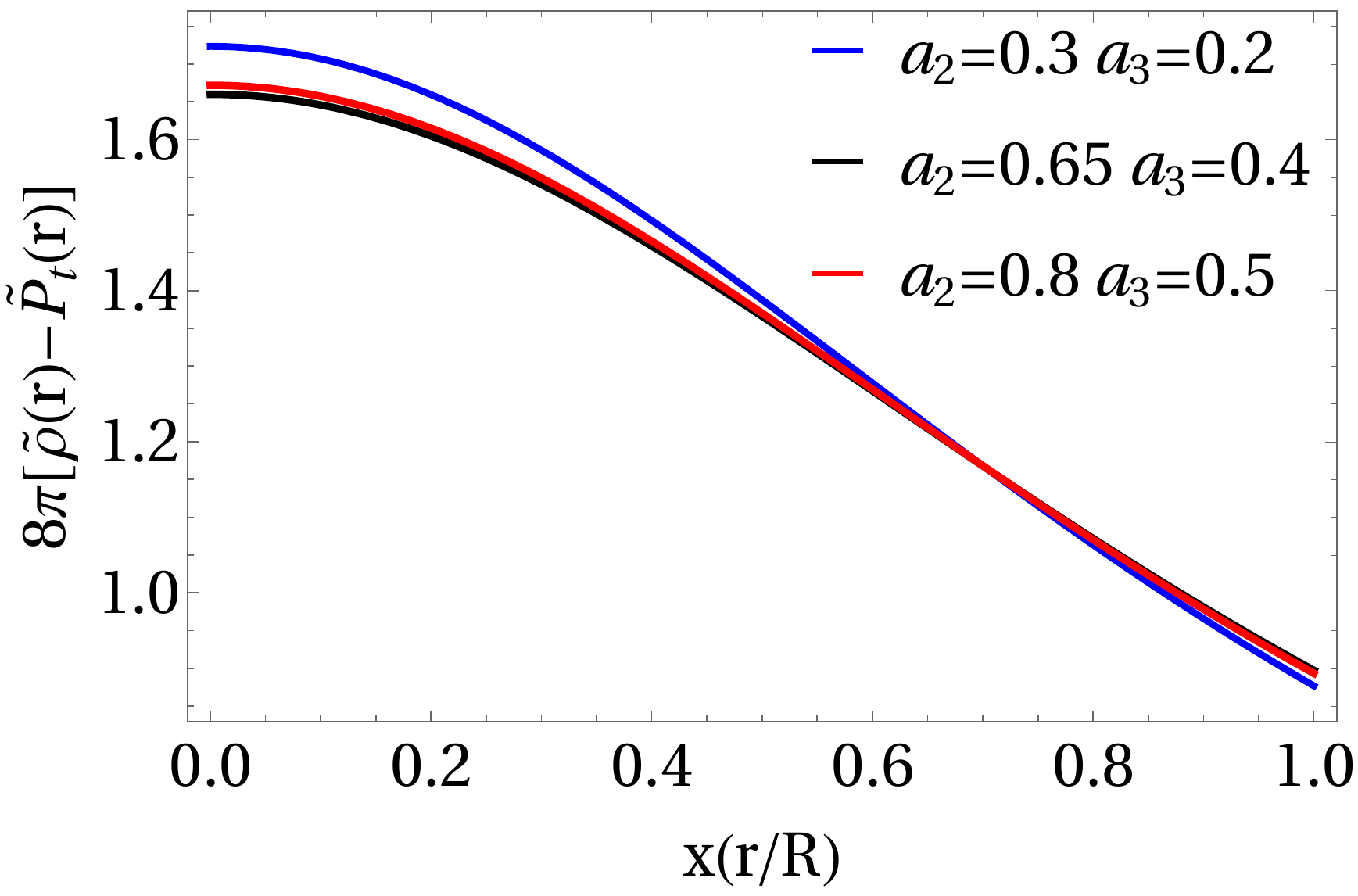}
        \label{dect4b}}                
    \caption{$\tilde{\rho}(r)-\tilde{P}_{t}(r)$ as a function of $r$ for Model 1: (a) u = 0.19803, (b)  u = 0.2035,  Model 2: (c) u = 0.19803, (d)  u = 0.2035,  Model 3: (e) u = 0.19803, (f)  u = 0.2035,  Model 4: (g) u = 0.19803, (h)  u = 0.2035.}
    \label{dec2}
  \end{center}
\end{figure}

In Figs. \ref{dec1} and \ref{dec2} it can be seen that all the solutions satisfy DEC for all the parameters involved.

\begin{figure}[ht!]
  \begin{center}
    \subfigure[]{
        \includegraphics[scale=0.223]{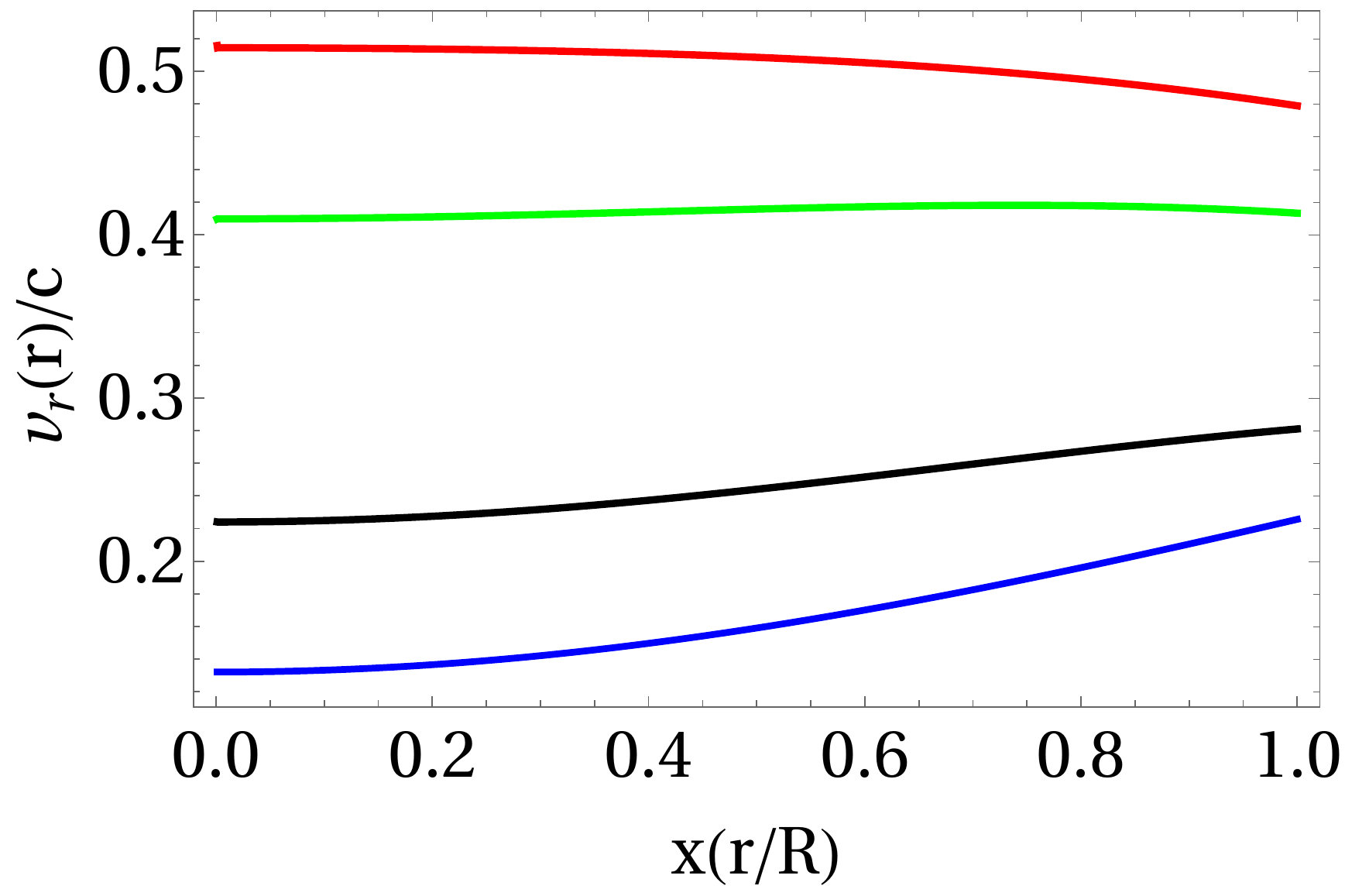}
        \label{causal1a}}
    \subfigure[]{
        \includegraphics[scale=0.223]{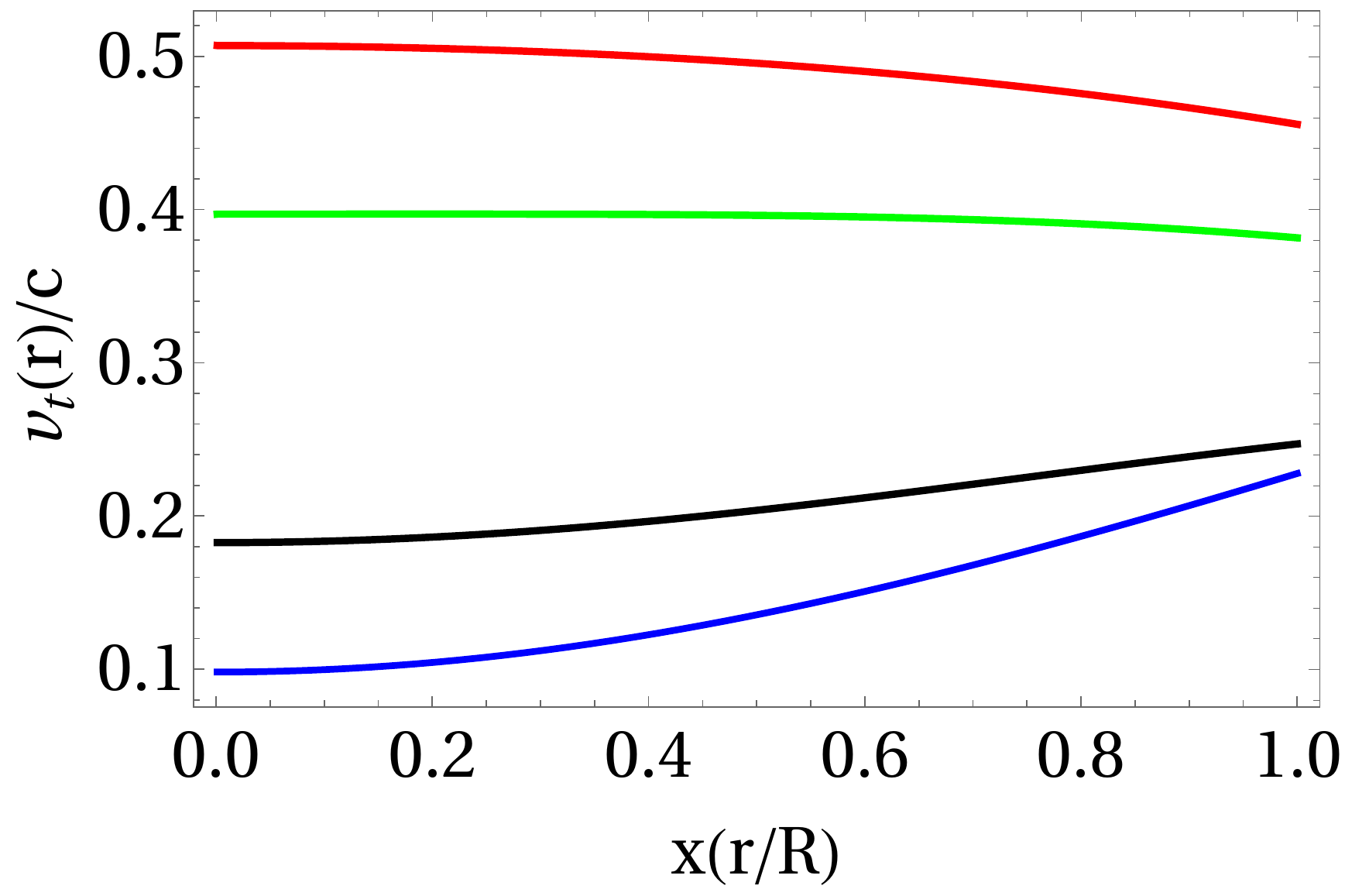}
        \label{causal1b}}
    \caption{Sound velocities as a function of $r$ for compactness factor $u = 0.19803$: (a) Radial velocity $v_{r}$ and (b) Tangential velocity $v_{t}$. Models 1, 2, 3 and 4 are identified with blue, black, red  and green line respectively. }
    \label{causal1}
  \end{center}
\end{figure}

\begin{figure}[ht!]
  \begin{center}
    \subfigure[]{
        \includegraphics[scale=0.223]{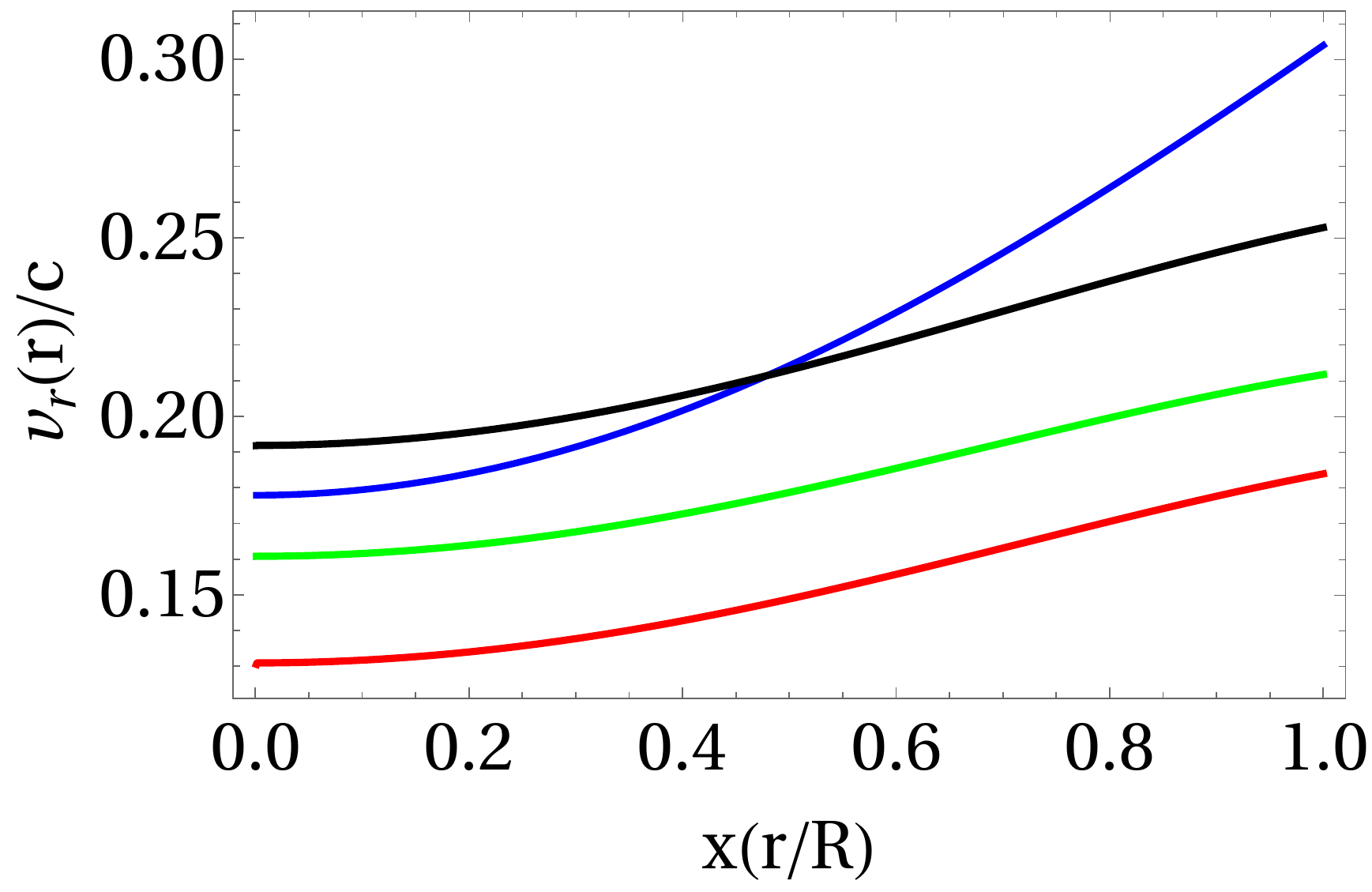}
        \label{causal2a}}
    \subfigure[]{
        \includegraphics[scale=0.223]{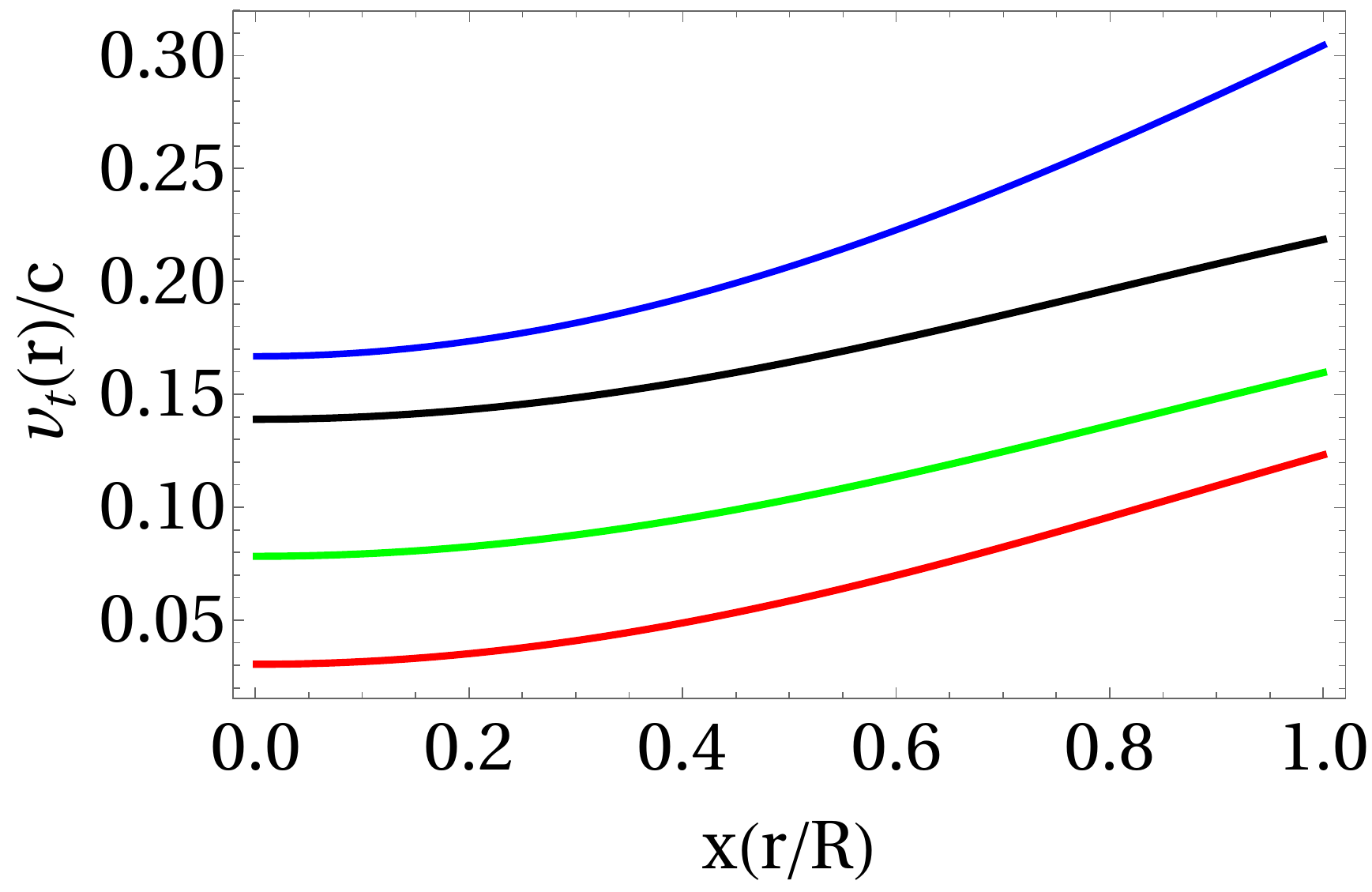}
        \label{causal2b}}
    \caption{Sound velocities as a function of $r$ for compactness factor $u = 0.2035$: (a) Radial velocity $v_{r}$ and (b) Tangential $v_{t}$. Models 1, 2, 3 and 4 are identified with blue, black, red  and green line respectively.}
    \label{causal2}
  \end{center}
\end{figure}

In figures \ref{causal1} and \ref{causal2} we show that the radial and tangential sound velocities are less than unity, as required (we are assuming $c=1$). 

\subsection{Redshift and density ratio}
In the previous section we have demonstrated that, based on the compactness parameter of both SMC X-1 and Cen X-3 in Table \ref{table1}, the four models satisfy the basic physical requirements to be considered as suitable interior configurations. Now, with the aim to to explore which model is more adequate  to describe the compact objects under consideration, in this section
we go a step further and study the redshift $Z=e^{-\nu/2}-1$ and the density ratio $\tilde{\rho}(0)/\tilde{\rho}(R)$ to each model and compare our results with the  values in Table \ref{table1}.

\begin{figure}[ht!]
  \begin{center}
    \subfigure[]{
        \includegraphics[scale=0.223]{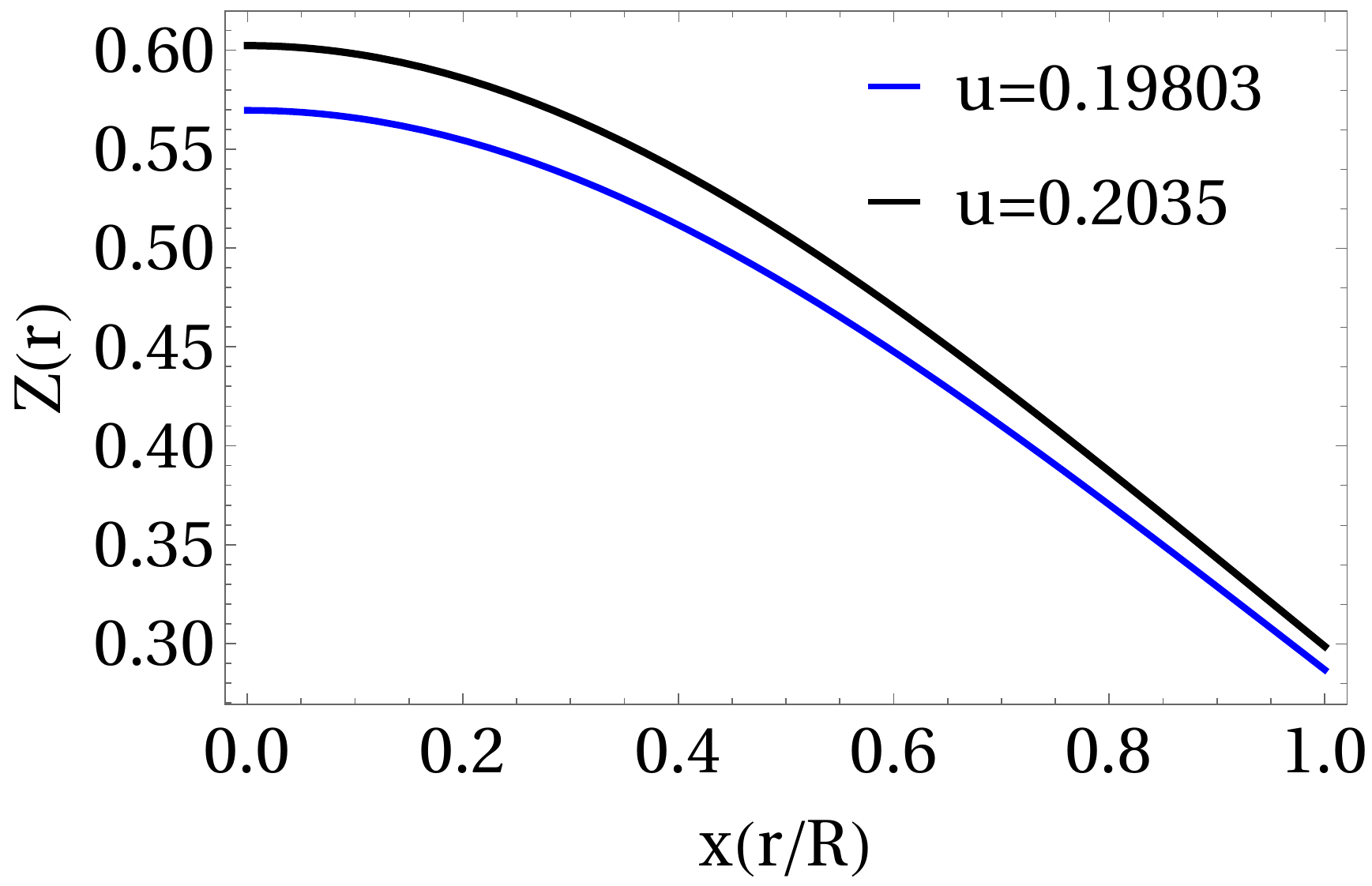}
        \label{redshift1}}
    \subfigure[]{
        \includegraphics[scale=0.223]{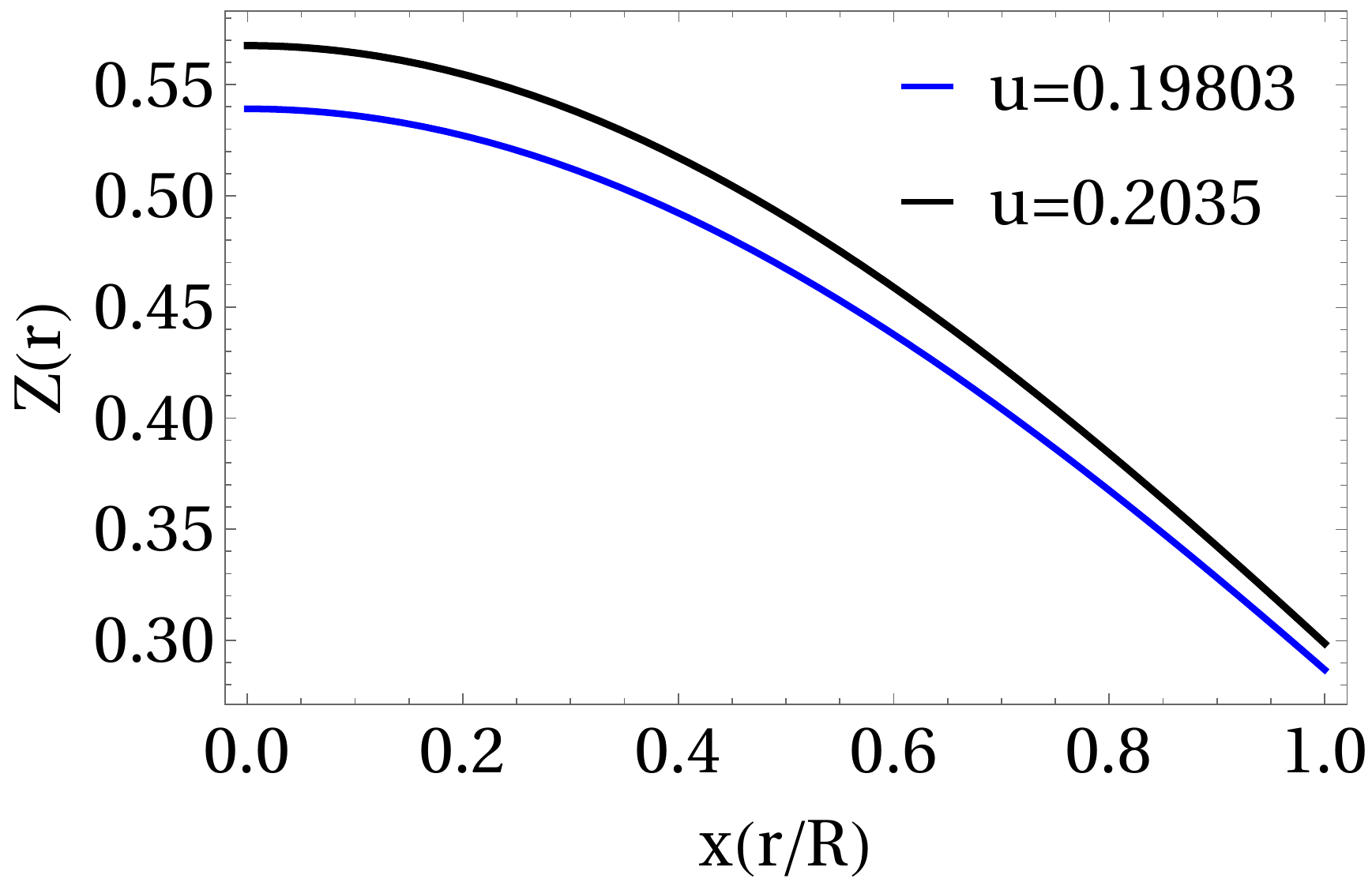}
        \label{redshift2}}
    \subfigure[]{
        \includegraphics[scale=0.223]{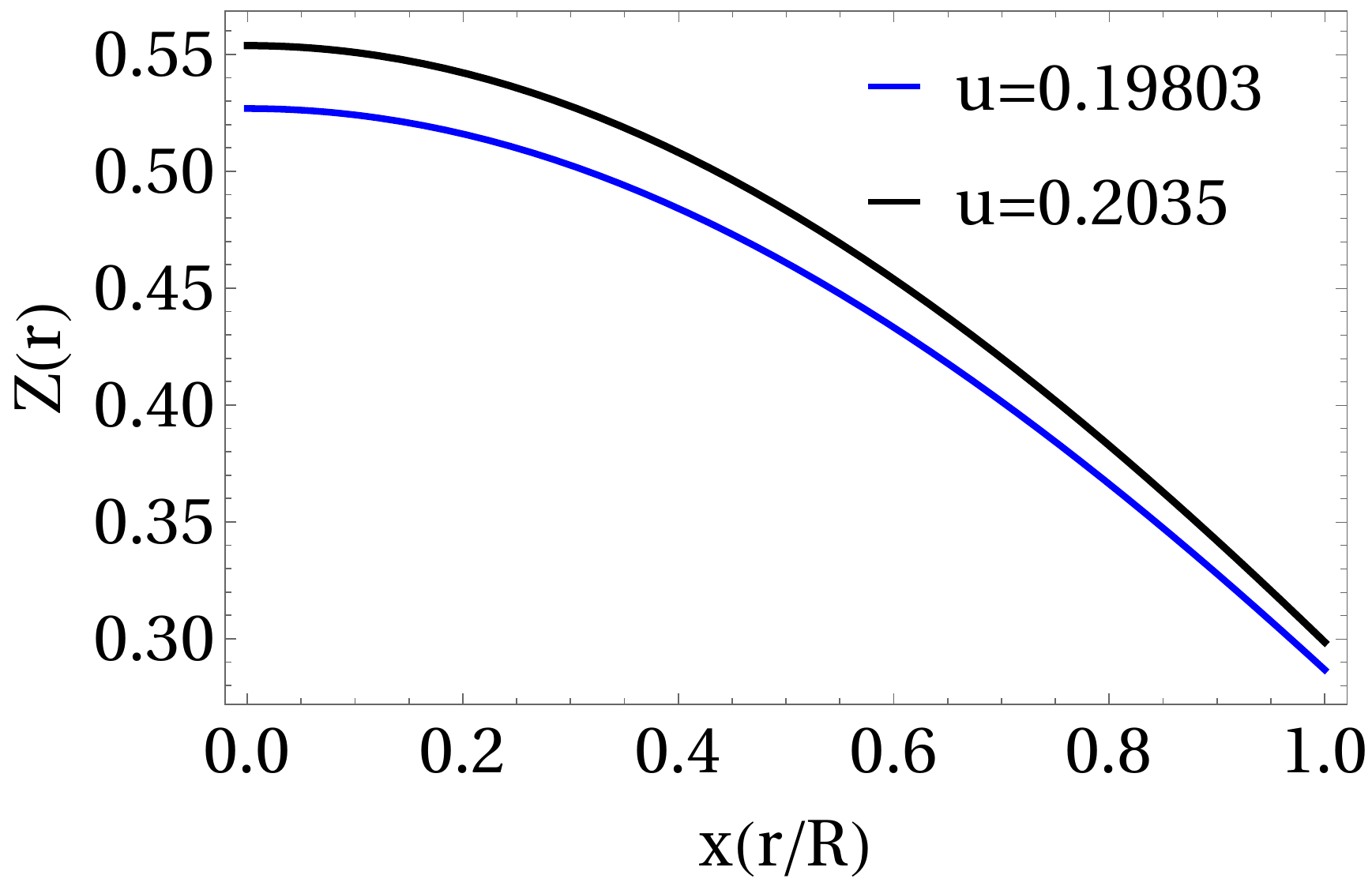}
        \label{redshift3}}
    \subfigure[]{
        \includegraphics[scale=0.223]{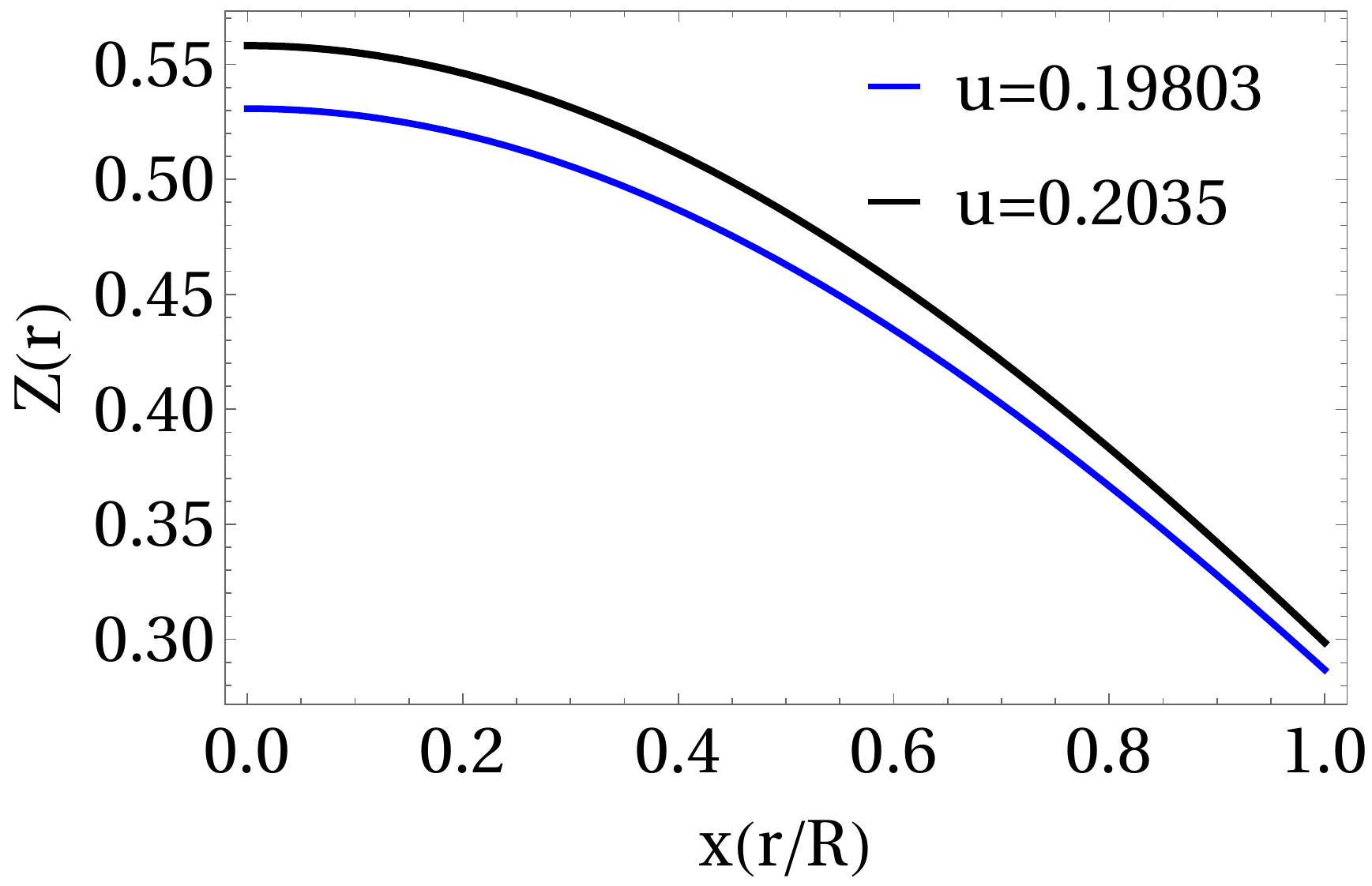}
        \label{redshift4}}
    \caption{$Z$ for Model 1 (a) Model 2 (b) Model 3 (c) and Model 4 (d)}
    \label{redshift}
  \end{center}
\end{figure}

In figure \ref{redshift} we show the redshift $Z$ as a function of the radial coordinate. Note that $Z$ decreases outward and its value at the surface is less than the universal bound for solutions satisfying the DEC, namely $Z_{bound}=5.211$.
\begin{table*}[htb!]
\small
\centering
\begin{tabular}{ccc}
\hline
Model & $\rho(0)/\rho(R)$ \\ \hline
Model 1 ($a_2=0.53$, $a_3=0.57$) & 2.62443 \\         
Model 2 ($a_2=2.7$, $a_3=1.6$) & 1.84825 \\
Model 3 ($a_2=0.8$, $a_3=0.2$) & 1.43124 \\         
Model 4 ($a_2=1.1$, $a_3=0.39$) & 1.53133 \\         
\hline         
\end{tabular}
\caption{Estimated values of the density ratio for SMC X-1 (u = 0.19803).}
\label{table2}
\end{table*}

\begin{table*}[htb!]
\small
\centering
\begin{tabular}{ccc}
\hline
Model & $\rho(0)/\rho(R)$ \\ \hline
Model 1 ($a_2=0.53$, $a_3=0.57$) & 2.49405 \\         
Model 2 ($a_2=0.3$, $a_3=0.2$)  & 1.97267 \\         
Model 3 ($a_2=0.31$, $a_3=0.2$) & 1.92661 \\         
Model 4 ($a_2=0.8$, $a_3=0.5$)  & 1.92086 \\        
\hline         
\end{tabular}
\caption{Estimated values of the density ratio for Cen X-3 (u = 0.2035).}
\label{table3}
\end{table*}

The values of the density ratio for SMC X-1 reported in \cite{maurya2018exact}
is $\tilde{\rho}(0)/\tilde{\rho}(R)\approx1.4659$. Now, from 
Table \ref{table2}, we appreciate that Models 3 and 4 fit accurately to SMC X-1. Similarly, $\tilde{\rho}(0)/\tilde{\rho}(R)\approx1.915$ for Cen X-3 as appears in \cite{prasad2019relativistic} so that Models 2, 3 and 4 are good candidates to describe this compact objects.

In summary, Models 3 and 4 might be considered as suitable solutions describing SMC X-1 while models 2, 3 and 4 are the solutions for Cen X-3.

\section{Conclusions}
In this work, we extended four isotropic models to aniso-\break tropic domains by Gravitational Decoupling. As a supplementary condition, we used a complexity factor which corresponds to a generalization of the obtained from the well--known Tolman IV solution. We verify the basic physical acceptability conditions; namely: i) the metric functions are regular at the origin. Furthermore $g_{tt}(0)=constant$ and $g^{rr}(0)=1$, ii) both the density energy and pressures are regular at the origin and decrease monotonously outward, iii) the solutions satisfy the dominant energy condition. All of these conditions were tested after imposing the compactness parameters of both  SMC X-1 and Cen X-3 systems. It is worth mentioning that, although all the solutions are well behaved, only some of them can be considered as suitable models for the compact objects under consideration based on the density ratio. More precisely, Models 3 and 4 are more appropriated to describe SMC X-1 while Models 2, 3 and 4 can be used for Cen X-3.\\

It should be interesting to explore the response of the system against perturbation. However, this and other points are under active consideration to future works.

\section{Acknowledgements}
 E.C acknowledges Decanato de Investigaci\'on y Creatividad, USFQ, Ecuador, for continuous support.
\newpage
\section{Appendix: Auxiliary functions}\label{appendix-m2}
\begin{eqnarray*}
\zeta&=&a_2 + a_3 r^2\\
\chi&=&\ln{\left(1+\frac{a_3}{a_2}r^2\right)}\\
\end{eqnarray*}

\begin{eqnarray*}
\eta_{1}&=&6 a_1 a_{2}^3 C^3 r^2+a_{2}^2 a_3 C \left(3 a_1 C r^2 \beta_{2}(r)+8 a_{3}^2\right)\nonumber\\
&& + a_2 a_{3}^2 C r^2 \left(8 a_{3}^2-a_1 \left(C^2 r^4 + 6\beta_{3}(r)\right)\right)\nonumber\\
&&-2 a_1 a_{3}^3 r^2\\
\eta_{2}&=&a_{2}^2 B \left(3 a_{3}^2-2 a_1 B r^2\right)-a_1 a_{3}^2 r^2\nonumber\\
&&\hspace{1.0cm} +a_2 a_3 B r^2 \left(3 a_{3}^2-a_1 \gamma_{2}(r)\right),
\end{eqnarray*}

\begin{eqnarray*}
\varrho_{1}&=& a_3 A\left(3 a_2 + a_3 r^2\right) + a_2 B \left(a_2 - a_3 r^2\right)\\
\varrho_{2}&=&8 a_2 a_{3}^3 C^2 \left(C r^2 \beta_{5}(r)+6\right) \zeta(r)^2+a_1 \mathcal{S}_{1}(r)\\
\varrho_{3}&=&3 a_2 a_{3}^2 B^2 \left(B r^2+3\right) \zeta(r)^2 + a_1 \mathcal{S}_{2}(r)
\end{eqnarray*}

\begin{eqnarray*}
\mathcal{P}_{1}&=&a_2 B^2 \left(5 a_1 r^2 + 8 a_3 \zeta(r)\right)- a_1 A B \left(a_2-5 a_3 r^2\right)\nonumber\\
&& -a_1 a_3 A^2\\
\mathcal{P}_{2}&=& 4 a_1 a_3 A B r^2 \left(2 a_2+a_3 r^2\right)\nonumber\\
&& + a_2 B^2 \bigg(  4 a_2 r^2 \left(a_1+4 a_{3}^2\right)+a_3 r^4 \left(a_1+8 a_{3}^2\right)\nonumber\\
&&\hspace{1.5cm} + 8 a_{2}^2 a_3 \bigg) - a_1 a_2 a_3 A^2.\\
\mathcal{P}_{3}&=&6 a_1 a_2 C\beta_{8}(r)(a_3-a_2 C)^2 \zeta(r)\chi(r)\nonumber\\
&& + 8 a_2 a_{3}^4 C \beta_{1}^3 \zeta(r) + a_3\beta_{8}(r)\bigg[-6 a_1 a_{2}^3 C^3 r^2\nonumber\\
&& + a_{2}^2 a_3 C \left(-3 a_1 C r^2 \beta_{9}(r)- 8 a_{3}^2 \right) \nonumber\\
&& + a_2 a_{3}^2 C r^2\left(a_1 \left(C r^2 \beta_{10}(r)-6\right)-8 a_{3}^2 \right)\nonumber\\ 
&& + 2 a_1 a_{3}^3 r^2\bigg],\\
\mathcal{P}_{4}&=&-18 a_1 a_2 C^2 (a_3-a_2 C)^2 \zeta(r)^2\chi(r)\nonumber\\
&& + 24 a_2 a_{3}^4 C^2 \chi(r)^2 + 6 a_{2}^2 a_{3}^2 C^2 r^2 \left(C r^2 \beta_{7}(r)+3\right)\nonumber\\
&& + a_1 a_3 \bigg[18 a_{2}^4 C^4 r^2+9 a_{2}^3 a_3 C^3 r^2 \left(3 C r^2-4\right)\nonumber\\
&& - a_2 a_{3}^3 \left(8 C^4 r^8+34 C^3 r^6+14 C r^2+1\right)\nonumber\\
&& \hspace{3.0cm}-6 a_{3}^4 C r^4\bigg].
\end{eqnarray*}
\begin{eqnarray*}
\mathcal{P}_{5}&=& a_3 r^2 \bigg[ a_1\gamma_{6}(r) \left(2 a_{2}^2 B^2+a_2 a_3 B \gamma_{2}(r)+a_{3}^2\right)\nonumber\\
&&\hspace{2.0cm} + 3 a_2 a_{3}^2 B^2 \gamma_{7}(r) \zeta(r)\bigg]\nonumber\\
&&\hspace{1.0cm} - 2 a_1 a_2 B \gamma_{6}(r)(a_2 B-a_3)\zeta(r)\chi(r).
\end{eqnarray*}
\begin{eqnarray*}
\mathcal{P}_{6}&=& 15 a_2 a_{3}^3 B^2 \zeta(r)^2+10 a_1 a_2 B^2 (a_2 B-a_3) \zeta(r)^2 \chi(r)\nonumber\\
&& + a_1 a_3\bigg[ -10 a_{2}^3 B^3 r^2+5 a_{2}^2 a_3 B^2 r^2 \gamma_{8}(r)-5 a_{3}^3 B r^4\nonumber\\
&& + a_2 a_{3}^2 \left(B r^2 \left(-9 B^2 r^4+B r^2-11\right)-1\right)\bigg],
\end{eqnarray*}

\begin{eqnarray*}
\mathcal{S}_{1}&=& 6 a_{2}^4 C^3 \beta_{4}(r)+3 a_{2}^3 a_3 C^2 \left(3 C r^2 \beta_{6}(r)+4\right)\nonumber\\
&& + 2 a_{2}^2 a_{3}^2 C \beta_{4}(r)\left(C r^2 \beta_{7}(r)+3\right) -2 a_{3}^4 r^2 \beta_{4}(r)\nonumber\\
&& + a_2 a_{3}^3 \left(C r^2 \left(C r^2 \left(C r^2 \beta_{3}(r) + 48\right)+4\right)+6\right)\\
\mathcal{S}_{2}&=&-2 a_{2}^3 B^2 \gamma_{3}(r)+a_{2}^2 a_3 B \left(B r^2 \gamma_{4}(r)-2\right)\nonumber\\
&& + a_2 a_{3}^2 \left(B r^2 \left(B r^2 \gamma_{5}(r)+3\right)+ 3 \right)\nonumber\\
&&\hspace{4.5cm} -a_{3}^3 r^2\gamma_{3}(r)
\end{eqnarray*}

\begin{eqnarray}
\beta_{1}&=&Cr^2+1\hspace{0.5cm} \beta_{2}=Cr^2-4 \hspace{0.5cm}
\beta_{3}=Cr^2-1\nonumber\\
\beta_{4}&=&3Cr^2-1 \hspace{0.5cm}
\beta_{5}=Cr^2+3\hspace{0.5cm}
\beta_{6}=3Cr^2-5\nonumber\\
\beta_{7}&=&Cr^2-9\hspace{0.5cm}
\beta_{8}=9Cr^2+1\hspace{0.5cm}
\beta_{9}=Cr^2-4\nonumber\\ 
\beta_{10}&=&Cr^2+6\nonumber\\
\beta_{11}&=&-6 a_{2}^3 C^3-3 a_{2}^2 a_3 C^2 \beta_{9}+ 2 a_{3}^3\nonumber\\
&&\hspace{2.0cm} +a_2 a_{3}^2 C \left(C r^2 \beta_{10}-6\right),\nonumber
\end{eqnarray}

\begin{eqnarray*}
\gamma_{1}&=&B r^2+1\hspace{0.5cm} 
\gamma_{2}= Br^2-2\hspace{0.5cm}
\gamma_{3}= Br^2-1\\ 
\gamma_{4}&=&5-3Br^2\hspace{0.5cm}
\gamma_{5}= Br^2+9\hspace{0.5cm}
\gamma_{6}=7Br^2+1\\
\gamma_{7}&=&Br^2-5\hspace{0.5cm} 
\gamma_{8}= 2-3Br^2.\\ 
\gamma_{9}&=&2 a_{2}^2 B^2+a_2 a_3 B \gamma_{2}+a_{3}^2.
\end{eqnarray*}

\end{document}